\documentclass[
reprint,
superscriptaddress,
amsmath,amssymb,
aps,
prb,
]{revtex4-2}

\usepackage{graphicx}
\usepackage{dcolumn}
\usepackage{bm}
\usepackage[colorlinks = true,
            linkcolor = blue,
            urlcolor  = blue,
            citecolor = blue,
            anchorcolor = blue]{hyperref}
\usepackage{hyperref}

\usepackage[dvipsnames]{xcolor}

\usepackage{color}
\begin{document}

\title[]{\large Linear and nonlinear vibrational excitation driven by molecular polaritons}

\author{Wenxiang Ying}
\email{wying3@sas.upenn.edu}
\affiliation{Department of Chemistry, University of Pennsylvania, Philadelphia, Pennsylvania 19104, USA}

\author{Carlos M. Bustamante}
\author{Franco P. Bonafé}
\affiliation{Max Planck Institute for the Structure and Dynamics of Matter and Center for Free-Electron Laser Science, Luruper Chaussee 149, Hamburg 22761, Germany}

\author{Richard Richardson}
\affiliation{Department of Physics, Arizona State University, Tempe, Arizona 85287, United States}

\author{Michael Ruggenthaler}
\affiliation{Max Planck Institute for the Structure and Dynamics of Matter and Center for Free-Electron Laser Science, Luruper Chaussee 149, Hamburg 22761, Germany}

\author{Maxim Sukharev}
\email{Maxim.Sukharev@asu.edu}
\affiliation{Department of Physics, Arizona State University, Tempe, Arizona 85287, United States}
\affiliation{College of Integrative Sciences and Arts, Arizona State University, Mesa, Arizona 85212, United States}

\author{Angel Rubio}
\email{angel.rubio@mpsd.mpg.de}
\affiliation{Max Planck Institute for the Structure and Dynamics of Matter and Center for Free-Electron Laser Science, Luruper Chaussee 149, Hamburg 22761, Germany}
\affiliation{Initiative for Computational Catalysis (ICC), Flatiron Institute, Simons Foundation,  162 5th Avenue, New York, NY 10010 USA}

\author{Abraham Nitzan}
\email{anitzan@sas.upenn.edu}
\affiliation{Department of Chemistry, University of Pennsylvania, Philadelphia, Pennsylvania 19104, USA}
\affiliation{School of Chemistry, Tel Aviv University, Tel Aviv 69978, Israel}


\begin{abstract}
Following our recent numerical study [arXiv:2601.16299 (2026)], we investigate vibrational excitation induced by transient optical driving in molecular ensembles strongly coupled to a cavity mode using the field-driven Holstein--Tavis--Cummings model. We analyze how pulsed excitation redistributes energy among electronic, photonic, and vibrational degrees of freedom in molecular polaritons. Vibrational dynamics are examined over a broad range of pulse durations and intensities within both the single-excitation approximation and a mean-field description of collective light--matter coupling. Despite their distinct formulations and microscopic descriptions, these two approaches yield consistent scaling relations for vibrational excitation. 
In particular, we disentangle linear and nonlinear contributions to vibrational excitation, which are reflected in distinct quadratic and quartic scaling behaviors with respect to the driving field amplitude (that is, linear and quadratic dependence on the incident pulse intensity). The microscopic origin of the nonlinear component is identified as a polariton-mediated intrapulse stimulated Raman-like process, enabled by a pulse spectral bandwidth large enough to overlap both upper and lower polaritons (rather than a conventional multi-pulse scheme). These results establish a unified framework for understanding vibrational excitation under pulsed polariton driving and provide guidance for the interpretation and control of ultrafast polariton experiments.
Discrepancies between the mean-field and single-excitation approaches under certain pulsed conditions are identified and analyzed.
\end{abstract}

\maketitle

\section{Introduction}
Strong light--matter coupling between molecular ensembles and confined electromagnetic modes gives rise to hybrid light--matter states -- polaritons -- whose properties differ fundamentally from those of bare molecular or photonic excitations. Over the past decade, molecular polaritons formed under vibrational~\cite{Ebbesen_angew_2016, Ebbesen_science_2019, Simpkins2023, Ebbesen_nanophotonics_2020, Ebbesen_Angew_2021, Hirai_2020, verdelli_2024} or electronic strong coupling~\cite{Zeng2023JACS, Lee2024JACS, Hutchinson2012ACIE, Munkhbat2018SA, Ng2015AOM, Mony2021AFM} have attracted considerable attention due to their ability to modify energy relaxation pathways, chemical reactivity, and spectroscopy. Despite extensive theoretical~\cite{Ribeiro_JPCL2018, Li_JCP2021, Li_NC2022, Elious_2023, Mukamel_PRL2023, Mondal_JCP2025, Yuen-Zhou_PRL2025, Yuen-Zhou_NC2025, Shi_JPCL2026} and experimental efforts~\cite{Watanabe_2020, Son_2022, Xiong_PNAS2018, Xiong_SciAdv2019, Xiong_Science2025, xiong_Science2020, Xiong_Science2022, Scholes_JPCL2020, Scholes_JPCL2025}, the real-time dynamics of molecular polaritons under external driving remain challenging to describe, owing to the strongly coupled and inherently nonequilibrium nature of electronic, photonic, and vibrational degrees of freedom (DOF).

Ultrafast laser pulses provide a versatile means to selectively populate polariton states and to probe their subsequent dynamics~\cite{Scholes_JPCL2021, Son_CPR2024, Xiong_ChemRev2024}. Depending on pulse duration, spectral bandwidth, and intensity, a single optical field can access qualitatively different dynamical behaviors, ranging from state-selective excitation to broadband excitation involving multiple polaritonic branches. When molecular vibrations are included, it is natural to inquire about their role in the ensuing dynamics. Specifically, how the optical energy injected into polaritonic states is converted into molecular vibrational excitation, and how does this process depend on the molecule-radiation field interaction, particularly under strong coupling conditions in a cavity environment.

Despite extensive studies of vibrational polaritons and vibronic coupling in cavities, a systematic understanding of the way molecular vibrations affect and respond to optical driving of electronic transitions remains incomplete.
Most existing theoretical work has focused either on spontaneous relaxation following polariton preparation~\cite{Lai_JCP2024, Hu_JCP2025, Lai_JCP2026} or on steady-state pumping under continuous-wave driving~\cite{Koessler2025CS, Li_JPCL2025}. 
Note that pulsed excitation has also been considered in quantum-dynamical simulations~\cite{Elious_2023, Mukamel_PRL2023, Mondal_JCP2025, Shi_JPCL2026}; however, these studies typically adopt the impulsive (short-pulse) limit, in which the detailed temporal profile of the driving field can be neglected. 
On the other hand, most realistic experimental scenarios involve ultrafast pulse excitation, where the finite temporal and spectral extent of the driving field can introduce additional dynamical pathways that are absent in a monochromatic excitation scheme. 
A particularly subtle aspect of pulsed excitation is that a single broadband optical pulse can simultaneously supply multiple frequency components, enabling Raman-like vibrational excitation without the need for distinct pump and Stokes fields~\cite{Tahara_RSI2016, Ferrante_NC2018, Kuramochi_JACS2021}. In this case, vibrational activation arises from intrapulse frequency mixing, whereby different spectral components of the same pulse coherently couple electronic and vibrational transitions. 
While such self-induced Raman-like processes have been discussed in other contexts, their role in polariton-mediated vibrational activation, especially under weak to moderate field strengths and transient driving, has not been systematically explored.

Theoretical descriptions of polariton dynamics often rely on different approximations whose regimes of validity are not always clear under nonequilibrium conditions. Two commonly employed frameworks are the single-excitation (SE) approximation~\cite{Lai_JCP2024, Hu_JCP2025, Lai_JCP2026, Chng_2024, Ying_mn24, Chng_NL2025, Ying_NC2025}, which explicitly resolves the polariton eigenstructure within a restricted Hilbert space, and mean-field (MF) approaches~\cite{Yuen-Zhou_PRL2025, Cui_JCP2022, Cui_JCP2023, Keeling_PRL2022, Hsieh_JCP2023, Hsieh_JPCL2023, Reichman_Nanoph2024, Li_CP2025, Hsieh_JCTC2025}, which treat light--matter coupling self-consistently while factorizing higher-order correlations. Although both approaches are widely used and were shown to yield consistent results in some applications, it is an open question whether they yield consistent predictions for vibrational excitation induced by pulsed fields, particularly beyond linear response.

In this work, we present a unified theoretical investigation of linear and nonlinear vibrational activation driven by molecular polaritons~\footnote{In this paper we use the terms ``activation'' and ``excitation' interchangeably. Because relaxation processes are not included in the analysis, the consequences of coupling between the polartonic states and molecular vibrations is often displayed below as Rabi-type oscillations between excitations in the polaritonic and the vibrational sub-systems.}. Using the field-driven Holstein--Tavis--Cummings (HTC) model~\cite{Arkajit_Chemrev_2023, Ying_ARPC2026}, we analyze vibrational excitation over a broad range of pulse durations and field intensities, and explicitly compare results obtained within the SE and MF frameworks. Despite their distinct formulations and accessible observables, both approaches reveal the same underlying scaling behaviors, allowing us to disentangle linear and nonlinear pathways of vibrational activation.
We show that vibrational excitation on the excited electronic and photonic states exhibits a leading quadratic dependence on the driving field amplitude (namely linear dependence on the incident pulse intensity), while vibrational activation on the ground electronic state displays a quartic, nonlinear scaling (that is, quadratic dependence on the incident intensity). We identify the microscopic origin of this nonlinear contribution as a polariton-mediated intrapulse stimulated Raman-like process, enabled by the finite spectral bandwidth of a single optical pulse. Importantly, this mechanism operates without requiring strong-field stimulated Raman gain or multi-color excitation schemes and persists under realistic levels of cavity loss and dissipation.
Meanwhile, discrepancies between the SE and MF approaches are also observed under certain pulsed conditions. In particular, MF dynamics does not capture the polaron decoupling effect~\cite{Herrera_Spano_PRL, herrera2018theory}.

By establishing transparent scaling relations and demonstrating their consistency across different theoretical descriptions, this work provides a unified framework for understanding vibrational activation under pulsed polariton driving. These findings are directly relevant for the interpretation of ultrafast polariton experiments and suggest new opportunities for coherent control of vibrational and vibronic dynamics in strongly coupled light--matter systems.


\section{Model and Methods}
\subsection{The Field-Driven Holstein--Tavis--Cummings Hamiltonian}
We first introduce the field-driven HTC Hamiltonian with a single cavity mode~\cite{Li2021ARPC, Arkajit_Chemrev_2023, Ying_ARPC2026}. Neglecting the dipole self-energy term and adopting the long-wavelength and rotating-wave approximations (RWA), the total Hamiltonian can be written as follows
\begin{equation}\label{eq:hamgen}
    \hat{H}_\mathrm{HTC}(t) = \hat{H}_\mathrm{M} + \hat{H}_\mathrm{cav} + \hat{H}_\mathrm{LM} + \hat{H}_\text{pulse}(t).  
\end{equation}
The molecular Hamiltonian $\hat{H}_\mathrm{M}$ consists of $N$ independent molecules, each described by two electronic (excitonic) levels $|g_{n}\rangle$ and $|e_{n}\rangle$ (for molecule $n$) and one (harmonic) vibration, as well as their interactions, 
$\hat{H}_\mathrm{M} = \sum_{n = 1}^{N} \hbar \omega_{0} \hat{\sigma}_n^+ \hat{\sigma}_n^- + \hbar \nu \sum_{n} \hat{b}^\dagger_{n} \hat{b}_{n} + \sum_{n} \hat{\sigma}_n^+ \hat{\sigma}_n^- \otimes c_\nu (\hat{b}_{n} + \hat{b}^\dagger_{n})$, 
where $\hat{\sigma}_n^-$ ($\hat{\sigma}_n^+$) is the annihilation (creation) operator for the excitation of the $n_\mathrm{th}$ molecule, with $\hat{\sigma}^+_m |g_{n }\rangle = \delta_{m,n} |e_{n}\rangle$ and $\hat{\sigma}^-_m |g_{n }\rangle = 0$. $\hat{b}_{n}$ ($\hat{b}^\dagger_{n}$) is the annihilation (creation) operator for the vibrational mode (with frequency $\nu$) on the $n_\mathrm{th}$ molecule, and $c_\nu$ represents the molecular vibronic coupling strength, sometimes represented by the Huang-Rhys factor $S = (c_{\nu} / \nu)^2$. 
Furthermore, $\hat{H}_\mathrm{cav} = \hbar \omega_\mathrm{c} \hat{a}^{\dagger} \hat{a}$ in Eq.~\ref{eq:hamgen} describes the photonic mode DOF of the cavity, where $\hat{a}$ ($\hat{a}^\dagger$) is the annihilation (creation) operator for the photon mode with frequency $\omega_\mathrm{c}$; and 
\begin{equation} \label{eq:HLM}
    \hat{H}_\mathrm{LM} = \hbar g_\mathrm{c} \sum_{n=1}^{N} \left(\hat{a}^{\dagger}\hat{\sigma}_n^- + \hat{a}\hat{\sigma}_n^+ \right)
\end{equation}
describes the light--matter (exciton-photon) interaction. The light--matter coupling strength is $g_\mathrm{c}=\sqrt{\frac{\omega_\mathrm{c} }{2 \hbar \epsilon {\mathcal V}}} \hat{\bf e} \cdot {\boldsymbol\mu}_{eg}$, where ${\boldsymbol\mu}_{eg}$ is the molecular transition dipole, $\hat{\bf e}$ is the cavity field polarization direction, $\epsilon$ is the permittivity inside the cavity (for vacuum,  $\epsilon = \epsilon_0$), and ${\mathcal V}$ is the effective cavity quantization volume. 
Note that when entering into the ultrastrong coupling regime ($\sqrt{N} g_\mathrm{c} / \omega_\mathrm{c} > 0.1$), one needs to incorporate the counter-rotating wave terms ($\hat{a}^{\dagger}\hat{\sigma}_n^+$ and $\hat{a}\hat{\sigma}_n^-$)  and dipole-self-energies incorporated in Eq.~\ref{eq:HLM} to accurately describe the light-matter interaction. Here, we restrict our parameters away from the ultrastrong coupling regime. 

Finally, $\hat{H}_\text{pulse}$ characterizes the external field pumping excitation of the cavity field~\cite{Aizpurua_acsnano2016}, 
\begin{align}
    \hat{H}_\text{pulse}(t) = i\hbar [\hat{a}^\dagger \cdot E_-(t) - \hat{a} \cdot E_+(t)].
\end{align}
Without loss of generality, we assume Gaussian pulses with the form
\begin{equation} \label{eq:E_pm}
    E_\pm(t) = (\lambda_F / 2) \cdot \exp[- \Delta^2 (t - t_0)^2 / 2] \cdot e^{\pm i\omega_\text{p} (t - t_0)},
\end{equation}
characterized by the field amplitude $\lambda_F$ (with the field intensity $I \propto \lambda_F^2$), which also determines the effective coupling strength between the cavity mode and the incident pulse; the pulse center time $t_0$ and carrier frequency $\omega_\text{p}$; and the pulse spectral width $\Delta$. 
We have found that driving the collective molecular dipole $\hat{\mu} = (1 / \sqrt{N}) \sum_{n=1}^{N} (\hat{\sigma}_n^+ + \hat{\sigma}_n^-)$ rather than the cavity field yields similar results for the present considerations.

\begin{figure}[htbp]
    \centering
    \includegraphics[width=1.0\linewidth]{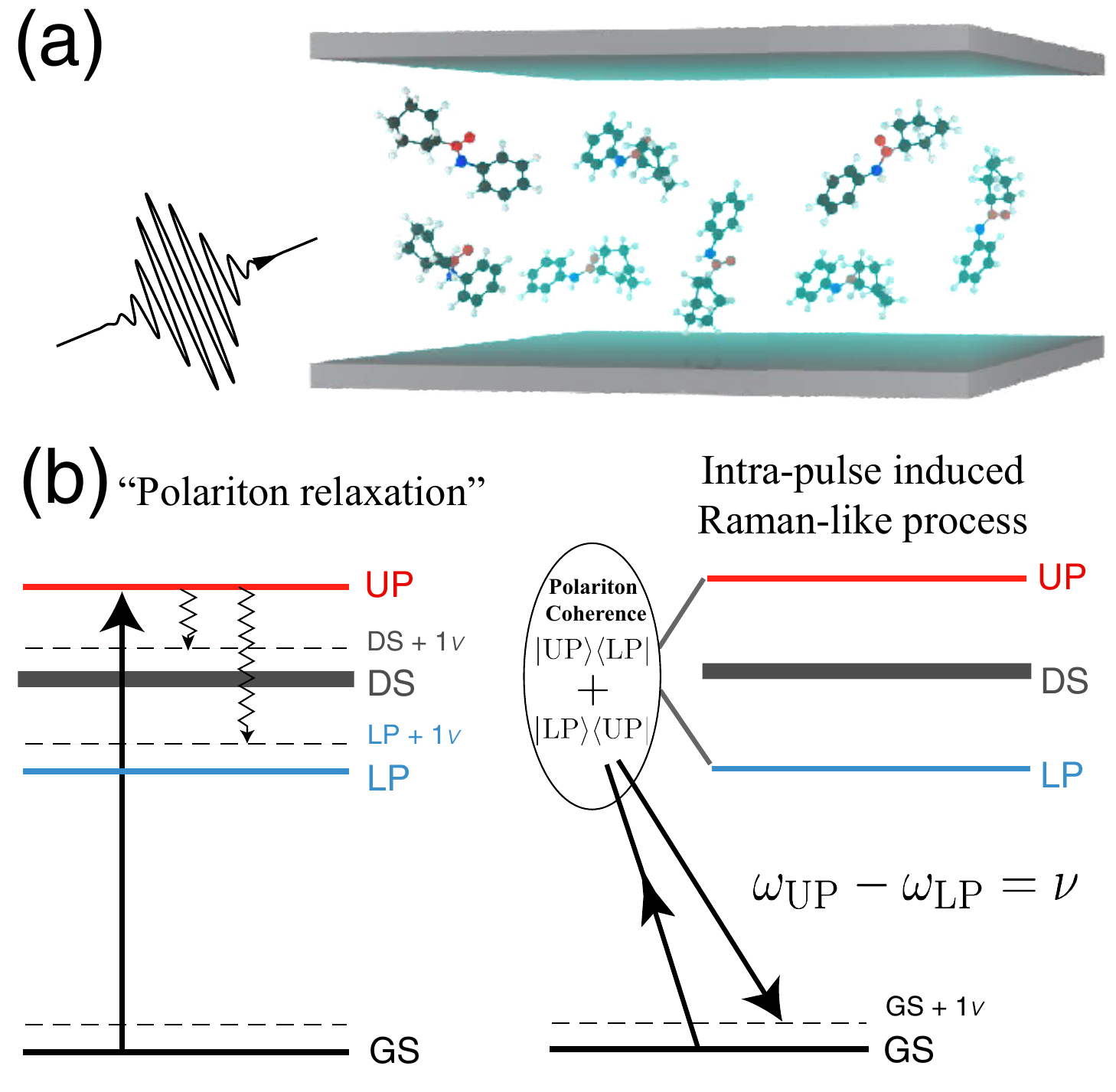}
    \caption{(a) Schematic illustration of field-driven molecular polariton dynamics under pulsed excitation, where an ultrafast optical field drives a molecular ensemble strongly coupled to a cavity mode. (b) Comparison of two distinct mechanisms for polariton-induced vibrational activation. Left: vibrational excitation arising from polariton relaxation processes, where population initially prepared in the upper polariton (UP) relaxes through dark states (DS) and lower polariton (LP) branches while generating vibrational quanta. Right: polariton-mediated intrapulse stimulated Raman-like process, in which the finite spectral bandwidth of a single optical pulse enables coherent frequency mixing between polaritonic states and creates UP--LP coherence, leading to vibrational excitation on the ground electronic state without requiring population relaxation or multi-pulse excitation schemes. }
    \label{fig:1}
\end{figure}

Figure~\ref{fig:1} presents a schematic overview of field-driven molecular polariton dynamics and highlights two conceptually distinct mechanisms for polariton-induced vibrational activation. As illustrated in Fig.~\ref{fig:1}a, a transient optical pulse drives a molecular ensemble strongly coupled to a confined cavity mode, creating nonequilibrium polariton states whose subsequent dynamics redistribute energy among electronic, photonic, and vibrational DOF.
Fig.~\ref{fig:1}b contrasts two different pathways by which vibrational excitation can arise in such systems. In the first mechanism, vibrational activation occurs through polariton relaxation, where population initially injected into the upper polariton relaxes to the dark states manifold and the lower polariton, accompanied by the emission of vibrational quanta. This process is inherently dissipative and relies on population transfer among polaritonic states. In contrast, the second mechanism corresponds to a polariton-mediated intrapulse stimulated Raman-like process, in which vibrational excitation is generated coherently by frequency mixing among different spectral components of a single broadband pulse (rather than a conventional multi-pulse process). Note that the resonance Raman scattering excitation spectra from exciton polaritons has also been studied in Ref.~\citenum{Rury_JPCC2019}. In this case, vibrational activation can occur directly on the ground electronic state without requiring prior population relaxation or multiple optical fields. The present work focuses on elucidating the mechanisms by which intramolecular vibronic coupling leads to excitation of molecular vibrations that resonate with the Rabi splitting associated with electronic strong coupling between the molecular system and optical cavity modes. 

\subsection{Single Excitation (SE) Approximation}
Consider first a system with just the two electronic levels (without vibrations and vibronic coupling) interacting with a cavity mode. It is described by the Tavis-Cummings (TC) Hamiltonian~\cite{tavis1968exact, Tavis_1969},
\begin{equation} \label{eq:hamtc}
    \hat{H}_\mathrm{TC} = \sum_{n = 1}^{N} \hbar \omega_{0} \hat{\sigma}_n^+ \hat{\sigma}_n^- + \hbar \omega_\mathrm{c} \hat{a}^{\dagger} \hat{a} + \hbar g_\mathrm{c} \sum_{n=1}^{N} \left(\hat{a}^{\dagger}\hat{\sigma}_n^- + \hat{a}\hat{\sigma}_n^+ \right).
\end{equation}
In the SE approximation this model is considered in the truncated Hilbert space, which is spanned by the zero photon-dressed ground state $|\text{G},0\rangle$ where all the molecules are in the ground state and no photon in the cavity, one photon-dressed ground state $|\text{G},1\rangle$ where all the molecules are in the ground state and one photon is in the cavity, and the single-molecule excited state $|\text{E}_n, 0\rangle$ where all the molecules are in the ground state except for the $n_\mathrm{th}$ molecule. These diabatic states are defined as $|\text{G},0 \rangle = |g_1\rangle\otimes...|g_{n }\rangle...\otimes|g_{N}\rangle\otimes|0\rangle$, $|\text{G},1 \rangle = |g_1\rangle\otimes...|g_{n }\rangle...\otimes|g_{N}\rangle\otimes|1\rangle$, and $|\text{E}_n, 0\rangle =|g_1\rangle\otimes...|e_{n}\rangle...\otimes|g_{N}\rangle\otimes|0\rangle$. Here, $|0\rangle$, $|1\rangle$ are the zero and one photon Fock states, which are eigenstates of the Hamiltonian of the cavity mode.
This approximation leads to a drastic reduction of the Hilbert-space dimension. Instead of the full light--matter Hilbert space, whose dimension scales as $N_\mathrm{c}\times 2^{N}$ when all molecular excitations are retained and $N_c$ is the number of photonic Fock states (in the SE approximation $N_c = 2$), the SE subspace contains only $N+1$ electronic--photonic states (one photon state and $N$ single-molecule excitations), together with the vibrational DOF.
As a result, the total Hilbert-space dimension in the SE approximation can be efficiently simulated by direct time-dependent Schr\"odinger equation (TDSE) propagation of the total wavefunction $|\Psi(t)\rangle$ for moderate values of $N$, 
\begin{align}
     i\hbar \frac{\partial}{\partial t} |\Psi(t)\rangle = \hat{H}_\text{HTC}(t) |\Psi(t)\rangle, 
\end{align}
enabling explicit resolution of vibronic dynamics with respect to the polariton and dark states. Details are presented in Sec.~III-A of the Supporting Information. 

In the model defined by Eq.~\ref{eq:hamtc}, all molecules are coupled identically to the cavity modes, making it possible to identify a collective ``bright'' excitonic state $|\mathrm{B}\rangle=\frac{1}{\sqrt{N}}\sum_{n=1}^{N} |\text{E}_n, 0\rangle$ that carries the entire coupling to the $|\text{G},1\rangle$ state through $\hat{H}_\mathrm{LM}$, generating polariton states $|\pm\rangle$ which are eigenstates of $\hat{H}_\mathrm{TC}$. These polariton states are expressed as follows~\cite{tavis1968exact, Arkajit_Chemrev_2023, Ying_ARPC2026}
\begin{subequations}\label{eq:TC-polariton}
\begin{align}
    |+\rangle &= \cos\Theta |\mathrm{B}\rangle  + \sin\Theta |\text{G},1\rangle\\
    |-\rangle &= -\sin\Theta |\mathrm{B}\rangle + \cos\Theta |\text{G},1\rangle, 
\end{align}
\end{subequations}
where the mixing angle is $\Theta = \frac{1}{2} \tan^{-1} \left[ \frac{2 \sqrt{N}g_\mathrm{c}}{\omega_\mathrm{c} - \omega_0} \right] \in [0, \frac{\pi}{2})$, and the corresponding energies are $\omega_\pm = \frac{\omega_0 + \omega_\text{c}}{2} \pm \frac{1}{2}\sqrt{(\omega_0 -\omega_\text{c})^2 + 4N g_\text{c}^2}$, where $\omega_+$ and $\omega_-$ correspond to the upper (UP) and the lower (LP) polaritons, respectively. 
The collective Rabi splitting
\begin{align} \label{eq:Rabi}
    \Omega = \omega_+ - \omega_- = \sqrt{(\omega_\text{c} - \omega_0)^2 + 4N g^2_\text{c}}
\end{align}
becomes $\Omega = 2\sqrt{N} g_\text{c}$ at resonance $\omega_\text{c} = \omega_0$. 
Furthermore, in the SE subspace the TC Hamiltonian admits one UP, one LP, and $N-1$ degenerate dark states, as detailed in Sec.~I of the Supporting Information. Note that even in the presence of coupling disorder, the bright state and polariton states remain well defined~\cite{Houdre_1996}.

\subsection{Mean-Field (MF) Approximation}
Another common approach to access the large-$N$ limit is the MF approximation that has been widely used in the context of collective light--matter coupling~\cite{Yuen-Zhou_PRL2025, Keeling_PRL2022, Cui_JCP2022, Cui_JCP2023, Hsieh_JCP2023, Hsieh_JPCL2023, Reichman_Nanoph2024, Li_CP2025, Hsieh_JCTC2025}. 
A general theoretical justification for the exactness of MF equations of motion (with respect to the collective observables) for open Dicke-type models in the thermodynamic limit ($N\to\infty$) has been established in Ref.~\citenum{Carollo_PRL2021}, with further extensions discussed in Ref.~\citenum{Carollo_NJP2023}. 
Within this framework, the validity of the MF approximation relies on the condition that the total number of excitations remains much smaller than $N$, such that collective light--matter coupling dominates while higher-order inter-exciton correlations are suppressed~\cite{Carollo_PRL2021, Carollo_NJP2023}. 
One caveat, however, is that MF treatments of the HTC model intrinsically assume a single-configuration (product state) wavefunction (or density matrix, see Eq.~\ref{eq:DM-decompose} below) and therefore neglect light--matter and intermolecular entanglement. As a result, MF treatments may miss important physical effects compared to the SE representation when both are considered in the weak excitation limit at which the SE approximation is valid.

The MF description is based on a product ansatz for the density matrix~\cite{Keeling_PRL2022},
\begin{equation} \label{eq:DM-decompose}
    \hat{\rho}(t) = \hat{\rho}_a (t) \bigotimes_{n=1}^{N} \hat{\rho}_n(t),
\end{equation}
where $\hat{\rho}_a(t)$ denotes the cavity field reduced density matrix (RDM), and $\hat{\rho}_n(t)$ denotes the $n_\text{th}$ molecular RDM. Eq.~\ref{eq:DM-decompose} neglects inter-molecular correlations while retaining the collective coupling to the cavity field. A further simplification replaces the quantum description of the radiation field by its classical representation in terms of a (complex) cavity field amplitude $\alpha (t)$. 
Furthermore, in the present analysis we assume that the size of the molecular subsystem is much smaller the the radiation wavelength, making it possible to disregard different molecular positions and the position dependence of $\alpha$.
Under this approximation, the dynamics of the molecular subsystem is governed by an effective single-emitter Hamiltonian that depends self-consistently on the cavity field amplitude $\alpha(t)$. The effective Hamiltonian reads as
\begin{align}
    \hat{H}^\text{eff}(\alpha(t)) &= \hbar \omega_0 |e\rangle \langle e| + \hbar\nu\,b^\dagger b + c_\nu\,|e\rangle\!\langle e| \otimes (b+b^\dagger) \notag\\
    &~~~ + \hbar g_\mathrm{c}\left(\alpha^{*}(t)|g\rangle\!\langle e| + \alpha(t)\,|e\rangle\!\langle g|\right).
\label{eq:Hmol}
\end{align}
A detailed derivation is provided in Section~II-A of the Supporting Information.
Note that a key assumption made here is the permutation symmetry among the molecules, which means no energetic or coupling disorders, so that all reduced molecular density matrices can be taken to be identical, $\hat{\rho}_n (t) \equiv \hat{\rho}_\mathrm{M} (t)$, spanned by a single set of molecular (vibronic) basis $\{|g/e, \nu\rangle\}$, and the equations of motion reduce to
\begin{equation} \label{eq:MF-1}
    \frac{\partial}{\partial t} \hat{\rho}_\text{M}(t) = -\frac{i}{\hbar} [\hat{H}^\text{eff}(\alpha(t)), \hat{\rho}_\text{M}(t)], 
\end{equation}
supplemented, when appropriate, by Lindblad terms~\cite{Keeling_PRL2022} describing electronic relaxation, pure dephasing, and vibrational damping. 
In the absence of environment-induced decoherence, the Liouville-von Neumann equation~\ref{eq:MF-1} reduces to a Schr\"odinger equation for single-molecule wavefunctions, consistent with the MF approximation adopted in Ref.~\citenum{Cui_JCP2023}, which is based on a time-dependent Hartree product wavefunction ansatz.

Furthermore, the equation of motion for the cavity field amplitude reads
\begin{equation} \label{eq:MF-2}
    \dot{\alpha}(t)
    = -\,i\,(\omega_\mathrm{c} - i\kappa/2)\,\alpha(t)
      - i g_\mathrm{c}\,N\,\sigma(t)\,+\,E_-(t),
\end{equation}
where $\sigma(t)=\mathrm{Tr}\!\left[\rho_\mathrm{M}(t)\,|g\rangle\!\langle e|\right]$ denotes the electronic coherence of a representative two-level system and $\kappa$ is the cavity loss rate. Eqs.~\ref{eq:MF-1} and \ref{eq:MF-2} provide a self-consistent description of collective light--matter dynamics. 
From a computational perspective, the numerical cost of solving these equations does not depend on the number of molecules, scaling as $\mathcal{O}(1)$ with respect to $N$. Details are provided in Sec.~II-B of the Supporting Information. 

\begin{figure*}[htbp]
    \centering
    \includegraphics[width=0.8\linewidth]{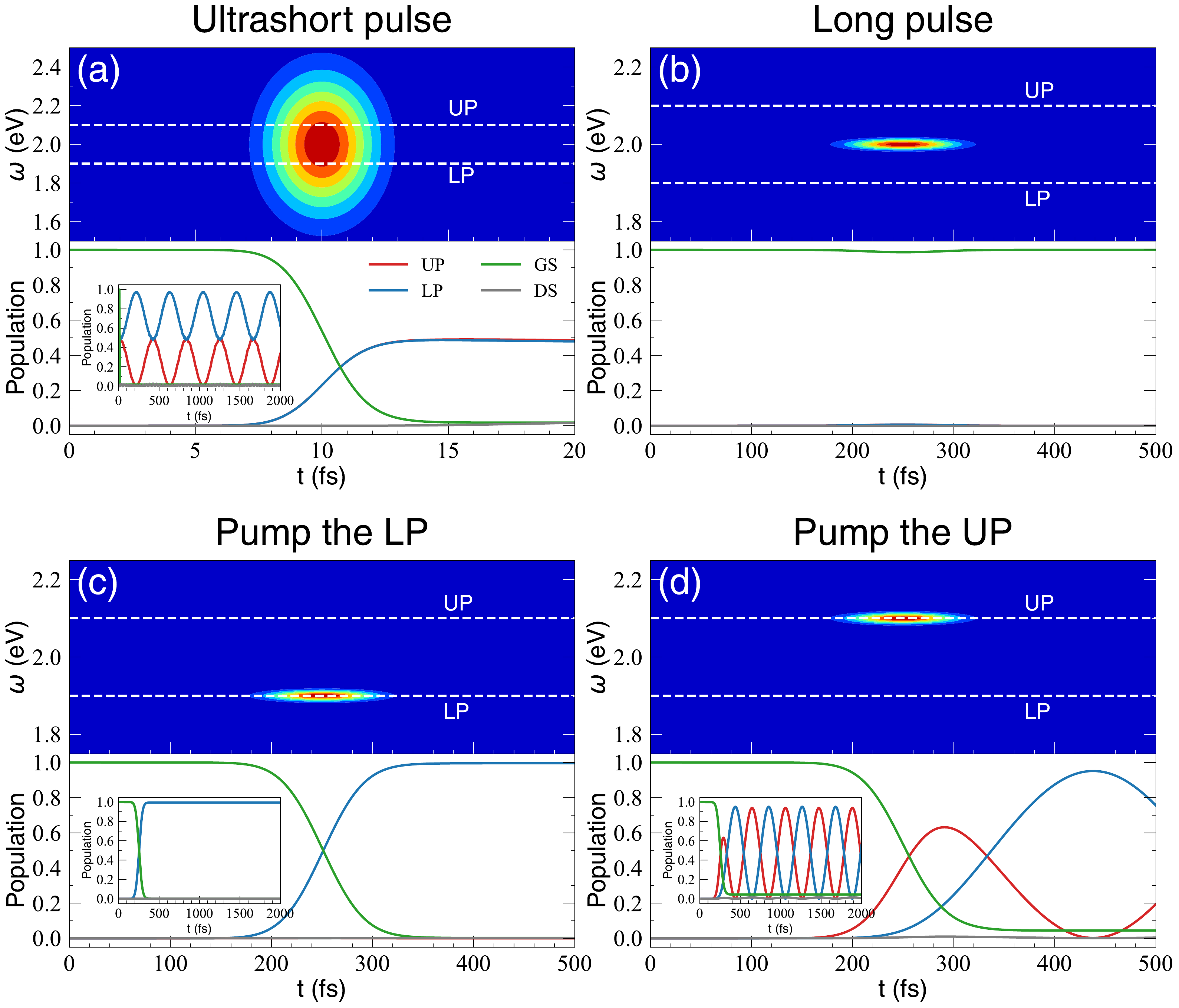}
    \caption{Field-driven polariton dynamics under the SE approximation. Time--frequency representations of the driving field using Wigner functions (upper half in each panel) and corresponding population dynamics (lower half in each panel) for different pulse conditions. Here, we fix the Rabi frequency to be resonant with the intramolecular vibrational mode, $\Omega=\nu=0.20~\mathrm{eV}$, such that the UP and the LP branches are located at $2.1$ and $1.9~\mathrm{eV}$, respectively (indicated by the white dashed lines in the upper half of each panel). The pulse conditions are: (a) Ultrashort pulse with bandwidth exceeding the Rabi splitting ($\Delta =$ 2.07 eV, or $\Delta^{-1}=2$~fs), simultaneously overlapping the UP and LP branches, amplitude $\lambda_F =$ 0.4 eV, and central frequency $\omega_\text{p} =$ 2.0 eV. Both polaritons are populated, generating a coherent UP--LP superposition and subsequent population exchange at frequency $c_\nu / \sqrt{N}$, as highlighted in the inset. (b) Long pulse with narrow spectral bandwidth ($\Delta = 0.083$ eV, or $\Delta^{-1}=50$~fs), amplitude $\lambda_F =$ 0.023 eV, and a carrier frequency $\omega_\mathrm{p}=$ 2.0 eV focused between the two polaritons; excitation is strongly suppressed in this spectral regime of dark modes. (c) Long pulse ($\Delta = 0.083$ eV, or $\Delta^{-1}=50$~fs) with amplitude $\lambda_F =$ 0.023 eV, resonantly tuned to the LP ($\omega_\mathrm{p}=$ 1.9 eV), resulting in selective LP excitation without appreciable UP population or UP--LP coherence. (d) Long pulse ($\Delta = 0.083$ eV, or $\Delta^{-1}=50$~fs) with amplitude $\lambda_F =$ 0.023 eV, resonantly tuned to the UP ($\omega_\mathrm{p}=$ 2.1 eV), followed by coherent population exchange between UP and LP at frequency $c_\nu / \sqrt{N}$ driven by polaritonic coupling, as shown in the inset.
    }
    \label{fig:2}
\end{figure*}

\section{Results and Discussion}
In this section, we investigate the quantum dynamics of the field-driven HTC Hamiltonian in Eq.~\ref{eq:hamgen} over a broad range of pulse durations and field intensities, using both the SE and MF approximations. 
For notational simplicity, we set $\hbar \equiv 1$ throughout.
The system is initialized in the global ground state $|\Psi(0)\rangle = |\text{G},0\rangle \bigotimes_{n=1}^N |0_n\rangle$ and driven by a transient Gaussian pulse described in Eq.~\ref{eq:E_pm} that excites the cavity field.
Unless otherwise stated, we use $N = 4$ in the SE calculations and $N = 10^4$ in the MF simulations to approximate the thermodynamic limit.
In both cases, the energy spacing between the molecular ground and excited electronic states, as well as the cavity frequency, are set to $2.0~\mathrm{eV}$, corresponding to the resonant light--matter condition.
Furthermore, the Rabi frequency (see expression in Eq.~\ref{eq:Rabi}) is chosen to be resonant with the intramolecular vibrational mode, $\Omega=\nu=0.20~\mathrm{eV}$, such that the upper and lower polariton branches are located at $2.1$ and $1.9~\mathrm{eV}$, respectively~\footnote{This implies different assumptions about light-molecule coupling $g_\mathrm{c}$, as different number of molecules $N$ are used in the SE and MF cases but a same Rabi splitting $\Omega$ is assigned. }.
The electron--vibration coupling strength is fixed to $c_\nu=20~\mathrm{meV}$ (corresponding to a horizontal shift of $R = \sqrt{2 c_{\nu}^2 / (\hbar \nu^3)}$ relative to the minimum of the ground electronic state).
Additional computational and model details are provided in Sec.~II-B (MF) and Sec.~III-A (SE) of the Supporting Information.

\begin{figure*}[htbp]
    \centering
    \includegraphics[width=0.8\linewidth]{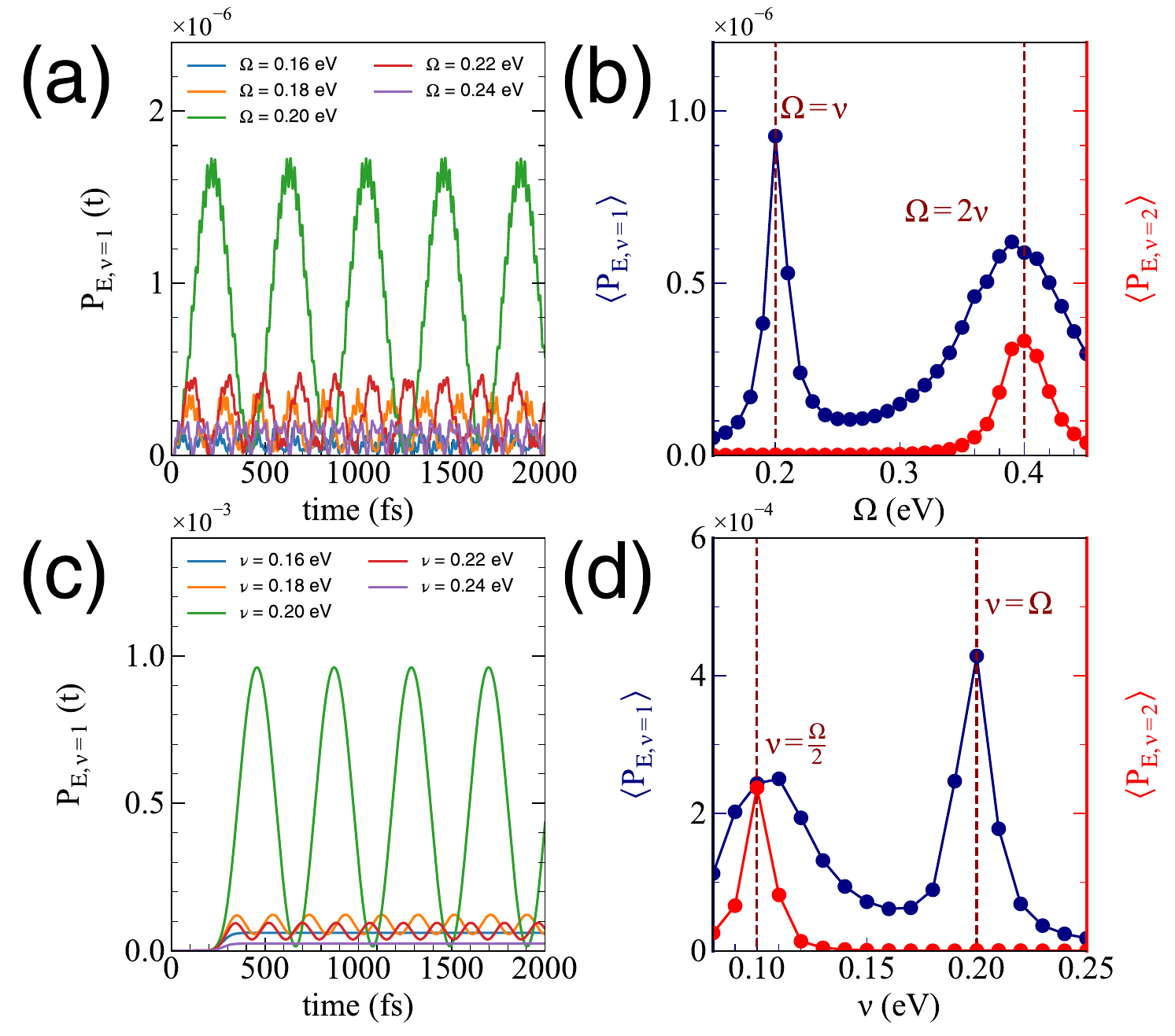}
    \caption{Field-driven vibronic dynamics under the SE approximation. Here, the Rabi frequency $\Omega$ and vibrational frequency $\nu$ are varied in order to examine the resonance behavior of the energy transfer. The field amplitude is $\lambda_F=$ 1 meV. 
    The upper panels (a-b) correspond to an ultrashort pulse condition and fix the vibrational frequency at $\nu=0.20$~eV, while the Rabi splitting $\Omega$ is varied by changing the single molecule coupling strength $g_\text{c}$; the UP and the LP are located at $\omega_+ = 2.0~\text{eV} + \Omega / 2$ and $\omega_- = 2.0~\text{eV} - \Omega / 2$, respectively. 
    (a) Vibrational population dynamics on the excited electronic and photonic states $P_{\text{E},\nu=1} (t)$. Pronounced oscillatory vibrational excitation is observed in the upper electronic states when $\Omega$ approaches $\nu$. (b) Time-averaged vibrational population $\langle P_{\text{E},\nu=1}\rangle$ in the upper electronic states displayed as a function of $\Omega$, showing a clear resonance peak at $\Omega=\nu$ (blue symbols). For comparison, $\langle P_{\text{E},\nu=2}\rangle$ (red symbols) exhibits a weaker response and a secondary feature near $\Omega=2\nu$. 
    The lower panels (c-d) correspond to a long pulse condition and fix the Rabi splitting $\Omega=0.20$~eV (so that the UP and the LP are located at $\omega_+ = 2.1~\text{eV}$ and $\omega_- = 1.9~\text{eV}$, respectively), while the vibrational frequency $\nu$ is varied.
    (c) Vibrational population dynamics on the excited electronic and photonic states $P_{\text{E},\nu=1} (t)$. The observation is similar to panel (a) but smoother. (d) Time-averaged vibrational population as a function of $\nu$, again revealing resonant enhancement when $\nu=\Omega$ for $\langle P_{\text{E},\nu=1}\rangle$ (blue symbols). A weaker secondary feature near $\nu=\Omega/2$ for $\langle P_{\text{E},\nu=2}\rangle$ (red symbols) is also seen.
    }
    \label{fig:3}
\end{figure*}

In what follows, we present the SE results only (throughout Figs.~\ref{fig:2}-\ref{fig:5} in the main text), while the MF results are all in Sec.~II of the Supporting Information. 

\subsection{Field-Driven Polariton Dynamics}
We first examine the field-driven molecular polariton dynamics with various driving fields. 
The driving field can be characterized in both the time- and frequency-domains, which are related through Fourier transform. While the time-domain representation highlights the temporal center and duration of the pulse, the frequency-domain representation reveals the spectral components spanned by the field. A mixed time--frequency representation that simultaneously captures both aspects is provided by the Wigner spectrum~\cite{Cohen_ProcIEEE1989, Hlawatsch_IEEE1992, Alonso_2011}. Using the positive-frequency component of the field $E_+(t)$ defined in Eq.~\ref{eq:E_pm}, the Wigner distribution is given by~\cite{Coker_JCTC2021}
\begin{align} \label{eq:Wigner_def}
    W(t, \omega) &= \int_{-\infty}^{+\infty} ds~ E_+\!\left(t+\frac{s}{2}\right)
    E_+^*\!\left(t-\frac{s}{2}\right) e^{-i\omega s} \notag\\
    &= \frac{\sqrt{\pi}\,\lambda_F^2}{2 \Delta}
    \exp\!\left[- \Delta^2 (t - t_0)^2 - \frac{(\omega-\omega_{\mathrm{p}})^2}{\Delta^2} \right],
\end{align}
which corresponds to a two-dimensional Gaussian distribution in the time--frequency plane. The marginals of $W(t,\omega)$ are directly related to the intensity profiles of the pulse in the time- and frequency-domains, respectively.
Two qualitatively distinct regimes can be identified depending on the pulse width relative to the Rabi splitting $\Omega$. An ultrashort pulse satisfying $\sqrt{2\pi}\Delta \gg \Omega$ is broad enough to overlap both upper and lower polaritons. In contrast, for a long narrow band pulse with $\sqrt{2\pi}\Delta \ll \Omega$, the field can selectively excite one polariton or none.

Figure~\ref{fig:2} presents field-driven polariton dynamics.
Fig.~\ref{fig:2}a shows the polariton dynamics induced by an ultrashort pulse with temporal width $\Delta^{-1}=2$~fs (corresponding to a spectral width $\Delta = 2.07$ eV), amplitude $\lambda_F =$ 0.4 eV, central frequency $\omega_\text{p} =$ 2.0 eV and central time $t_0 = 10$ fs. The upper panel displays the corresponding Wigner spectrum, whose broad frequency support overlaps both the UP and the LP branches. As a result, both polaritons are simultaneously excited, as seen in the population dynamics shown in the lower panel. The inset highlights the subsequent coherent population exchange between UP and LP at long times, reflecting the generation of polaritonic coherence~\footnote{Alternatively, this process may be viewed as a fast generation of population in the excited state of the free molecule, followed by oscillations between this state and the cavity mode. } 
Figs.~\ref{fig:2}b-d presents the corresponding dynamics for a long pulse with temporal width $\Delta^{-1}=50$~fs (corresponding to a spectral width $\Delta = 0.083$ eV) and carrier frequencies $\omega_\mathrm{p}=$ 2.0, 1.9 and 2.1~eV that selectively overlaps the dark modes, the lower polariton and the upper polariton, respectively. The pulse amplitude is $\lambda_F =$ 0.023 eV, central time is $t_0 = 250$ fs. 
For $\omega_\mathrm{p}=$ 2.0 eV, the narrow spectral bandwidth overlaps the dark modes and excitation is strongly suppressed (see Fig.~\ref{fig:2}b). 
When the lower polariton is excited (Fig.~\ref{fig:2}c), energetic constraint inhibits population transfer to the upper polariton or to the dark states. In contrast, when the upper polariton is excited (Fig.~\ref{fig:2}d), coherent population transfer between the UP and LP is observed, as shown in the inset. This population exchange is mediated by the collective phonon mode (see Eq.~S6 of the Supporting Information). 

Note that in realistic situations, relaxation phenomena (such as exciton annihilation, vibrational relaxation, cavity loss) will affect both the polariton states and the vibrational modes, but they are disregarded in the calculation shown here.
In what follows we address the resulting dynamics in the vibrational space. 

\subsection{Polariton Induces Vibrational Excitation}
\begin{figure*}[htbp]
    \centering
    \includegraphics[width=1.0\linewidth]{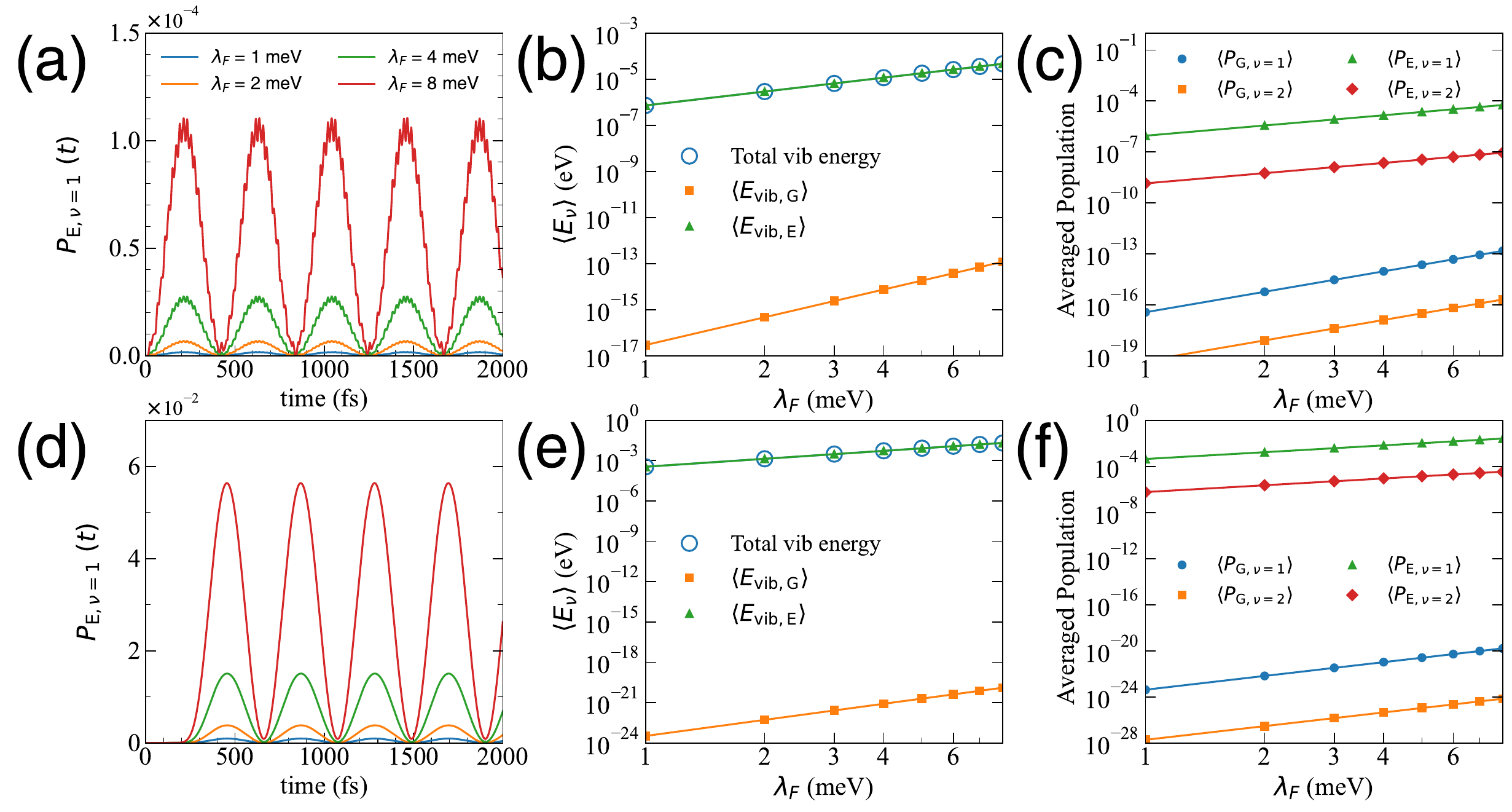}
    \caption{Polariton-vibrational energy exchange under the SE approximation. Here, the vibrational frequency is taken identical to the Rabi frequency, $\Omega = \nu = 0.20$ eV (so that the UP and the LP are located at $\omega_+ = 2.1~\text{eV}$ and $\omega_- = 1.9~\text{eV}$, respectively), while the field amplitude $\lambda_F$ varies. 
    The upper panels (a-c) correspond to an ultrashort pulse condition.
    (a) Time-dependent vibrational population on the excited electronic and photonic states $P_\mathrm{E,\nu=1}(t)$ for different field amplitudes $\lambda_F$. The vibrational response increases strongly with field amplitude, with pronounced oscillatory behavior. (b,c) Time-averaged vibrational energies (b) and vibrational population (c) as functions of the field amplitude $\lambda_F$. The total and excited-state vibrational energies exhibit a quadratic dependence on $\lambda_F$, while the ground-state vibrational population shows a quartic scaling, reflecting distinct linear and nonlinear excitation channels. 
    The lower panels (d-f) correspond to a long pulse condition.
    (d) The same as panel (a) except with long-pulse excitation. In this spectrally selective regime, vibrational excitation exhibits coherent oscillations primarily on the excited electronic and photonic state. (e,f) Time-averaged vibrational energies (e) and vibrational populations (f) as functions of $\lambda_F$. Power-law scaling behavior similar to that of the ultrashort-pulse case is observed, although with reduced overall amplitudes due to the absence of broadband polaritonic coherence. }
    \label{fig:4}
\end{figure*}

\begin{figure*}[htbp]
    \centering
    \includegraphics[width=1.0\linewidth]{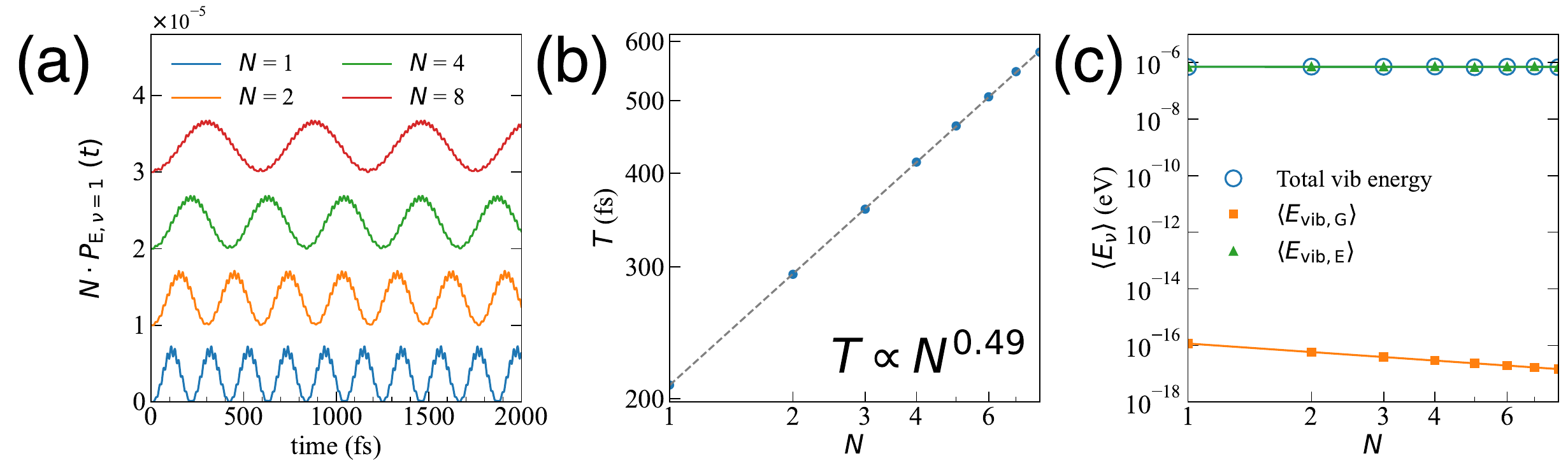}
    \caption{Effect of the number of molecules $N$ on field-driven vibrational activation under the SE approximation. The system is studied under the resonant condition $\Omega=\nu =$ 0.20 eV (so that the UP and the LP are located at $\omega_+ = 2.1~\text{eV}$ and $\omega_- = 1.9~\text{eV}$, respectively). An ultrashort pulse is used, with the field amplitude fixed as $\lambda_F=1$~meV. (a) Time-dependent vibrational populations in the excited electronic and photonic states multiplied by $N$, $N \cdot P_{\mathrm{E},\nu=1}(t)$, for different values of $N$. (b) Oscillation period (extracted from panel (a)) as a function of $N$, plotted in the log-log scale. The linear fitting indicates $T \propto \sqrt{N}$. (c) Time-averaged vibrational energies as functions of $N$. The total vibrational energy (open circles) is dominated by contributions from the excited electronic and photonic states $\langle E_\mathrm{vib, E} \rangle$ (triangle symbols), which remains independent with $N$; while $\langle E_\mathrm{vib, G} \rangle$ (square symbols) decreases as $1/N$.  }
    \label{fig:5}
\end{figure*}

We next examine the process described in Fig.~\ref{fig:2} from the perspective of the vibrational motion, in which use the same set of parameters as above except for a much smaller $\lambda_F$ to ensure a weak-field regime. 
Specifically, we restrict our attention to the pulsed field defined in Eq.~\ref{eq:E_pm} and consider two fixed excitation conditions: 
(1) a short-time, broadband pulse with temporal width $\Delta^{-1} = 2~\mathrm{fs}$ (corresponding to a spectral width $\Delta = 2.07~\mathrm{eV}$) and central frequency $\omega_{\mathrm p} = 2.0~\mathrm{eV}$, positioned midway between the upper and lower polaritons, hereafter referred to as the \emph{ultrashort pulse}~\footnote{The ultrashort pulse contains roughly one optical cycle, but this does not preclude the use of the RWA. The validity of the RWA is primarily controlled by the light--matter coupling strength rather than the pulse bandwidth; in the perturbative regime considered here, counter-rotating terms contribute only at higher order and remain negligible. While the broad bandwidth enables short-time excitation, it does not imply that off-resonant or counter-rotating processes dominate the dynamics.}; and 
(2) a long-time, spectrally narrow pulse with temporal width $\Delta^{-1} = 50~\mathrm{fs}$ (corresponding to a spectral width $\Delta = 0.083~\mathrm{eV}$) and central frequency $\omega_{\mathrm p} = 2.1~\mathrm{eV}$, chosen to selectively excite the upper polariton branch, hereafter referred to as the \emph{long pulse}~\footnote{We note that although a pulse with a temporal width of 50~fs would not typically be considered long in ultrafast spectroscopy, we refer to it as a ``long pulse'' throughout this work for convenience. }.
These pulse parameters are employed throughout Figs.~\ref{fig:3}--\ref{fig:5}; for brevity, we will simply denote them as the ultrashort and long pulses, respectively.

It is instructive to examine the energy transfer behavior as a function of detuning between the Rabi frequency $\Omega$ and the vibration frequency $\nu$. 
Figure~\ref{fig:3} illustrates resonant vibrational excitation induced by field-driven polariton dynamics. Fig.~\ref{fig:3}a show the time-dependent vibrational population (per molecule) on the excited electronic and photonic states
\begin{align}
    P_{\mathrm{E},\nu=1}(t) =\frac{1}{N}\langle \Psi(t)|\hat{\Pi}_{\mathrm{E},\nu = 1}|\Psi(t)\rangle,
\end{align}
under ultrashort-pulse excitation, where the projection operator $\hat{\Pi}_{\mathrm{E},\nu} = (|\mathrm{G},1\rangle\langle \mathrm{G},1|
+ \sum_{n} |\mathrm{E}_n,0\rangle\langle \mathrm{E}_n,0|) \otimes \sum_{j=1}^{N} |\nu_j = 1\rangle\langle \nu_j = 1|$. See details in Sec.~III-B of the Supporting Information.
In this regime, the vibrational frequency is fixed at $\nu=0.20$~eV while the Rabi splitting $\Omega$ is varied by changing the single molecule coupling strength $g_\text{c}$; and the UP and the LP are located at $\omega_+ = 2.0~\text{eV} + \Omega / 2$ and $\omega_- = 2.0~\text{eV} - \Omega / 2$, respectively. 
Pronounced oscillatory vibrational excitation is observed on the excited electronic and photonic states when $\Omega$ approaches specific resonance conditions, whereas vibrational population in the ground electronic state remains comparatively much smaller ($\sim 10^{-15}$, see Fig.~S10 of the Supporting Information).
The corresponding resonance structure is summarized in Fig.~\ref{fig:3}b, which displays the time-averaged vibrational population in the excited electronic and photonic states per molecule $\langle P_{\mathrm{E},\nu=1}\rangle$ (see definition in Sec.~III-B of the Supporting Information) as a function of $\Omega$. The quantity $\langle P_{\mathrm{E},\nu=1}\rangle$ exhibits a clear resonance peak at $\Omega=\nu$ (that is, UP $\leftrightarrow$ LP $+\,1\nu$), corresponding to resonant energy exchange between the UP and the LP accompanied by the creation of a single vibrational quantum. A second resonance peak appears near $\Omega=2\nu$, which can be attributed to higher-order vibronic processes involving dark states, corresponding to transitions of the form UP $\leftrightarrow$ DS $+\,1\nu$, or equivalently, UP $\leftrightarrow$ LP $+\,2\nu$ mediated by dark states manifold. Consistent with this interpretation, the two-phonon population $\langle P_{\mathrm{E},\nu=2}\rangle$ (red symbols) displays a pronounced resonance at $\Omega=2\nu$, confirming its higher-order vibrational character.

Figs.~\ref{fig:3}c--d present the analogous results obtained under long-pulse excitation. In this case, the Rabi splitting is fixed at $\Omega=0.20$~eV (so that the UP and the LP are located at $\omega_+ = 2.1~\text{eV}$ and $\omega_- = 1.9~\text{eV}$, respectively), while the vibrational frequency $\nu$ is varied. Fig.~\ref{fig:3}c show that vibrational excitation is strongly suppressed in the time domain except near resonance. The corresponding resonance structure is summarized in Fig.~\ref{fig:3}d, where enhanced vibrational population are observed when $\nu = \Omega$ and $\Omega / 2$, respectively, being consistent with Fig.~\ref{fig:3}b.

It is important to note that the resonance condition at $\Omega = 2\nu$ relies on the assumption of negligible energetic disorder, under which the dark states manifold remains degenerate and supports a well-defined vibronic resonance. In the presence of disorder, the dark states manifold is no longer strictly dark nor energetically degenerate, but instead forms a broadened quasi-continuum. Consequently, one might expect transitions of the type UP $\rightarrow$ DS $+\,\nu$ become irreversible and spectrally broadened, leading to a smearing or suppression of the sharp $\Omega = 2\nu$ resonance feature. This is left for future inspections. In contrast, the primary resonance condition at $\Omega = \nu$, which is associated with coherent upper--lower polariton energy exchange, remains robust against disorder due to its collective and delocalized polaritonic character.

\subsection{Polariton-Mediated Intrapulse Stimulated Raman-like Process}
We further investigate the dependence of vibrational energies and populations on the pulse amplitude $\lambda_F$, which reveals a polariton-mediated intrapulse stimulated Raman-like excitation mechanism. Here we fix the resonance condition of $\Omega = \nu = 0.20$ eV (so that the UP and the LP are located at $\omega_+ = 2.1~\text{eV}$ and $\omega_- = 1.9~\text{eV}$, respectively). 

Figure~\ref{fig:4} summarizes the field-amplitude scaling behavior of vibrational activation under pulsed polariton driving. The driving field amplitude $\lambda_F$ is varied from 1 to 8~meV, remaining in the weak-excitation regime where the SE approximation is valid. Fig.~\ref{fig:4}a shows the time-dependent vibrational population in the excited electronic and photonic states, $P_{\mathrm{E},\nu=1}(t)$, under ultrashort-pulse excitation. Figs.~\ref{fig:4}b and~\ref{fig:4}c show the corresponding time-averaged vibrational energies and populations as functions of $\lambda_F$ on logarithmic scales, together with the associated linear fits (see Sec.~III-B of the Supporting Information). 
Figs.~\ref{fig:4}d--f present analogous results obtained under long-pulse excitation.
For both ultrashort- and long-pulse excitation, the vibration energy on the excited electronic and photonic state $\langle E_\mathrm{vib,E} \rangle$ is contributing dominantly to the total vibration energy compared to that on the ground electronic state $\langle E_\mathrm{vib,G} \rangle$, as shown in Figs.~\ref{fig:4}b and~\ref{fig:4}e. 
In contrast, the vibrational populations in the ground electronic state $\langle P_\mathrm{G,\nu} \rangle$ is negligibly small. This indicates that at this field amplitude, the linear process is dominating, while the nonlinear process is much weaker. 
Furthermore, in the long-pulse regime, the numerical value of the vibration population in the ground electronic state is much smaller compared to that of the ultrashort-pulse case (by comparing panel (c) and panel (f) for $\langle P_\mathrm{G,\nu} \rangle$), reflecting the reduced role of broadband polaritonic coherence under spectrally selective driving.

Despite these quantitative differences, both pulse regimes exhibit robust and well-defined scaling relations with respect to the driving field amplitude $\lambda_F$. Specifically, the vibrational energy and population in the excited electronic and photonic states scale quadratically with the field amplitude, $\langle E_{\mathrm{vib,E}}\rangle,\,\langle P_{\mathrm{E},\nu}\rangle \propto \lambda_F^2$ (or equivalently, $\propto I$ in terms of the field intensity), reflecting a leading-order linear response to the applied field, as schematically illustrated in the left panel of Fig.~\ref{fig:1}b. In contrast, the vibrational energy and population in the ground electronic state scale quartically, $\langle E_{\mathrm{vib,G}}\rangle,\,\langle P_{\mathrm{G},\nu}\rangle \propto \lambda_F^4$ (or equivalently, $\propto I^2$ in terms of the field intensity), characteristic of a higher-order nonlinear excitation pathway.
This quartic scaling can be naturally interpreted as arising from a Raman-type process, in which vibrational excitation on the ground electronic state is generated through virtual electronic excitation and de-excitation within a single optical pulse, as illustrated in the right panel of Fig.~\ref{fig:1}b. Importantly, this mechanism corresponds to a \emph{single-pulse, self-induced intrapulse stimulated Raman-like process}, enabled by the finite spectral bandwidth of the driving field, rather than a conventional multi-color or multi-pulse Raman scheme. As a consequence, vibrational excitation on the ground electronic state remains finite even in the long-pulse limit (but much weaker than the linear process), provided that the pulse bandwidth does not vanish.

\subsection{Collectivity and Scaling with the Number of Molecules}
In this section, we examine the collective nature of vibrational activation and its scaling with the number of molecules $N$. All results are obtained under the resonant condition $\Omega=\nu=0.20$~eV (so that the UP and the LP are located at $\omega_+ = 2.1~\text{eV}$ and $\omega_- = 1.9~\text{eV}$, respectively) with an ultrashort pulse excitation, where the field amplitude is fixed as $\lambda_F=1$~meV. Within this protocol, we explore the few-molecule regime up to $N=8$.

Figure~\ref{fig:5} explores the dependence of field-driven vibrational activation on the number of molecules $N$. 
Fig.~\ref{fig:5}a show the time-dependent vibrational population in the excited electronic and photonic states of the molecular subsystem, $N \cdot P_{\mathrm{E},\nu=1}(t)$~\footnote{Alternatively, one can present the total vibration energy on the electronic excited states, $N \sum_\nu \nu P_\mathrm{E, \nu}(t)$, which looks very similar because this result is dominated by the $\nu = 1$ term. } plotted against $N$. These curves are vertically displaced for clarity. 
The oscillating behavior corresponds to Rabi oscillations between the UP and the vibrationally excited lower polariton, LP + $1\nu$. The period of these oscillations and the average vibrational energy in the ground and excited electronic and photonic state of the molecular subsystem are shown as functions of $N$ in Figs.~\ref{fig:5}b and \ref{fig:5}c, respectively. The following observations can be made:
\begin{itemize}
    \item For the weak incident field used here (for which the 1-exciton approximation is valid), energy transfer to the vibrational DOF of the molecular subsystem is dominated by the UP $\leftrightarrow$ LP $+\,1\nu$ process that leave the molecule in the excited electronic state. The Raman-like process (right panel of Fig.~\ref{fig:1}b) that leaves the molecule in the ground electronic state is seen, however with much smaller probability (Fig.~\ref{fig:5}c). 
    \item The total energy absorbed by the system reflects the integrated intensity (fluence) of the incident pulse and does not depend on $N$ (recall that the incident field is coupled to the cavity mode). The fact that the amplitude of the Rabi oscillations (Fig.~\ref{fig:5}a) and the average vibrational energy imparted in the excited electronic and photonic state (Fig.~\ref{fig:5}c) do not depend on $N$ reflects the fact an $N$-independent fraction of the energy absorbed by the system is periodically exchanged with the vibrational subsystem. Obviously the exchanged vibrational energy per molecule scales as $1/N$. It is interesting to note that the (extremely small) vibrational excitation seen in the ground electronic state falls with increasing $N$, reflecting the fact that the amplitude ($\sim 1/\sqrt{N}$) of the molecular bright state eventers in higher order in the transition probability for non-linear Raman-like transition. 
    \item The fact that the period of the Rabi oscillation scales like $T(N) \sim \sqrt{N}$ (Fig.~\ref{fig:5}b) reflects the well-known polaron decoupling effect~\cite{Herrera_Spano_PRL, herrera2018theory} --- the coupling between the molecular electronic bright mode (hence the electronic polaritons) and molecular vibrations scales like $1 / \sqrt{N}$ (see Sec.~III-B and Eq.~S6 of the Supporting Information). 
\end{itemize}

Taken together, these results indicate that the total vibration energy under resonant polariton driving is governed primarily by collective light--matter coupling rather than by the absolute number of molecules in the ensemble. The corresponding results for long-pulse excitation are presented in Fig.~S11 of the Supporting Information, where a similar scaling behavior is observed.
In contrast, the rate of vibrational excitation, as characterized by the oscillation period, decreases with increasing $N$, reflecting the reduction of effective polariton--phonon coupling strength in larger ensembles. As a consequence, in the large-$N$ limit vibrational activation can become increasingly slow and may be strongly suppressed by dissipation channels, when present.

\subsection{Mean-Field Dynamics and Connection to FDTD Simulations}
We conclude by comparing the SE and MF descriptions of field-driven polariton dynamics, and clarifying their connection to classical electrodynamical simulations such as the finite-difference time-domain (FDTD) approach.

For field-driven polariton dynamics, the MF results are presented in Fig.~S2 in the Supporting Information. Both the SE and MF approaches capture the essential features of polariton excitation within the HTC model. The SE framework explicitly resolves polariton eigenstates and dark states manifold, allowing direct access to polariton populations and light--matter coherences in addition to exciton and photon populations. In contrast, the MF description is formulated in terms of collective excitonic and photonic expectation values and does not explicitly resolve polariton or dark states. When expressed in terms of exciton and photon populations, SE and MF results are consistent under long-pulse excitation, where polaritonic coherence plays a limited role. Under ultrashort-pulse excitation, however, the two approaches differ quantitatively, reflecting the presence of significant UP--LP coherence that is explicitly retained in the SE description but only indirectly encoded in the MF dynamics.

For polariton induced vibration excitation, the corresponding MF results are shown in Fig.~S3 in the Supporting Information. Both SE and MF descriptions capture the essential physics of vibrational excitation governed by the Rabi splitting, although the resonance manifests in different observables. In the SE framework, resonant vibrational activation appears primarily in the vibrational population in the excited electronic and photonic states $\langle P_{\mathrm{E},\nu}\rangle$. In contrast, within the MF description the resonance is predominantly reflected in the vibrational population in the ground electronic state $\langle P_{\mathrm{g},\nu}\rangle$ (see definition in Sec.~II-B of the Supporting Information). It is important to emphasize that the quantities $\langle P_{\mathrm{G},\nu}\rangle$ and $\langle P_{\mathrm{E},\nu}\rangle$ in the SE approximation are not directly comparable to $\langle P_{\mathrm{g},\nu}\rangle$ and $\langle P_{\mathrm{e},\nu}\rangle$ in the MF framework, as the two approaches employ distinct reduced descriptions of the electronic and photonic DOF and yield different microscopic dynamics. Nevertheless, both approaches consistently identify the Rabi splitting as the key energy scale governing resonant vibrational activation. Moreover, we show that light--matter beating in the MF dynamics plays a crucial role in enabling resonant vibrational excitation, regardless of a quantum or classical description of the nuclei (see Figs.~S4-S6 in the Supporting Information). We further analyze the impact of cavity loss ($\kappa \neq 0$) on vibrational excitation (Fig.~S7 and S13 in the Supporting Information). 

For the polariton-mediated intrapulse stimulated Raman-like process, the MF results are presented in Fig.~S8 in the Supporting Information. Both SE and MF descriptions reproduce identical field-amplitude scaling relations for the averaged vibrational energies and populations, independent of whether ultrashort or long pulses are employed. In both approaches, vibrational excitation is dominated by contributions from the electronically (and photonic) excited states and scales quadratically with the driving field amplitude $\lambda_F$, while a much weaker electronic ground state contribution persists and exhibits quartic scaling. This agreement demonstrates that the observed scaling behavior is a robust feature of polariton-mediated intrapulse stimulated Raman-like process and does not depend sensitively on the specific theoretical framework used to describe the driven light--matter system.

Regarding the dependence on the number of molecules $N$, the MF dynamics exhibits no explicit $N$-dependence in the oscillation period of $P_{g, \nu=1}(t)$ or total vibration energy.
See Fig.~S9 of the Supporting Information. 
This discrepancy reflects a structural limitation of mean-field factorization~\cite{Cui_JCP2022}: MF does not capture polaron decoupling (and the associated $c_\nu / \sqrt{N}$ renormalization), leading to an $N$-independent oscillation period in the weak-excitation regime.

Taken together, the SE and MF approaches describe different sets of physical observables but encode the same underlying Rabi-driven physics under pulsed excitation, including identical polaritonic energy scales and vibronically induced resonances. In the SE framework, vibrational activation can be interpreted in terms of transitions between polaritonic eigenstates, such as processes involving energy exchange between the UP and the LP accompanied by vibrational excitation. In contrast, the MF description does not admit a state-resolved picture of such transitions. Instead, the same resonant vibrational response emerges from Raman-like frequency mixing between the cavity field and the collective molecular polarization, driven by the bandwidth of the optical pulse.
The essential formal distinction between the two approaches is that the MF description does not explicitly retain polariton eigenstates or light--matter correlations. Consequently, it cannot capture genuine UP--LP coherence arising from polaritonic superpositions, but only a classical analogue manifested as beating between the MF normal modes. 

Finally, we note that the MF dynamics considered here is closely related to FDTD simulations based on coupled Maxwell--quantum dynamics~\cite{Sukharev_JCTC2025, Bustamante_JCTC2025, Sidler_Nanoph2025, bustamante_2026}. In such approaches, the cavity is modeled explicitly through its dielectric environment and geometry, providing a realistic description of the electromagnetic field. The molecular ensemble can be described either by an effective model Hamiltonian such as the shifted harmonic oscillator model employed in the present work, and accurate wavepacket dynamics can be performed~\cite{Sukharev_JCTC2025}; or by electronic-structure approaches with the inclusion of atomistic details, for example, the time-dependent density-functional tight-binding (TD-DFTB) methods associated with Ehrenfest dynamics for the (classical) nuclei~\cite{Bonafe_JCTC2020, Bustamante_JCTC2025, Sidler_Nanoph2025, bustamante_2026}.
The present MF formulation provides a natural bridge between quantum electrodynamics models of molecular polaritons and experimentally relevant Maxwell-based descriptions of light--matter interaction.

\section{Conclusion} \label{Sec:conclusion}
In this work, we have presented a comprehensive theoretical investigation of vibrational activation induced by transient optical driving in molecular ensembles strongly coupled to a cavity mode. Using the field-driven HTC model, we systematically analyzed how pulsed excitation redistributes energy among electronic, photonic, and vibrational DOF across a wide range of pulse durations, intensities, and system sizes. By employing the SE and the MF description of collective light--matter coupling in parallel, and comparing their predictions, we obtain a consistent theoretical picture of vibrational activation under nonequilibrium polariton dynamics.
A central theoretical result of this study is the identification of the linear and nonlinear vibrational excitation pathways, characterized by distinct power-law scaling with respect to the driving field amplitude. Vibrational excitation on the excited electronic and photonic states scales quadratically with the pulse amplitude $\lambda_F$, while the ground electronic state contribution exhibits a nonlinear quartic scaling. We demonstrated that this nonlinear quartic scaling behavior originates from a polariton-mediated intrapulse stimulated Raman-like process, enabled by the finite bandwidth of a single optical pulse. Unlike conventional Raman schemes that rely on multiple colors or sequential pulses, the present mechanism operates as a self-induced, single-pulse process and persists under both ultrashort and spectrally selective long-pulse excitation.

Our analysis further clarified the role of polaritonic coherence in mediating vibrational dynamics. Broadband ultrashort pulses generate coherent superpositions of the UP and the LP, leading to pronounced UP--LP beating and enhanced vibrational activation when the Rabi splitting resonates with molecular vibrational frequencies. In contrast, long-pulse excitation suppresses broadband polariton coherence but still supports resonant vibrational activation through spectrally selective polariton driving. These results underscore the importance of controlling the pulse bandwidth and central frequency in order to selectively access coherence-dominated or population-dominated regimes of polariton dynamics, and thereby tailor the resulting vibrational activation. A direct comparison between the SE and MF descriptions demonstrates that, despite their distinct formulations and sets of accessible observables, both approaches capture the same underlying Rabi-driven physics and yield identical scaling relations for vibrational activation, providing consistent support for the stimulated Raman mechanism. 
Nevertheless, discrepancies between SE and MF descriptions are observed under certain narrow band excitation conditions (such as Figs.~\ref{fig:3}c-d vs. Figs.~S3e-h), as well as the $N$-dependence of the oscillation period (Fig.~\ref{fig:5} vs. Fig.~S9). 
On the other hand, the MF framework provides a direct conceptual bridge to classical electrodynamical simulations based on Maxwell+quantum dynamics approaches. This connection facilitates direct comparison between microscopic cavity quantum electrodynamics models and experimentally realistic cavity geometries and electromagnetic environments.
More importantly, the predicted resonant vibrational activation shall be experimentally probed via time-resolved vibrational spectroscopy, such as transient absorption or Raman measurements sensitive to population changes in specific vibrational modes.

From an applications perspective, the ability to induce and control vibrational excitation through tailored polariton dynamics opens new opportunities for the control of vibronic processes in strongly coupled systems. 
The stimulated Raman mechanism identified here suggests that vibrational population can be selectively enhanced or suppressed by engineering pulse bandwidth, amplitude, and detuning relative to polaritonic energy scales, without requiring strong-field or multi-pulse protocols. Such control strategies are directly relevant for ultrafast pump--probe, transient absorption, and multidimensional spectroscopies of molecular polaritons, as well as for cavity-modified photochemistry where vibrational activation plays a key role in reaction pathways~\cite{Chang_JACS2026}.
More broadly, our results provide a transparent framework for interpreting ultrafast polariton experiments and for designing optical protocols that exploit collective light--matter coupling to steer vibrational and vibronic dynamics. Extensions of this work to include realistic molecular and cavity dissipation, anharmonic vibrations, multimode cavities, and strong electronic correlations will further deepen our understanding of nonequilibrium polariton chemistry and pave the way toward rational, coherence-based control of molecular function in optical cavities.

\section*{$\blacksquare$ Associated Content}

\subsection*{Supporting Information}
The Supporting Information is available free of charge at [url]. 

The details on dark states derived from the HTC Hamiltonian; mean-field quantum dynamics results of the field-driven HTC model, including derivation for the MF equations of motion, computational details, field-driven polariton dynamics, resonant Rabi-driven vibrational excitation, effect of cavity loss, scaling behaviors with respect to the driving field amplitude, and $N$-independence of the vibrational population oscillation period; additional information on the dynamics within the SE subspace, including details on the numerical simulations, linear fitting results in Figs.~4 and 5 of the main text, non-resonant behavior of the vibrational population in the ground electronic state, effect of the number of molecules $N$ under a long-pulse excitation, resonance peak shift due to vibronic coupling induced frequency renormalization, and effect of cavity loss. 

\section*{$\blacksquare$ Acknowledgments}
This work was supported by the European Research Council under ERC-2024-SyG-101167294; UnMySt. C. M. Bustamante thanks the Alexander von Humboldt-Stiftung for the financial support from the Humboldt Research Fellowship. M.S. acknowledges support by the Office of Naval Research, Grant No. N000142512090 and the Air Force Office of Scientific Research under Grant No. FA9550-25-1-0096. F.P.B. acknowledges financial support from the European Union’s Horizon 2020 research and innovation program under the Marie Sklodowska-Curie Grant Agreement no. 895747 (NanoLightQD).

\section*{$\blacksquare$ Author Information}



Complete contact information is available at: [url]

\subsection*{Notes}
The authors declare no competing financial interest.


\section*{$\blacksquare$ References}
\bibliography{Ref}

\end{document}


\title[]{\large Supporting Information for \\
Linear and nonlinear vibrational excitation driven by molecular polaritons}

\author{Wenxiang Ying}
\email{wying3@sas.upenn.edu}
\affiliation{Department of Chemistry, University of Pennsylvania, Philadelphia, Pennsylvania 19104, USA}

\author{Carlos M. Bustamante}
\author{Franco P. Bonafé}
\affiliation{Max Planck Institute for the Structure and Dynamics of Matter and Center for Free-Electron Laser Science, Luruper Chaussee 149, Hamburg 22761, Germany}

\author{Richard Richardson}
\affiliation{Department of Physics, Arizona State University, Tempe, Arizona 85287, United States}

\author{Michael Ruggenthaler}
\affiliation{Max Planck Institute for the Structure and Dynamics of Matter and Center for Free-Electron Laser Science, Luruper Chaussee 149, Hamburg 22761, Germany}

\author{Maxim Sukharev}
\email{Maxim.Sukharev@asu.edu}
\affiliation{Department of Physics, Arizona State University, Tempe, Arizona 85287, United States}
\affiliation{College of Integrative Sciences and Arts, Arizona State University, Mesa, Arizona 85212, United States}

\author{Angel Rubio}
\email{angel.rubio@mpsd.mpg.de}
\affiliation{Max Planck Institute for the Structure and Dynamics of Matter and Center for Free-Electron Laser Science, Luruper Chaussee 149, Hamburg 22761, Germany}
\affiliation{Initiative for Computational Catalysis (ICC), Flatiron Institute, Simons Foundation,  162 5th Avenue, New York, NY 10010 USA}

\author{Abraham Nitzan}
\email{anitzan@sas.upenn.edu}
\affiliation{Department of Chemistry, University of Pennsylvania, Philadelphia, Pennsylvania 19104, USA}
\affiliation{School of Chemistry, Tel Aviv University, Tel Aviv 69978, Israel}

\maketitle
\tableofcontents

\newpage
\section{Dark states derived from the Holstein--Tavis--Cummings Hamiltonian}

In addition to the two polariton eigenstates $|\pm\rangle$ defined in Eq.~7 of the main text, the single-excitation (SE) manifold of the Holstein--Tavis--Cummings (HTC) Hamiltonian contains $N-1$ dark states $|\mathrm{D}_k\rangle$, which are given by~\cite{tavis1968exact,TaoLi_2022}
\begin{equation}
\label{eq:dark}
|\mathrm{D}_k\rangle
=\frac{1}{\sqrt{N}}\sum_{n=1}^{N} \exp\!\left(-2\pi i\,\frac{n k}{N}\right)\,|\mathrm{E}_n,0\rangle,
\qquad k=1,\dots,N-1.
\end{equation}
These dark states are orthogonal to the collective bright exciton and remain degenerate at the bare excitonic energy $\hbar\omega_0$. 
Together with the polariton states $|\pm\rangle$, they form the complete polariton (Tavis--Cummings) basis of the SE manifold~\cite{tavis1968exact}.

To express the HTC Hamiltonian in this basis, we further perform a discrete Fourier transform of the vibrational annihilation (creation) operators,
\begin{equation}
\hat{b}_{k}
=\frac{1}{\sqrt{N}}\sum_{n=1}^{N}
\exp\!\left(2\pi i\,\frac{n k}{N}\right)\hat{b}_{n},
\end{equation}
with the inverse transform defined analogously. 
In the polariton basis, the HTC Hamiltonian (without the external field driving term) can be written as
\begin{equation}
\hat{H}_{\mathrm{HTC}}
=\hat{H}_{\mathrm{TC}}+\hat{h}_{\mathrm{vib}}+\hat{H}_{\mathrm{ex\text{-}ph}},
\end{equation}
where the Tavis--Cummings Hamiltonian becomes diagonal,
\begin{equation}
\hat{H}_{\mathrm{TC}}
=\hbar\omega_{+}|+\rangle\langle+|
+\hbar\omega_{-}|-\rangle\langle-|
+\hbar\omega_{0}\sum_{k=1}^{N-1}|\mathrm{D}_k\rangle\langle\mathrm{D}_k|,
\end{equation}
the vibrational Hamiltonian reads
\begin{equation}
\hat{h}_{\mathrm{vib}}
=\hbar\nu\sum_{n}\hat{b}_n^\dagger\hat{b}_n
=\hbar\nu\sum_{k}\hat{b}_k^\dagger\hat{b}_k,
\end{equation}
and the exciton--phonon coupling term takes the form
\begin{align}
\hat{H}_{\mathrm{ex\text{-}ph}}
&=\sum_{n}|\mathrm{E}_n,0\rangle\langle\mathrm{E}_n,0|
\otimes c_\nu(\hat{b}_n+\hat{b}_n^\dagger) \notag\\
&=\Big[\cos^2\Theta\,|+\rangle\langle+|
+\sin^2\Theta\,|-\rangle\langle-|
-\dfrac{\sin(2\Theta)}{2}\big(|+\rangle\langle-|+|-\rangle\langle+|\big)\Big]
\otimes \frac{c_\nu}{\sqrt{N}}(\hat{b}_0+\hat{b}_0^\dagger) \notag\\
&\quad +\cos\Theta\Big[
\sum_{k=1}^{N-1}
|\mathrm{D}_k\rangle\langle+|
\otimes \frac{c_\nu}{\sqrt{N}}(\hat{b}_k+\hat{b}_{-k}^\dagger)
+\mathrm{h.c.}\Big] \notag\\
&\quad -\sin\Theta\Big[
\sum_{k=1}^{N-1}
|\mathrm{D}_k\rangle\langle-|
\otimes \frac{c_\nu}{\sqrt{N}}(\hat{b}_{-k}+\hat{b}_k^\dagger)
+\mathrm{h.c.}\Big] \notag\\
&\quad +\sum_{k=1}^{N-1}\sum_{j=1}^{N-1}
|\mathrm{D}_k\rangle\langle\mathrm{D}_j|
\otimes \frac{c_\nu}{\sqrt{N}}(\hat{b}_{-j+k}+\hat{b}_{j-k}^\dagger).
\label{eq_HTC_HSB}
\end{align}
Here, $\Theta$ denotes the polariton mixing angle as defined in the main text. 
In Eq.~\ref{eq_HTC_HSB}, vibrational mode indices outside the range $\{1,\dots,N-1\}$ are understood modulo $N$, reflecting discrete translational symmetry of the molecular ensemble; for example, $\hat{b}_{-k}\equiv\hat{b}_{N-k}$.
Eq.~\ref{eq_HTC_HSB} explicitly shows that direct transitions between the polariton states $|\pm\rangle$ and the dark-state manifold $\{|\mathrm{D}_k\rangle\}$ are mediated solely by vibrational degrees of freedom (DOF). 
In particular, these transitions become resonant when the corresponding polaritonic energy splittings match the vibrational quantum $\hbar\nu$, providing the microscopic origin of UP--LP coupling and dark-state--assisted vibronic dynamics discussed in the main text.

\newpage
\section{Mean-Field Dynamics}
In this section, we present mean-field (MF) quantum dynamics results of the field-driven HTC model, including derivation for the MF equations of motion, computational details, field-driven polariton dynamics, resonant Rabi-driven vibrational activation, effect of cavity loss, scaling behaviors with respect to the driving field amplitude, and $N$-independence of the vibrational population oscillation period.

\subsection{Derivation for the MF equations of motion} \label{sec:EOM-MF}
We follow Ref.~\citenum{Keeling_PRL2022} to derive the mean-field approximation and equations of motion.  
We begin by assuming a mean-field product ansatz for the density matrix,
\begin{equation} \label{eq:DM_factorize}
\hat{\rho}(t) = \hat{\rho}_a(t) \bigotimes_{n=1}^{N} \hat{\rho}_n(t),
\end{equation}
where $\hat{\rho}_a(t)$ denotes the cavity field reduced density matrix (RDM), and $\hat{\rho}_n(t)$ denotes the $n_\text{th}$ molecular RDM.
The system evolves according to the Liouville-von Neumann equation,
\begin{equation} \label{eq:Lvn}
    \frac{\partial}{\partial t} \hat{\rho}(t) = -\frac{i}{\hbar}[\hat{H}, \hat{\rho}(t)],
\end{equation}
with the field-driven HTC Hamiltonian
\begin{align}
    \hat{H}(t) &= \hbar \omega_\text{c} \hat{a}^\dagger \hat{a} + \hat{H}_\text{pulse}(t) + \sum_{n=1}^{N} \left[\hat{H}_n + \hbar g_\mathrm{c} \left(\hat{a}^{\dagger}\hat{\sigma}_n^- + \hat{a}\hat{\sigma}_n^+ \right) \right], 
\end{align}
where $\hat{H}_n(t)$ is the individual molecular Hamiltonian for molecule $n$, reading as
\begin{align}
    \hat{H}_n &= \hbar \omega_{0} \hat{\sigma}_n^+ \hat{\sigma}_n^- + \hbar \nu \hat{b}^\dagger_{n} \hat{b}_{n} + \hat{\sigma}_n^+ \hat{\sigma}_n^- \otimes c_\nu (\hat{b}_{n} + \hat{b}^\dagger_{n}). 
\end{align}
And $\hat{H}_\text{pulse}(t) = i\hbar [\hat{a}^\dagger \cdot E_-(t) - \hat{a} \cdot E_+(t)]$ characterizes the external-field driving of the cavity mode.

By taking partial trace on both sides of Eq.~\ref{eq:Lvn}, one obtains the equations of motion for the RDMs as follows,
\begin{align}
    \frac{\partial}{\partial t} \hat{\rho}_a(t) &= -\frac{i}{\hbar}\,\mathrm{Tr}_{\otimes n}[\hat{H}(t), \hat{\rho}(t)], \label{eq:EOM_rhoa_0} \\
    \frac{\partial}{\partial t} \hat{\rho}_n(t) &= -\frac{i}{\hbar}\,\mathrm{Tr}_{a,\otimes m \neq n}[\hat{H}(t), \hat{\rho}(t)], \label{eq:EOM_rhon_0} 
\end{align}
respectively. Evaluating the partial traces yields a set of coupled mean-field equations. For the photon mode DOF, it yields
\begin{align} \label{eq:EOM_rhoa}
    \frac{\partial}{\partial t} \hat{\rho}_a(t) &= -\frac{i}{\hbar}\,[\hbar \omega_\text{c} \hat{a}^\dagger \hat{a} - i\hbar \hat{a} \cdot E_+(t) + i\hbar \hat{a}^\dagger \cdot E_-(t), \hat{\rho}_a(t)] - \frac{i}{\hbar} \sum_{n=1}^{N} \hbar g_\mathrm{c} \left(\mathrm{Tr}_n [\hat{a}^{\dagger} \hat{\sigma}_n^-, \hat{\rho}_a(t) \otimes \hat{\rho}_n(t)] + \text{h.c.} \right) \notag\\
    &= -i \omega_\text{c} \,[\hat{a}^\dagger \hat{a}, \hat{\rho}_a(t)] - E_+(t) [\hat{a}, \hat{\rho}_a(t)] + E_-(t) [\hat{a}^\dagger, \hat{\rho}_a(t)] \notag\\
    &~~~ - \frac{i}{\hbar} \sum_{n=1}^{N} \hbar g_\mathrm{c} \left([\hat{a}^{\dagger}, \hat{\rho}_a(t)]  \times \mathrm{Tr}_n [\hat{\rho}_n(t)\hat{\sigma}_n^-] + \bcancel{\hat{a}^{\dagger} \hat{\rho}_a(t) \mathrm{Tr}_n [\hat{\sigma}_n^-, \hat{\rho}_n(t)]} + \text{h.c.} \right) \notag\\
    &= -i \omega_\text{c} \,[\hat{a}^\dagger \hat{a}, \hat{\rho}_a(t)] - E_+(t) [\hat{a}, \hat{\rho}_a(t)] + E_-(t) [\hat{a}^\dagger, \hat{\rho}_a(t)] - i g_\mathrm{c} [\hat{a}^{\dagger}, \hat{\rho}_a(t)] \sum_{n=1}^{N} \sigma_n (t) - i g_\mathrm{c} [\hat{a}, \hat{\rho}_a(t)] \sum_{n=1}^{N} \sigma^*_n (t),
\end{align}
where $\sigma_n (t) := \mathrm{Tr}_n [\hat{\rho}_n(t)\hat{\sigma}_n^-]$, and $\sigma^*_n (t)$ denotes the complex conjugation.  
Eq.~\ref{eq:EOM_rhoa} can be further simplified as an algebraic equation of the time-dependent complex cavity field amplitude $\alpha (t) := \mathrm{Tr}[\hat{a} \hat{\rho}_a(t)]$. Associating $\hat{a}$ from the left then taking trace on both sides of Eq.~\ref{eq:EOM_rhoa}, one obtains
\begin{align}
    &\frac{\partial}{\partial t} \mathrm{Tr}[\hat{a} \hat{\rho}_a(t)] = -i \omega_\text{c} \,\mathrm{Tr}\{ \hat{a}[\hat{a}^\dagger \hat{a}, \hat{\rho}_a(t)]\} - E_+(t) \mathrm{Tr}\{ \hat{a} [\hat{a}, \hat{\rho}_a(t)] \} + E_-(t) \mathrm{Tr}\{ \hat{a} [\hat{a}^\dagger, \hat{\rho}_a(t)] \} \notag\\
    &~~~~~~~~~~~~~~~~~~~ - i g_\mathrm{c} \mathrm{Tr}[\hat{a} [\hat{a}^{\dagger}, \hat{\rho}_a(t)]\} \sum_{n=1}^{N} \sigma_n (t) - i g_\mathrm{c} \mathrm{Tr}[\hat{a} [\hat{a}, \hat{\rho}_a(t)]\} \sum_{n=1}^{N} \sigma^*_n (t) \notag\\
    &~~~~~~ = -i \omega_\text{c} \,\mathrm{Tr}\{ \hat{a}\hat{a}^\dagger \hat{a} \hat{\rho}_a(t) - \hat{a} \hat{\rho}_a(t) \hat{a}^\dagger \hat{a} \} - E_+(t) \bcancel{\mathrm{Tr}[\hat{a}^2 \hat{\rho}_a(t) - \hat{a} \hat{\rho}_a(t) \hat{a} ]} + E_-(t) \mathrm{Tr}\{ \hat{a} \hat{a}^\dagger \hat{\rho}_a(t) - \hat{a} \hat{\rho}_a(t) \hat{a}^\dagger\} \notag\\
    &~~~~~~~~~~~~~~~~~~~ - i g_\mathrm{c} \mathrm{Tr}[\hat{a} \hat{a}^{\dagger} \hat{\rho}_a(t) - \hat{a} \hat{\rho}_a(t) \hat{a}^{\dagger} \} \sum_{n=1}^{N} \sigma_n (t) - i g_\mathrm{c} \bcancel{\mathrm{Tr}[\hat{a}^2 \hat{\rho}_a(t) - \hat{a} \hat{\rho}_a(t) \hat{a} ]} \sum_{n=1}^{N} \sigma^*_n (t) \notag\\
    &~~~~~~ = -i \omega_\text{c} \,\mathrm{Tr}\{ (\hat{a}^\dagger\hat{a} + 1) \hat{a} \hat{\rho}_a(t) - \hat{a} \hat{\rho}_a(t) \hat{a}^\dagger \hat{a} \} + E_-(t) \mathrm{Tr}[(\hat{a}^\dagger\hat{a} + 1) \hat{\rho}_a(t) - \hat{a} \hat{\rho}_a(t) \hat{a}^{\dagger} ] \notag\\
    &~~~~~~~~~~~~~~~~~~~ - i g_\mathrm{c} \mathrm{Tr}[(\hat{a}^\dagger\hat{a} + 1) \hat{\rho}_a(t) - \hat{a} \hat{\rho}_a(t) \hat{a}^{\dagger} \} \sum_{n=1}^{N} \sigma_n (t) \notag\\
    &~~~~~~ = -i \omega_\text{c}\mathrm{Tr} [\hat{a} \hat{\rho}_a(t)] -i \omega_\text{c} \,\bcancel{\mathrm{Tr}\{\hat{a}^\dagger\hat{a} \hat{a} \hat{\rho}_a(t) - \hat{a} \hat{\rho}_a(t) \hat{a}^\dagger \hat{a} \}} +  E_-(t) \mathrm{Tr}[\hat{\rho}_a(t)] +  E_-(t) \bcancel{\mathrm{Tr}[\hat{a}^\dagger\hat{a} \hat{\rho}_a(t) - \hat{a} \hat{\rho}_a(t) \hat{a}^{\dagger} ]} \notag\\
    &~~~~~~~~~~~~~~~~~~~ - i g_\mathrm{c} \mathrm{Tr}[\hat{\rho}_a(t)] \sum_{n=1}^{N} \sigma_n (t) - i g_\mathrm{c} \bcancel{\mathrm{Tr}[\hat{a}^\dagger\hat{a} \hat{\rho}_a(t) - \hat{a} \hat{\rho}_a(t) \hat{a}^{\dagger}]} \sum_{n=1}^{N} \sigma_n (t) \notag\\
    &~~~~~~ = -i \omega_\text{c}\mathrm{Tr} [\hat{a} \hat{\rho}_a(t)] + E_-(t) - i g_\mathrm{c} \sum_{n=1}^{N} \sigma_n (t), 
\end{align}
which is just
\begin{align} \label{eq:EOM_alpha_1}
    \dot{\alpha}(t)
    = -\,i\,\omega_\mathrm{c}\,\alpha(t)
      - i g_\mathrm{c} \sum_{n=1}^{N} \sigma_n (t) + E_-(t). 
\end{align}
By assuming the permutation symmetry among all the molecules, $\sigma_n (t) \to \sigma (t)$ and $\sum_{n=1}^{N} \to N$, and further include phenomenological cavity loss (with loss rate $\kappa$), the equations of motion in Eq.~\ref{eq:EOM_alpha_1} becomes
\begin{align} \label{eq:EOM_alpha_2}
    \dot{\alpha}(t)
    = -\,i\,(\omega_\mathrm{c} - i\kappa/2)\,\alpha(t)
      - i g_\mathrm{c}\,N\,\sigma(t) + E_-(t), 
\end{align}
which is just Eq.~12 of the main text. 

For the molecular DOF, Eq.~\ref{eq:EOM_rhon_0} reads as
\begin{align} \label{eq:LvN_MF}
    \frac{\partial}{\partial t} \hat{\rho}_n(t) &= -\frac{i}{\hbar} [\hat{H}_n, \hat{\rho}_n(t)] - i g_\mathrm{c} \mathrm{Tr} [\hat{a}^\dagger \hat{\rho}_a(t)] [\hat{\sigma}_n^-, \hat{\rho}_n(t)] - i g_\mathrm{c} \mathrm{Tr} [\hat{a} \hat{\rho}_a(t)] [\hat{\sigma}_n^+, \hat{\rho}_n(t)] \notag\\
    &= -\frac{i}{\hbar} [\hat{H}_n, \hat{\rho}_n(t)] - i g_\mathrm{c} \alpha^*(t) [\hat{\sigma}_n^-, \hat{\rho}_n(t)] - i g_\mathrm{c} \alpha(t) [\hat{\sigma}_n^+, \hat{\rho}_n(t)] \notag\\
    &= -\frac{i}{\hbar} [\hat{H}^\text{eff}_n(\alpha(t)), \hat{\rho}_n(t)],
\end{align}
where $\hat{H}^\text{eff}_n(\alpha(t)) := \hat{H}_n + \hbar g_\mathrm{c} [\alpha^*(t) \hat{\sigma}_n^- + \alpha(t) \hat{\sigma}_n^+]$ is the effective Hamiltonian. 
Further assuming the permutation symmetry among all the molecules, $\hat{\rho}_n(t) \to \hat{\rho}_\text{M}(t)$, $\hat{\sigma}_n^- \to |g\rangle \langle e|$, $\hat{\sigma}_n^+ \to |e\rangle \langle g|$, where $\{|g\rangle, |e\rangle\}$ is a single set of molecular electronic state basis. Furthermore, $\hat{b}_n \to \hat{b}$ and $\hat{b}^\dagger_n \to \hat{b}^\dagger$. As a result, the effective Hamiltonian reads as
\begin{align} \label{eq:Heff}
    \hat{H}^\text{eff}_n(\alpha(t)) \to \hat{H}^\text{eff}(\alpha(t)) &= \hbar \omega_0 |e\rangle \langle e| + \hbar\nu\,b^\dagger b + c_\nu\,|e\rangle\!\langle e| \otimes (b+b^\dagger) + \hbar g_\mathrm{c}\left(\alpha^{*}(t)|g\rangle\!\langle e| + \alpha(t)\,|e\rangle\!\langle g|\right), 
\end{align}
which is just Eq.~10 of the main text. And the equations of motion for $\hat{\rho}_\text{M}(t)$ reads as 
\begin{align}
    \frac{\partial}{\partial t} \hat{\rho}_\text{M}(t) &= -\frac{i}{\hbar} [\hat{H}^\text{eff}(\alpha(t)), \hat{\rho}_\text{M}(t)], 
\end{align}
which is just Eq.~11 of the main text. 
One can further add driving and dissipation terms when necessary -- such as Lindblad terms for electronic relaxation, pure dephasing, and vibrational damping. 

In the absence of environment-induced decoherence, the Liouville--von Neumann equation in Eq.~\ref{eq:LvN_MF} reduces to a Schr\"odinger equation for single-molecule wavefunctions,
\begin{align} \label{eq:Schrodinger_MF}
    i \hbar \frac{\partial}{\partial t} |\psi_n(t)\rangle
    = \hat{H}_n^{\mathrm{eff}}(\alpha(t)) \, |\psi_n(t)\rangle,
\end{align}
which is numerically more efficient to solve. Moreover, the resulting Schr\"odinger equation governed by the effective Hamiltonian $\hat{H}_n^{\mathrm{eff}}(\alpha(t))$ is fully consistent with the MF approximation adopted in Ref.~\citenum{Cui_JCP2023}, which is based on a time-dependent Hartree product wavefunction ansatz,
\begin{align} \label{eq:TDH}
    |\Psi(t)\rangle = |\alpha(t)\rangle \prod_{n=1}^N |\psi_n(t)\rangle ,
\end{align}
where $|\alpha(t)\rangle$ is a coherent state satisfying $\hat{a}|\alpha(t)\rangle = \alpha(t)|\alpha(t)\rangle$, and each $|\psi_n(t)\rangle$ represents the vibronic state of molecule $n$. Treating $\alpha(t)$ as a dynamical variable then leads to Eq.~\ref{eq:EOM_alpha_2}.
Furthermore, by assuming the permutation symmetry among all the molecules, one can further reduce Eq.~\ref{eq:Schrodinger_MF} into a single equation, with the effective Hamiltonian of Eq.~\ref{eq:Heff}. 

We note that an important limitation of the time-dependent Hartree product wavefunction ansatz in Eq.~\ref{eq:TDH} (or equivalently, the density matrix factorization in Eq.~\ref{eq:DM_factorize}) is its intrinsically single-configuration nature, which enforces a factorized description of the cavity field and molecular DOF. As a consequence, genuine light--matter and intermolecular entanglement---central to the quantum definition of polariton states---cannot be captured within this mean-field framework. While the MF approach correctly reproduces polariton energies and the associated Rabi splitting through collective coupling, it does so at the level of classical normal-mode hybridization rather than quantum-entangled eigenstates. As such, even in the weak-pulse and large-$N$ limits---where mean-field descriptions are commonly assumed to be reliable---the MF approach may still miss important physics captured by the single-excitation representation, owing to its inability to describe explicit light--matter entanglement and state-to-state correlations.

\subsection{Computational details}

\subsubsection{Details on mean-field quantum dynamics propagation}
We perform quantum dynamics simulations by numerically integrating the MF equations of motion derived from the HTC model, 
\begin{subequations} \label{eq:MF-EOM}
\begin{align}
    \frac{\partial}{\partial t} \hat{\rho}_\text{M}(t) &= -\frac{i}{\hbar} [\hat{H}^\text{eff}(\alpha(t)), \hat{\rho}_\text{M}(t)], \\
    \dot{\alpha}(t) &= -\,i\,(\omega_\mathrm{c} - i\kappa/2)\,\alpha(t) - i g_\mathrm{c}\,N\,\sigma(t) + E_-(t),
\end{align}
\end{subequations}
where $\sigma(t) := \mathrm{Tr}\!\left[\hat{\rho}_\mathrm{M}(t)\,|g\rangle\!\langle e|\right]$, as detailed in Sec.~\ref{sec:EOM-MF}. Within the MF approximation, the cavity field is represented by a complex coherent amplitude $\alpha(t)$, while the molecular subsystem is described by a reduced vibronic density matrix $\hat{\rho}_{\mathrm{M}}(t)$ defined on a single-molecule Hilbert space spanned by $\{|g,\nu\rangle,|e,\nu\rangle\}$ with $\nu=0,\dots,N_\nu-1$ and evolved by an effective Hamiltonian $\hat{H}^\text{eff}(\alpha(t))$ (see Eq.~\ref{eq:Heff}), exploiting permutation symmetry in the large-$N$ limit (numerically we set $N = 10^4$).
The above coupled equations of motion for $\alpha(t)$ and $\rho_\mathrm{M}(t)$ are solved self-consistently in the time domain. The initial conditions are chosen as $\alpha(0)=0$ and $\rho_\mathrm{M}(0)=|g, \nu=0\rangle\langle g, \nu=0|$, corresponding to an empty cavity and molecules prepared in their electronic and vibrational ground states, respectively. 

For the molecular DOF, we choose $\hbar \omega_0 =$ 2.0 eV, $\hbar\nu =$ 0.20 eV, and $c_\nu =$ 20 meV unless otherwise specified. For the cavity DOF, we always keep the resonance condition $\hbar\omega_\text{c} = \hbar\omega_0 =$ 2.0 eV. And we set $\hbar\Omega =$ 0.20 eV unless otherwise specified. 
Time-dependent driving by a Gaussian optical pulse is included through the cavity field coupling. Here we consider two scenarios if not declared separately: an ultrashort pulse with temporal width $\Delta^{-1} = 2$ fs centered at $t_0 = 10$ fs, and a long pulse with temporal width $\Delta^{-1} = 50$ fs centered at $t_0 = 250$ fs. We use a field amplitude $\lambda_F =$ 1 meV by default unless otherwise specified (the same as in the main text). 
Furthermore, optional Lindblad terms can be incorporated to account for cavity loss, electronic relaxation, pure dephasing, and vibrational damping.

All MF simulations are carried out using a fourth-order Runge--Kutta (RK-4) scheme with a fixed time step of $\Delta t = 0.05$~fs. For ultrashort-pulse excitation, the total propagation time is $t_\text{max}=20$~ps in order to capture long-time vibrational dynamics and coherence-induced oscillations. For long-pulse excitation, a shorter propagation time of $t_\text{max}=2$~ps is sufficient due to the absence of long-lived oscillatory behavior. Numerical convergence with respect to vibrational Hilbert-space truncation is verified by retaining up to $N_\nu=4$ vibrational quanta per mode near the primary resonance condition $\Omega=\nu$, and up to $N_\nu=6$ near the secondary resonance $\Omega=2\nu$, with no qualitative changes observed upon further increase of the vibrational cutoff.

\subsubsection{Observables of interest}
From the resulting trajectories, we can compute electronic and photonic populations, vibrational populations resolved by vibronic states, and vibrational energies either per molecule or for the full ensemble. 
All molecular expectation values are evaluated as $\langle \hat{O} \rangle(t)=\mathrm{Tr}\!\left[\hat{\rho}_{\mathrm{M}}(t)\hat{O}\right]$, where the trace is over the single-molecule vibronic space. 
For an ensemble of $N$ identical, uncorrelated molecules, extensive quantities ({\it e.g.}, total vibrational energy) are obtained by multiplying the single-molecule result by $N$.
We note that, within the MF framework, linear optical response functions such as absorption spectra can also be computed using standard weak-probe or linear-response protocols. In the present work, however, we focus exclusively on nonequilibrium vibrational dynamics induced by pulsed excitation. In particular, we are interested in the following observables, \\

\paragraph{Electronic and photonic populations.}
The electronic excited-state population and cavity photon number are computed as
\begin{equation} \label{eq:Pex_Pph}
P_{\mathrm{ex}}(t)=\mathrm{Tr}\!\left[\hat{\rho}_{\mathrm{M}}(t)\,|e\rangle\langle e|\right], 
\qquad
P_{\mathrm{ph}}(t)=|\alpha(t)|^2,
\end{equation}
where $|e\rangle\langle e|$ acts on the electronic two-level subspace. 
Equivalently, $P_{\mathrm{ph}}(t)$ is the photon number associated with the coherent cavity field. Representative results are shown in Fig.~\ref{fig:s2}. 

\paragraph{Vibrational energies on the ground and excited electronic states.}
Let $\hat{n}=b^\dagger b$ denote the vibrational number operator, and $\hat{P}_\mathrm{g}=|g\rangle\langle g|$, $\hat{P}_\mathrm{e}=|e\rangle\langle e|$ the electronic projectors. 
The mean vibrational occupations are evaluated as
\begin{equation}
\langle \hat{n} \rangle(t)=\mathrm{Tr}\!\left[\hat{\rho}_{\mathrm{M}}(t)\hat{n}\right],\qquad
\langle \hat{n} \rangle_\mathrm{g}(t)=\mathrm{Tr}\!\left[\hat{\rho}_{\mathrm{M}}(t)\hat{P}_\mathrm{g}\hat{n}\right],\qquad
\langle \hat{n} \rangle_\mathrm{e}(t)=\mathrm{Tr}\!\left[\hat{\rho}_{\mathrm{M}}(t)\hat{P}_\mathrm{e}\hat{n}\right].
\end{equation}
The corresponding vibrational energies (measured relative to the zero-point energy) are given by
\begin{equation} \label{eq:Evib_MF}
E_{\mathrm{vib}} (t)=\hbar\nu \langle \hat{n} \rangle(t),\qquad
E_{\mathrm{vib,g}} (t)=\hbar\nu \langle \hat{n} \rangle_\mathrm{g}(t),\qquad
E_{\mathrm{vib,e}} (t)=\hbar\nu \langle \hat{n} \rangle_\mathrm{e}(t),
\end{equation}
for a single molecule. The total ensemble vibrational energies are obtained by multiplying these expressions by $N$. 

\paragraph{Vibrational populations resolved by vibronic states.}
To resolve vibrational populations on each electronic state, we introduce the projectors
\begin{equation}
\hat{\Pi}_{\mathrm{g},\nu}=|g,\nu\rangle\langle g,\nu|,\qquad 
\hat{\Pi}_{\mathrm{e},\nu}=|e,\nu\rangle\langle e,\nu|,\qquad \nu=0,\dots,N_\nu-1.
\end{equation}
The corresponding vibronic populations are evaluated as
\begin{equation} \label{eq:Pge_MF}
P_{\mathrm{g},\nu}(t)=\mathrm{Tr}\!\left[\hat{\rho}_{\mathrm{M}}(t)\hat{\Pi}_{\mathrm{g},\nu}\right],\qquad
P_{\mathrm{e},\nu}(t)=\mathrm{Tr}\!\left[\hat{\rho}_{\mathrm{M}}(t)\hat{\Pi}_{\mathrm{e},\nu}\right].
\end{equation}
Representative examples are shown in Figs.~\ref{fig:s3}a, \ref{fig:s3}b, \ref{fig:s3}e, \ref{fig:s3}f. 

\subsubsection{Time-averaging}
Furthermore, time-averaged observables are obtained by averaging the corresponding time-dependent quantities over a post-pulse time window $[t_{\mathrm{p}},t_\text{max}]$, chosen such that the external driving field has sufficiently decayed. 
In practice, we set $t_{\mathrm{p}}=100~\mathrm{fs}$ for the ultrashort-pulse excitation and $t_{\mathrm{p}}=1000~\mathrm{fs}$ for the long-pulse excitation, respectively.
Specifically, the time-averaged total ensemble vibrational energies are evaluated as
\begin{align}
\langle E_{\mathrm{vib}} \rangle &= N\,\frac{1}{t_\text{max}-t_{\mathrm{p}}}\int_{t_{\mathrm{p}}}^{t_\text{max}}\!dt\,E_{\mathrm{vib}} (t), \\
\langle E_{\mathrm{vib},\mathrm{g/e}} \rangle &= N\,\frac{1}{t_\text{max}-t_{\mathrm{p}}}\int_{t_{\mathrm{p}}}^{t_\text{max}}\!dt\,E_{\mathrm{vib},\mathrm{g/e}} (t),
\end{align}
where the expressions of $E_{\mathrm{vib}} (t)$ and $E_{\mathrm{vib},\mathrm{g/e}} (t)$ are given in Eq.~\ref{eq:Evib_MF}. 
Representative examples of the resulting vibrational energy dynamics are shown in Figs.~\ref{fig:s8}c and \ref{fig:s8}g.
Similarly, the time-averaged vibrational populations resolved by electronic state and vibrational quantum number are computed as
\begin{equation}
\langle P_{\mathrm{g/e},\nu} \rangle
= \frac{1}{t_\text{max}-t_{\mathrm{p}}}\int_{t_{\mathrm{p}}}^{t_\text{max}}\!dt\,P_{\mathrm{g/e},\nu}(t),
\end{equation}
where the expressions of $P_{\mathrm{g/e},\nu}(t)$ are given in Eq.~\ref{eq:Pge_MF}. Representative examples of these quantities are shown in Figs.~\ref{fig:s3}c, \ref{fig:s3}d, \ref{fig:s3}g, \ref{fig:s3}h, \ref{fig:s8}d, and \ref{fig:s8}h.

\subsection{Field-driven polariton dynamics}

\begin{figure}
    \centering
    \includegraphics[width=0.6\linewidth]{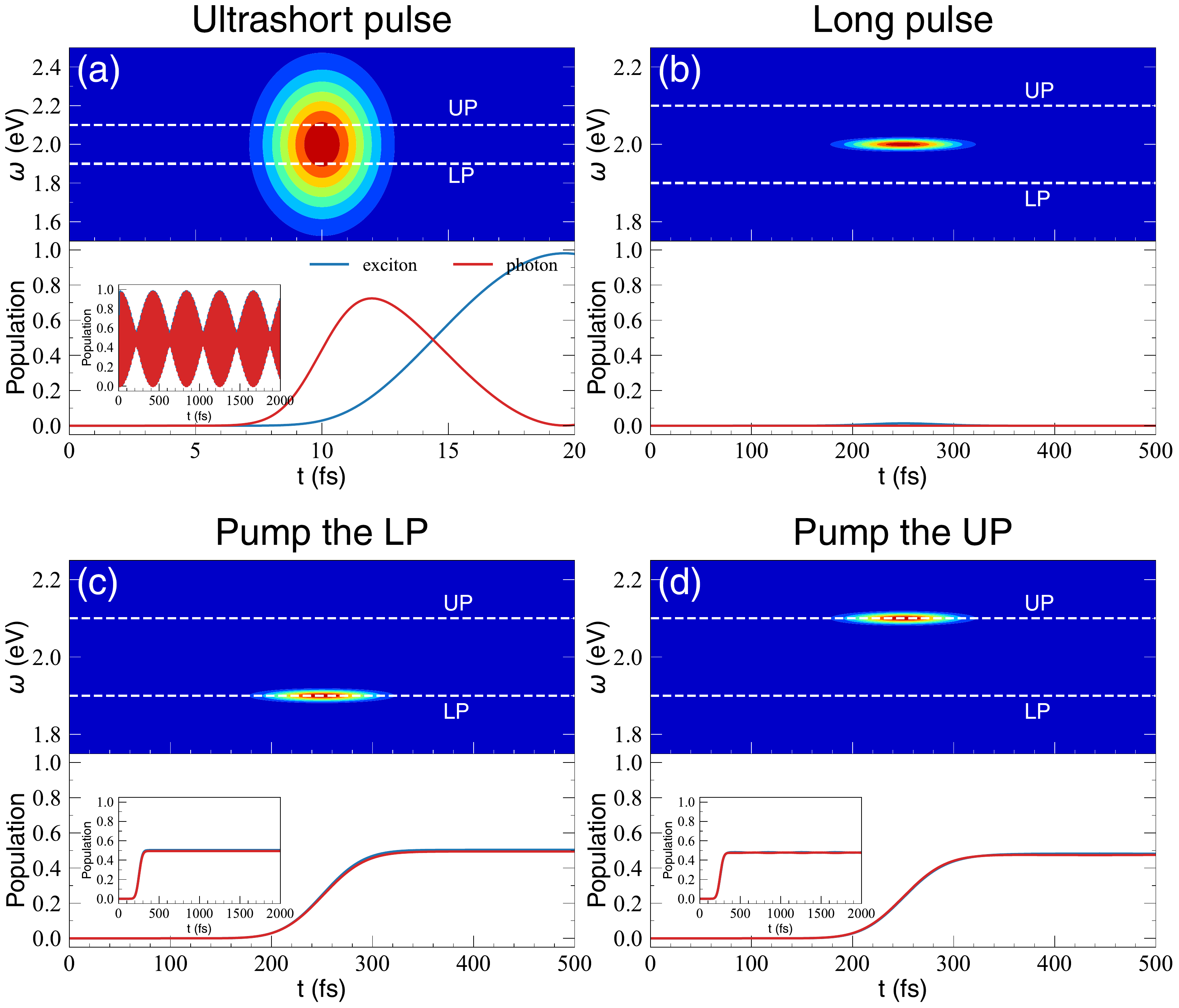}
    \caption{The same as Fig.~2 of the main text but switched to the exciton-photon number representation. }
    \label{fig:s1}
\end{figure}

\begin{figure}
    \centering
    \includegraphics[width=0.6\linewidth]{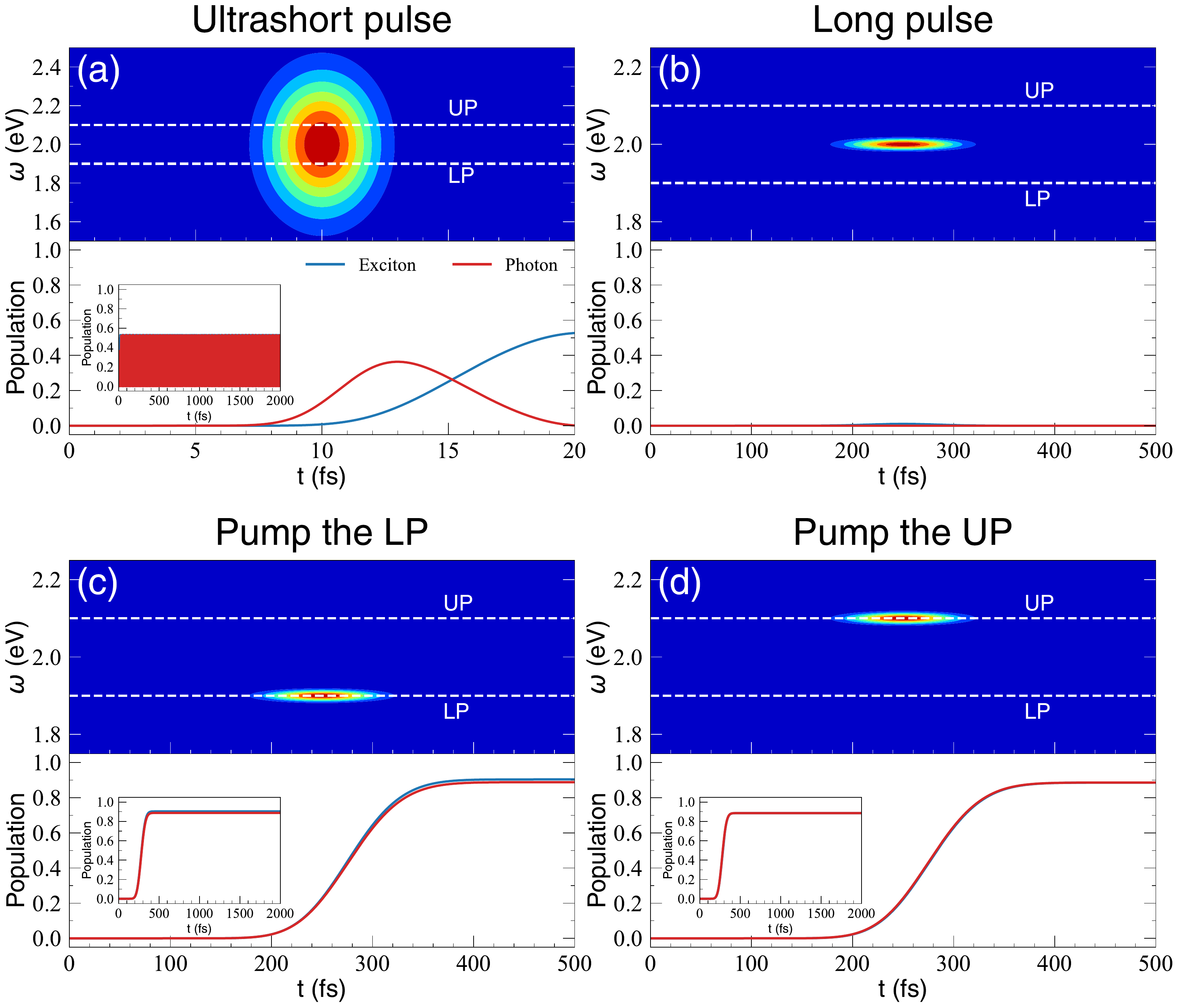}
    \caption{The same as Fig.~\ref{fig:s1} but obtained with MF dynamics. The field amplitudes are: (a) $\lambda_F =$ 0.2 eV, (b)-(d) $\lambda_F =$ 0.02 eV. }
    \label{fig:s2}
\end{figure}

Following Sec.~III-A of the main text, we explore the field-driven molecular polariton dynamics with various driving fields under the MF representation. 

\subsubsection{Mean-field normal modes}
Within the MF framework, true quantum polariton eigenstates are absent due to the neglect of light--matter entanglement. Instead, the relevant excitations correspond to the \emph{normal modes} of the coupled cavity--polarization dynamics. In the weak-excitation regime, 
and neglecting dissipation and nuclei motion, the linearized MF equations for the cavity amplitude $\alpha(t)$ and collective coherence $\sigma(t)$ read (c.f. Eq.~\ref{eq:MF-EOM})
\begin{equation}
\dot \alpha = -i\omega_\mathrm{c} \alpha - i g_{\mathrm c}N\sigma, 
\qquad
\dot \sigma = -i\omega_0 \sigma - i g_{\mathrm c} \alpha.
\end{equation}
Seeking normal-mode solutions of the form $\alpha(t),\sigma(t)\propto e^{-i\omega t}$ leads to the eigenvalue problem
\begin{equation}
\omega 
\begin{pmatrix}
\alpha \\
\sigma
\end{pmatrix}
=
\begin{pmatrix}
\omega_\mathrm{c} & Ng_{\mathrm c} \\
g_{\mathrm c} & \omega_0
\end{pmatrix}
\begin{pmatrix}
\alpha \\
\sigma
\end{pmatrix},
\end{equation}
which yields the characteristic equation
\begin{equation}
(\omega_\mathrm{c}-\omega)(\omega_0-\omega) - g_{\mathrm c}^2 N = 0.
\end{equation}
The resulting normal-mode frequencies are therefore
\begin{equation} \label{eq:MF-nm}
\omega_\pm
=
\frac{\omega_\mathrm{c}+\omega_0}{2}
\pm
\sqrt{g_{\mathrm c}^2 N + \frac{(\omega_\mathrm{c}-\omega_0)^2}{4}}.
\end{equation}
On resonance ($\omega_\mathrm{c}=\omega_0$), this reduces to $\omega_\pm=\omega_\mathrm{c} \pm g_{\mathrm c}\sqrt{N}$, giving a Rabi splitting $\Omega=2g_{\mathrm c}\sqrt{N}$. These MF normal-mode energies coincide with the quantum polariton eigenenergies obtained in the single-excitation Tavis--Cummings model. 

\subsubsection{Numerical results in the exciton--photon number basis}
Since the MF description is formulated solely in terms of collective excitonic and photonic DOF, a direct comparison with the SE dynamics requires expressing the SE results in the same representation. 
To this end, Figure~\ref{fig:s1} presents the same field-driven dynamics as Fig.~2 of the main text, but projected onto the exciton--photon number basis. 
Figure~\ref{fig:s2} shows the corresponding results obtained directly from MF dynamics under similar driving conditions, where the field amplitudes are $\lambda_F =$ 0.1 eV for panel (a), and $\lambda_F =$ 0.01 eV for panels (b)-(d), respectively.
In terms of exciton and photon populations, one sees that under spectrally narrow excitation (Figs.~\ref{fig:s2}b-d), the long pulse predominantly excites one of the two MF normal modes in Eq.~\ref{eq:MF-nm}, and the MF results show good qualitative agreement with the SE description, mimicking selective excitation of a polariton branch in the SE theory.
Nevertheless, the resulting vibration excitation dynamics under MF description does not agree with SE (absence of the resonance effect), as shown later in Fig.~\ref{fig:s3}. 

Under ultrashort-pulse excitation, pronounced differences emerge in the light--matter beating dynamics by comparing Fig.~\ref{fig:s2}a with Fig.~\ref{fig:s1}a insets. 
In the SE dynamics, broadband excitation simultaneously populates both upper and lower polariton branches, thereby generating genuine UP--LP coherence at the quantum-state level. 
As a consequence, the population dynamics exhibits two distinct temporal features (see inset of Fig.~\ref{fig:s1}a): rapid oscillations at the light--matter coupling frequency $\Omega$ (or in terms of period $T = 2\pi / \Omega$), superimposed on a slower envelope (with temporal width of approximately 400~fs, which is exactly the UP--LP oscillation period $T = 2 \pi \sqrt{N} / c_\nu$) originating from coherent interference between the two polaritonic eigenstates.
By contrast, the corresponding MF result in Fig.~\ref{fig:s2}a displays only a single dominant oscillation frequency associated with collective light--matter exchange $\Omega$, and the long-time envelope observed in Fig.~\ref{fig:s1}a is absent. 
This qualitative difference reflects the fact that, within the MF framework, polariton coherence does not exist as an explicit superposition of quantum eigenstates. Because this beating does not correspond to operator-level UP--LP coherence, interference between distinct polaritonic branches does not emerge as a separate slow timescale in the MF population dynamics. Instead, coherence is encoded only at the level of classical coupled oscillators, that both MF normal modes in Eq.~\ref{eq:MF-nm} are excited under a broadband pulse excitation, leading to beatings between the MF normal modes and exhibiting as a single dominant oscillation frequency associated with collective light--matter exchange $\Omega$ shown in Fig.~\ref{fig:s2}a. 
This is a classical analogy of the actual quantum polariton coherence. Later we will show that this classical light--matter beating dynamics at frequency $\Omega$ is crucial for resonant vibration excitation (analogous to the quantum SE results in Fig.~3 of the main text), see Fig.~\ref{fig:s5} and related discussions. 

\subsection{``Polariton'' induces vibration excitation} \label{sec:MF-resonance}
Following Sec.~III-B of the main text, we explore the resulting vibration excitation under field-driven polariton dynamics. 
Again, within the MF framework, true quantum polariton eigenstates are absent due to the lack of light--matter entanglement. Instead, the optical response is governed by the normal modes of the coupled cavity--polarization dynamics. The concept of ``polaritons'' under the MF framework thus refers to the normal modes from linearized MF dynamics (see Eq.~\ref{eq:MF-nm}). 
We show below that the MF normal modes beating is crucial for driving vibration excitation. 

\subsubsection{MF normal modes beating drives vibration excitation}
Under broadband excitation, both MF normal modes with frequencies $\omega_\pm$ (see Eq.~\ref{eq:MF-nm}) are populated, leading to temporal beating at the frequency difference $\Omega = \omega_+ - \omega_-$. This beating manifests in the excited-state population $P_{\mathrm{ex}}(t)$ (see Eq.~\ref{eq:Pex_Pph}) as an oscillatory component in short-time regime (before the vibronic response significantly feeds back on the light--matter oscillation)
\begin{equation} \label{eq:MF-static_oscillate}
P_{\mathrm{ex}}(t) \sim P + \delta P \cos(\Omega t),
\end{equation}
which acts as a time-dependent driving force on the vibrational coordinate.
To see this explicitly, consider the MF vibronic Hamiltonian for a single molecule,
\begin{equation}
\hat H_{\nu} 
= \nu \hat b^\dagger \hat b 
+ c_\nu |e\rangle\langle e| (\hat b + \hat b^\dagger),
\end{equation}
where the vibrational mode may be treated either quantum mechanically or classically. 
For quantum nuclei, the Heisenberg equation for the expectation value 
$Q(t) \equiv \langle \hat b + \hat b^\dagger \rangle$ gives
\begin{equation}
\ddot Q(t) + \nu^2 Q(t) 
= - 2 c_\nu \nu\, P_{\mathrm{ex}}(t),
\end{equation}
which is identical to the classical driven oscillator equation. 
Substituting the oscillatory component of $P_{\mathrm{ex}}(t)$ yields
\begin{equation} \label{eq:driven-oscillator}
\ddot Q + \nu^2 Q 
= - 2 c_\nu \nu \, \delta P \cos(\Omega t).
\end{equation}
Thus, when $\nu \approx \Omega$, the vibration experiences resonant driving, resulting in enhanced population of the $\nu=1$ level. 
Therefore, although MF does not contain operator-level UP--LP coherence, the classical beating between MF normal modes provides an effective oscillatory force at the Rabi frequency, producing a sharp resonance in the vibrational response. This mechanism constitutes the mean-field counterpart of the quantum polariton-branch-changing resonance captured in the SE theory.

\begin{figure}[htbp]
    \centering
    \includegraphics[width=1.0\linewidth]{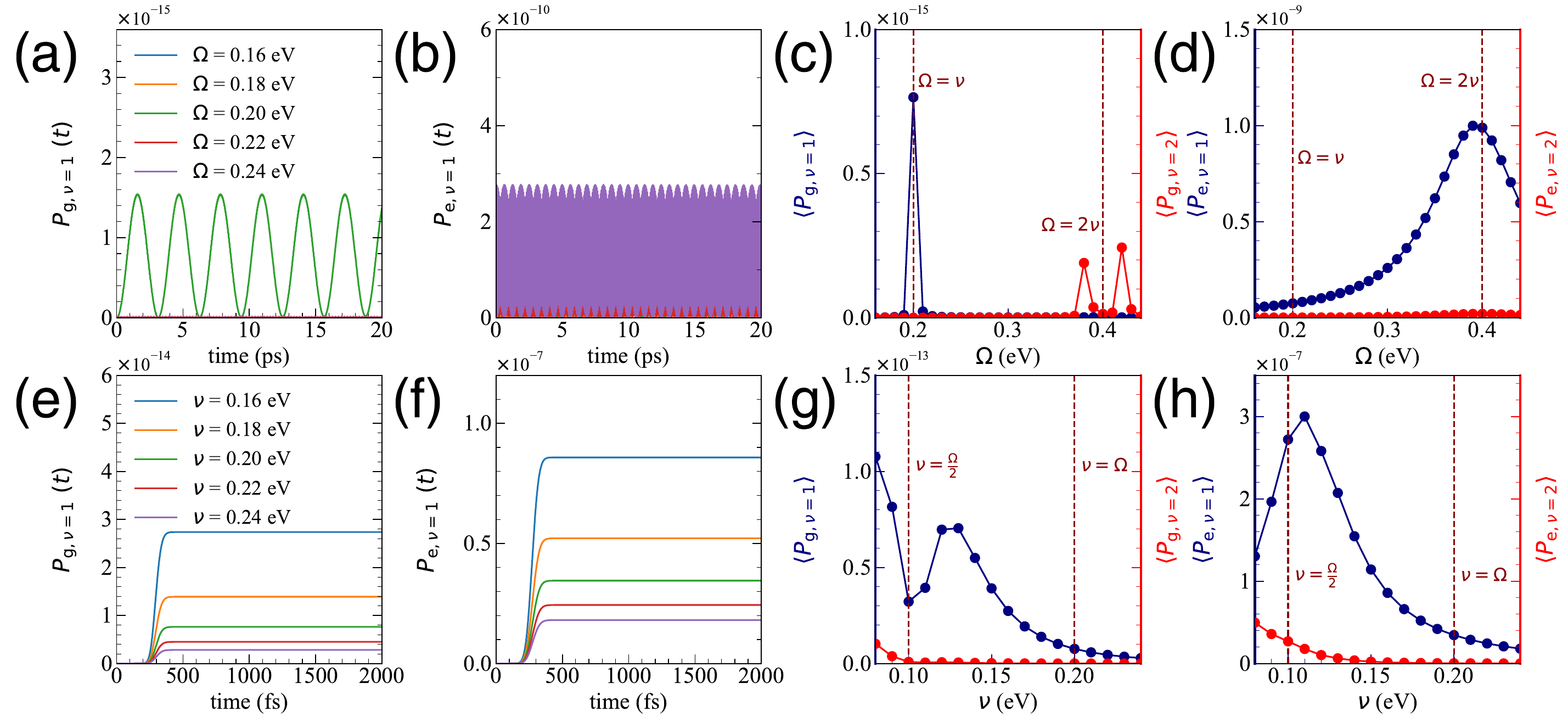}
    \caption{The same as Fig.~3 of the main text (and Fig.~\ref{fig:s10}), but obtained with MF dynamics. }
    \label{fig:s3}
\end{figure}

To demonstrate the above mechanism, Figure~\ref{fig:s3} presents the field-driven vibronic dynamics under ultrashort [panels (a)-(d)] and long pulse excitation [panels (e)-(h)], respectively. The figure corresponds directly to Fig.~3 of the main text but is obtained using the MF dynamics. Detailed analyses are as following.

\subsubsection{Broadband excitation drives both MF normal modes -- beating occurs} 
Under an ultrashort, broadband pulse excitation (that both MF normal modes are excited), the MF normal mode beating acts as a time-dependent driving force on the vibrational coordinate, as discussed in the previous section. One observes that the resonance of the ground-state vibrational response $P_{\mathrm{g},\nu=1}$ is extremely sharp (due to the nature of resonant driven oscillator), as shown in Fig.~\ref{fig:s3}a and \ref{fig:s3}c. 
This feature represents a mean-field analogue of the quantum resonant vibrational excitation of $P_{\mathrm{E},\nu=1}$ reported in Fig.~3 of the main text, which corresponds to the resonant UP $\leftrightarrow$ LP$+1\nu$ transition in the SE description.

In contrast, the resonance feature of $\langle P_{\mathrm{e},\nu=1}\rangle$ around $\Omega = 2\nu$ in Fig.~\ref{fig:s3}d is significantly broader than that of $\langle P_{\mathrm{g},\nu=1}\rangle$ in Fig.~\ref{fig:s3}c. 
Within the SE framework, this resonance can be interpreted as the UP $\leftrightarrow$ DS$+1\nu$ transition, where the dense manifold of dark states leads to an effectively broadened response. 
The much sharper peak observed in $\langle P_{\mathrm{g},\nu=1}\rangle$ reflects a more selective vibrational excitation channel that is not accompanied by a large degeneracy of electronic configurations.
Furthermore, $\langle P_{\mathrm{e},\nu=2}\rangle$ in Fig.~\ref{fig:s3}c exhibits a pair of splitted peak around $\Omega = 2\nu$. 

\subsubsection{Resonance peak shift due to vibronic coupling induced frequency renormalization} 
We emphasize that imposing the bare resonance condition 
$\Omega = \nu$ does not guarantee true vibrational resonance once vibronic coupling is included. 
The coupling to vibrational sidebands renormalizes the effective bright-sector normal-mode frequencies through a vibrational self-energy, so that the relevant oscillation frequency governing the collective dynamics becomes
\begin{equation}
\Omega_{\mathrm{eff}} \neq \Omega .
\end{equation}
As a result, even when $\Omega=\nu$ is enforced at the bare level, the system may in fact be slightly detuned from resonance. 
This renormalization can be characterized by an effective detuning
\begin{equation}
\Delta_{\mathrm{eff}} = \Omega_{\mathrm{eff}} - \nu,
\end{equation}
which controls the long-time vibrational dynamics.

\begin{figure}[htbp]
    \centering
    \includegraphics[width=1.0\linewidth]{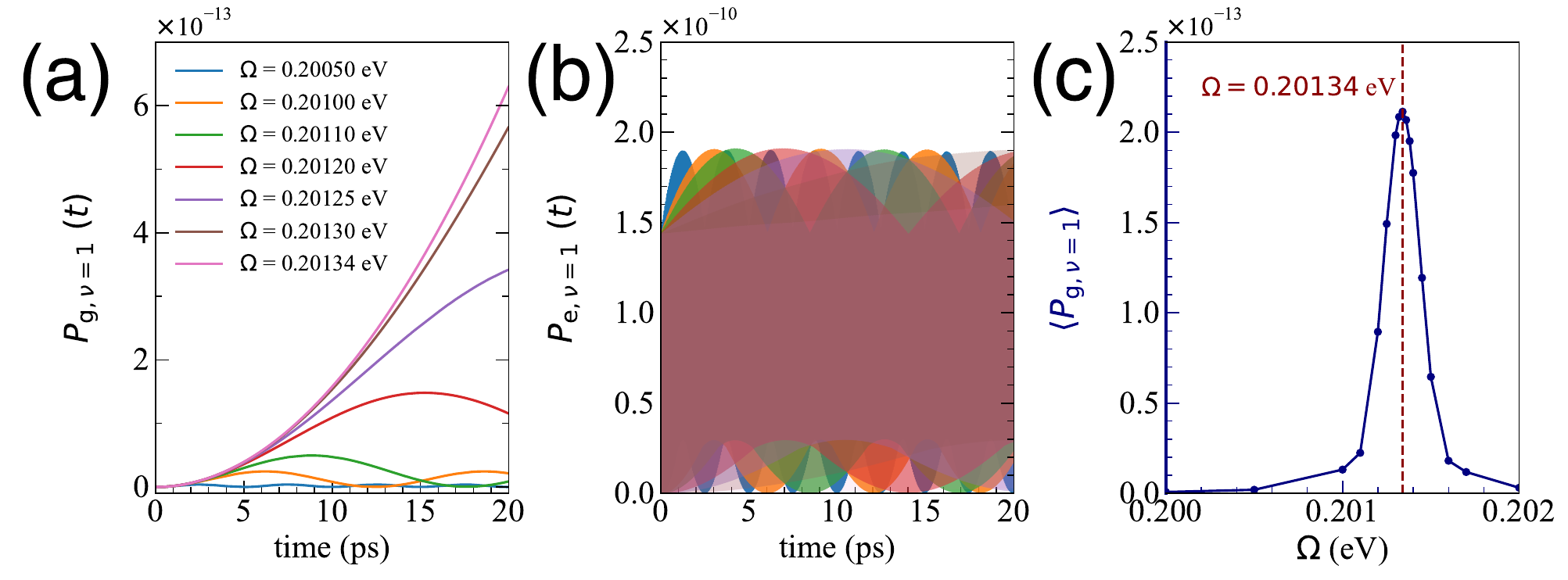}
    \caption{Fine tuning of the Rabi splitting to reach the renormalized resonance condition. (a) Dynamics of the vibrational population in the ground electronic state, $P_\mathrm{g,\nu=1}$, for several values of the bare Rabi splitting $\Omega$ in the vicinity of resonance. As $\Omega$ approaches $0.20134~\mathrm{eV}$, the vibrational response evolves from slowly modulated beating to monotonic secular growth, indicating restoration of the true resonance condition. (b) Corresponding excited-state vibrational population $P_{\mathrm e,\nu=1}(t)$. While the oscillatory envelope changes with $\Omega$, the overall amplitude remains comparable across the scanned values. (c) Time-averaged vibrational population $\langle P_{\mathrm g,\nu=1}\rangle$ as a function of $\Omega$, exhibiting a sharp maximum near $\Omega=0.20134~\mathrm{eV}$ (red dashed line), consistent with the renormalized resonance condition.
}
    \label{fig:s4}
\end{figure}

Physically, the renormalization arises from virtual coupling to vibronic sidebands. 
Although the cavity is tuned to the $0$--$0$ electronic transition, the bright-sector normal modes (located at $\omega_\pm \simeq \omega_0 \pm \Omega/2$) couple perturbatively to the first vibrational sideband near $\omega_0+\nu$. 
These off-resonant processes generate a vibrational self-energy that shifts the effective collective oscillation frequency from $\Omega$ to $\Omega_{\mathrm{eff}}$.
A semi-quantitative estimate can be obtained by integrating out the first vibronic sideband within second-order perturbation theory. 
In the displaced-oscillator (polaron) picture, the excited-state vibrational eigenstates are 
$|\tilde n\rangle = D(d)|n\rangle$, 
with displacement $d=c_\nu/\nu$ and Huang--Rhys factor $S=d^2=(c_\nu/\nu)^2$. 
The light--matter coupling between $|g,0\rangle$ and $|e,\tilde n\rangle$ is weighted by Franck--Condon overlaps, 
$\langle 0|D(d)|0\rangle=e^{-S/2}$ for the $0$--$0$ line and 
$\langle 0|D(d)|1\rangle=-\sqrt{S}\,e^{-S/2}$ for the first sideband. 
Accordingly, the effective collective coupling to the $0$--$1$ sideband scales as
\begin{equation}
V_{01}
\sim \sqrt{N}\hbar g_{\mathrm c}\sqrt{S}\,e^{-S/2}
= \frac{\hbar\Omega}{2}\sqrt{S}\,e^{-S/2},
\qquad (S\ll 1).
\end{equation}
Since the relevant virtual processes couple the polariton-like pole at 
$\omega_+ \simeq \omega_0 + \Omega/2$ 
to the vibronic sideband at $\omega_0+\nu$, 
the appropriate energy denominator is the polariton--sideband detuning
\begin{equation}
\Delta_{+,01} \approx \hbar(\nu - \Omega/2),
\end{equation}
rather than $\hbar\nu$. 
Second-order perturbation theory then yields a self-energy correction
\begin{equation}
\delta E 
\sim \frac{|V_{01}|^2}{\hbar(\nu - \Omega/2)}
\sim \frac{\hbar\,\Omega^2 S\,e^{-S}}{4(\nu - \Omega/2)} \sim 1~\mathrm{meV},
\label{eq:MF-detuning}
\end{equation}
for the present parameters ($\Omega\simeq\nu$ and $S\ll 1$).
Later, we numerically extracted the effective detuning in MF simulations as $\hbar\Delta_{\mathrm{eff}}\approx 1.34~\mathrm{meV}$ (see Fig.~\ref{fig:s4}c), being consistent with the perturbative analyses above. 
Importantly, $\Omega_{\mathrm{eff}}$ should be interpreted as an effective collective oscillation frequency governing the bright-sector response, rather than as a single eigenvalue of the full vibronic Hamiltonian.

The consequences of this renormalization are clearly seen in the vibrational dynamics described by the driven-oscillator equation
\begin{equation}
\ddot Q(t)+\nu^2 Q(t)
=
F_1 \cos(\Omega_{\mathrm{eff}} t),
\label{eq:driven-oscillator-2}
\end{equation}
where $Q(t)=\langle \hat b+\hat b^\dagger\rangle$ and 
$F_1=-2c_\nu\nu\,\delta P$. The resulting vibrational dynamics are analyzed as following. \\

\noindent
\textbf{(i) Apparent resonance: $\Omega=0.20~\mathrm{eV}$ (actually off-resonant).}

Although the bare condition $\Omega=\nu$ is imposed, vibronic renormalization yields 
$\Omega_{\mathrm{eff}} \neq \nu$. 
Writing $\Delta_\mathrm{eff}=\Omega_{\mathrm{eff}}-\nu$, the solution of Eq.~\eqref{eq:driven-oscillator-2} becomes
\begin{equation}
Q(t)
=
\frac{F_1}{\nu^2-\Omega_{\mathrm{eff}}^2}
\Big[\cos(\Omega_{\mathrm{eff}} t)-\cos(\nu t)\Big]
\propto
\sin(\nu t)\,
\sin\!\left(\frac{\Delta_\mathrm{eff} t}{2}\right).
\end{equation}
Thus the vibrational amplitude exhibits a slow beating envelope with characteristic period
\begin{equation}
T_{\mathrm{beat}} \approx \frac{2\pi}{|\Delta_\mathrm{eff}|},
\end{equation}
leading to the long-time modulation of $P_{\mathrm g,\nu=1}(t)$ observed in Fig.~\ref{fig:s3}a.

\medskip
\noindent
\textbf{(ii) Accurate resonance: $\Omega\sim 0.20134~\mathrm{eV}$.}

When the optical splitting is tuned such that 
$\Omega_{\mathrm{eff}}=\nu$, 
Eq.~\ref{eq:driven-oscillator-2} becomes exactly resonant:
\begin{equation}
\ddot Q + \nu^2 Q
=
F_1 \cos(\nu t).
\end{equation}
In the absence of dissipation, the solution contains a secular term,
\begin{equation}
Q(t)\simeq
\frac{F_1}{2\nu}\, t \sin(\nu t),
\end{equation}
implying quadratic growth of vibrational energy,
\begin{equation}
E_v(t)\propto t^2,
\end{equation}
and hence a monotonic increase of $P_{\mathrm g,\nu=1}(t)$.

Figure~\ref{fig:s4} demonstrates that fine tuning the bare Rabi splitting restores the accurate resonance condition predicted by the driven-oscillator analysis. 
As discussed above, vibronic coupling renormalizes the effective bright-sector oscillation frequency from $\Omega$ to $\Omega_{\mathrm{eff}}$, so that imposing the bare condition $\Omega=\nu$ does not in general yield exact resonance. 
By increasing $\Omega$ slightly to $0.20134~\mathrm{eV}$, the effective condition $\Omega_{\mathrm{eff}}=\nu$ is satisfied, in quantitative agreement with the perturbative estimate in Eq.~\ref{eq:MF-detuning}. 
The numerical results therefore confirm the theoretical analysis presented above.
Fig.~\ref{fig:s4}a shows that near the apparent resonance ($\Omega\approx 0.20~\mathrm{eV}$), the vibrational population exhibits long-time beating envelopes characteristic of a small effective detuning. 
As $\Omega$ approaches $0.20134~\mathrm{eV}$, the beating progressively disappears and the population increases monotonically, consistent with the secular growth expected under exact resonance in the lossless driven-oscillator model.

Fig.~\ref{fig:s4}b indicates that the excited-state vibrational population $P_{\mathrm e,\nu=1}(t)$ undergoes corresponding modifications in its temporal envelope as $\Omega$ is tuned, reflecting the change in effective detuning. 
However, the overall oscillation amplitude remains of the same order of magnitude across the scanned values, indicating that the resonance primarily affects phase coherence and long-time energy accumulation rather than instantaneous excitation strength.

Fig.~\ref{fig:s4}c summarizes the resonance behavior by plotting the time-averaged vibrational population as a function of $\Omega$. 
A pronounced and extremely sharp maximum appears near $\Omega=0.20134~\mathrm{eV}$, marking the renormalized resonance condition. 
It should be noted, however, that under exact resonance and in the absence of dissipation, the vibrational population grows monotonically in time and does not remain bounded during the simulation time -- the numerical simulation truncates at $t_\text{max} = 20$ ps. 
Therefore, the peak shown in panel~(c) should be interpreted qualitatively as an indicator of the resonance location rather than as a strictly converged steady-state maximum.

In summary, the qualitative difference observed in Fig.~\ref{fig:s4} originates from vibronic renormalization of the collective bright-sector splitting, and is very sensitive to frequency fine tuning near resonance due to the nature of driven oscillators. 
Although the bare condition $\Omega = 0.20~\mathrm{eV}$ appears resonant, vibronic self-energy corrections shift the effective oscillation frequency to $\Omega_{\mathrm{eff}} \neq \nu$, resulting in a small but finite detuning and the emergence of slow beating envelopes in the vibrational dynamics. 
By contrast, tuning to $\Omega = 0.20134~\mathrm{eV}$ restores the true resonance condition $\Omega_{\mathrm{eff}} = \nu$, leading to secular (monotonic) vibrational activation in the lossless limit.
We note that a similar resonance peak shift (due to frequency renormalization) is also observed in the SE subspace simulations, but not exhibiting such a sharp resonance as observed in MF. See Fig.~\ref{fig:s12}. 

Last but not the least, we emphasize that in realistic situations where cavity and molecular losses are present, the idealized sharp resonance is broadened. 
Dissipation effectively convolves the intrinsic (delta-function--like) resonance with a Lorentzian envelope of finite linewidth, smoothing the singular secular response into a bounded steady-state peak. 
Under such conditions, the resonance maximum typically appears close to the bare condition $\Omega \approx \nu$, since the small renormalization becomes comparable to or smaller than the dissipative linewidth. 
Therefore, while the lossless simulations reveal the subtle vibronic shift of the resonance condition, in experimentally relevant dissipative regimes the peak position is less sensitive to this fine detuning and appears effectively centered at $\Omega=\nu$. The effect of cavity loss is discussed later in Sec.~\ref{sec:MF-loss}. 

\subsubsection{Narrowband excitation drives one MF normal mode only -- beating is absent}
Under a long-pulse excitation, the resonance behavior of $\langle P_{\mathrm{e},\nu}\rangle$ in Fig.~\ref{fig:s3}h remains qualitatively similar to that in Fig.~\ref{fig:s3}d, whereas $\langle P_{\mathrm{g},\nu}\rangle$ no longer exhibits a pronounced resonance structure, as shown in Fig.~\ref{fig:s3}g. 
This indicates that while the electronic DOF are well captured by collective variables in the large-$N$ limit, local vibrational excitation does not necessarily follow from the collective electronic response.

The underlying reason is that, under long, narrow band excitation, the system is prepared quasi-adiabatically such that only a single MF normal mode $\omega_+$ is selectively populated. In this regime, Rabi oscillations are typically damped and the system relaxes to a steady-state population without exhibiting exciton--photon beating (manifested as the oscillating component in Eq.~\ref{eq:MF-static_oscillate} vanishes), as illustrated in Figs.~\ref{fig:s2}c and \ref{fig:s2}d. Consequently, the MF driving force acting on the vibrational coordinate (see Eq.~\ref{eq:driven-oscillator}) becomes effectively static, thereby suppressing selective vibrational activation. In other words, local vibrational modes are activated only when they experience a resonant, time-dependent electronic population.
In contrast, a short pulse excites both MF normal modes simultaneously, generating a beating at the MF Rabi frequency. This beating induces oscillations of the excited-state population at $\Omega$ and can be viewed as a classical analogue of UP--LP coherence, which transiently drives resonant vibrational activation when $\nu=\Omega$.

\subsubsection{Light--matter beating restores resonant vibration excitation}
\begin{figure}[htbp]
    \centering
    \includegraphics[width=1.0\linewidth]{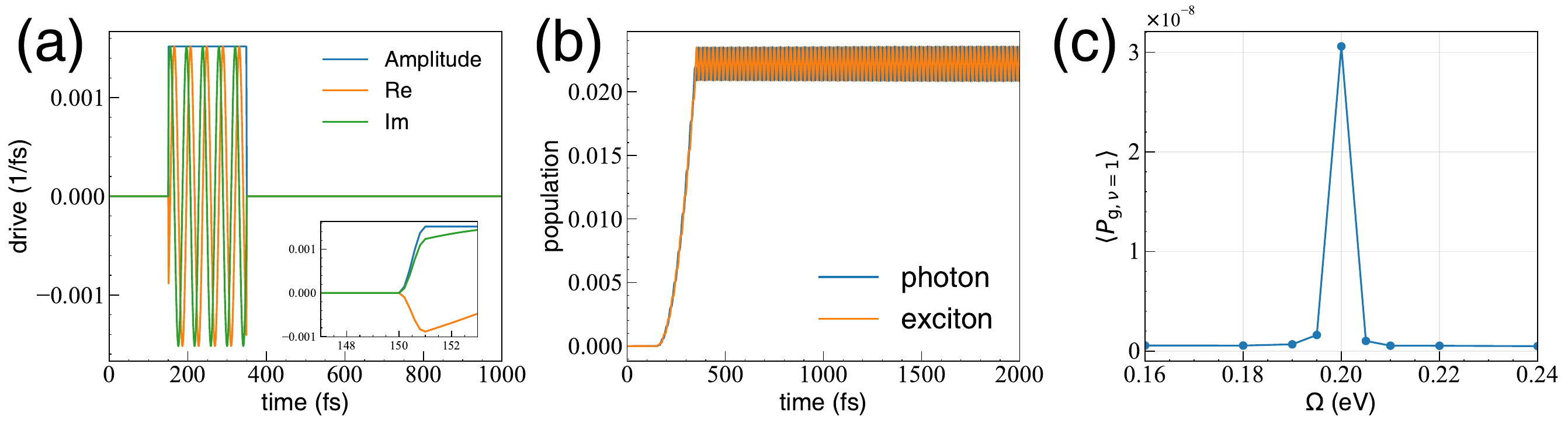}
    \caption{Mean-field resonant vibrational activation induced by short-turn-on excitation. (a) Time profile of the driving field, showing a sharp Tukey-window turn-on ($\sim$1~fs; inset) followed by a long pulse duration ($\sim$200~fs), which excites both MF normal modes simultaneously. (b) Resulting mean-field photon and exciton populations, exhibiting clear oscillations at the MF Rabi frequency. (c) Time-averaged vibrational population in the ground electronic state $\langle P_{\mathrm{g},\nu=1} \rangle$ as a function of the Rabi splitting $\Omega$, showing a pronounced resonance when $\Omega \sim \nu$. }
    \label{fig:s5}
\end{figure}

To further demonstrate this mechanism, Figure~\ref{fig:s5} presents MF dynamics under a short-turn-on, long-duration pulse that is commonly used in Maxwell-based simulations~\cite{bustamante_2026}. A sharp Tukey-window turn-on ($\sim$1~fs, see Fig.~\ref{fig:s5}a inner panel) is employed to simultaneously excite both MF normal modes, while the total pulse duration is kept long ($\sim$200~fs) to sustain the driven dynamics. To be specific, the cavity is driven by a time-dependent classical field with a finite-duration envelope,
\begin{equation}
    E_-(t) = (\lambda_F / 2) \cdot w(t) \cdot e^{- i\omega_\text{p} (t - t_0)},
\end{equation}
where $\lambda_F$ is the field amplitude, $\omega_\text{p}$ is the pulse carrier frequency, and $w(t)$ is a Tukey window that controls the pulse turn-on and turn-off. The Tukey window is defined over a total pulse duration $T_d$ centered at $t_0$ as
\begin{equation}
w(t)=
\begin{cases}
0, & t < t_0-\dfrac{T_d}{2} \ \text{or}\ t > t_0+\dfrac{T_d}{2}, \\[6pt]
\tfrac{1}{2}\!\left[1+\cos\!\left(\pi\!\left(\tfrac{2(t-t_0)}{\zeta T_d}+1\right)\right)\right],
& 0 \le x < \dfrac{\zeta}{2}, \\[6pt]
1, & \dfrac{\zeta}{2} \le x \le 1-\dfrac{\zeta}{2}, \\[6pt]
\dfrac{1}{2}\!\left[1+\cos\!\left(\pi\!\left(\tfrac{2(t-t_0)}{\zeta T_d}-\dfrac{2}{\zeta}+1\right)\right)\right],
& 1-\dfrac{\zeta}{2} < x \le 1,
\end{cases}
\end{equation}
where $x=(t-t_0+T_d/2)/T_d \in [0,1]$ is the normalized time coordinate and $\zeta\in[0,1]$ controls the fraction of the pulse devoted to cosine tapers. In the limit $\zeta\to 0$, the envelope approaches a square pulse, while finite $\zeta$ introduces a smooth turn-on and turn-off that suppresses spectral leakage.
In this work, a very sharp taper ($\zeta \ll 1$) is used, corresponding to an effective turn-on time of order $\sim$1~fs, while the total pulse duration $T_d$ is kept long ($\sim$200~fs). This protocol simultaneously excites both mean-field normal modes during the initial turn-on, generating a coherent beating at the MF Rabi frequency $\Omega$ that persists throughout the pulse duration, as is shown in Fig.~\ref{fig:s5}b.
This MF beating-induced population oscillation serves as a classical analogue of UP--LP coherence and transiently drives resonant vibrational activation when $\nu=\Omega$, as shown in Fig.~\ref{fig:s5}c. This resonance behavior is very similar to Fig.~\ref{fig:s3}c where a short Gaussian pulse is applied.
As a consequence, the pulse duration -- regardless of long or short, is not a key index for vibration excitation, but the MF normal modes beating instead (which provides the driving force), as demonstrated in Eq.~\ref{eq:driven-oscillator}. 

\subsubsection{Classical nuclei still works}
We further demonstrate that the resonant vibrational activation does not depend on whether the nuclear DOF are treated quantum mechanically or classically, as Eq.~\ref{eq:driven-oscillator} does not care about whether the vibration mode is quantum or classical. Within the MF framework, the effective single-molecule Hamiltonian with classical nuclei reads
\begin{align}
    \hat{H}^\mathrm{eff}(\alpha(t))
    &= \hbar \omega_0 |e\rangle\!\langle e|
    + \frac{\nu}{2}\big(P^2+Q^2\big)
    + c_\nu |e\rangle\!\langle e|\, Q
    + \hbar g_\mathrm{c}\!\left[\alpha^{*}(t)|g\rangle\!\langle e|+\alpha(t)|e\rangle\!\langle g|\right],
\end{align}
where $Q$ and $P$ denote the classical vibrational coordinate and momentum, respectively. Their dynamics follows Hamilton’s equations of motion,
\begin{equation}
    \dot Q(t)=\nu P(t), \qquad
    \dot P(t)=-\nu Q(t)-c_\nu\, P_{\mathrm{ex}}(t),
\end{equation}
with $P_{\mathrm{ex}}(t)$ the exciton population defined in Eq.~\ref{eq:Pex_Pph}. Furthermore, the vibration energy of the classical nuclei oscillator is just $E_{\mathrm{vib}}(t)=\hbar\nu \left[Q^2(t)+P^2(t)\right] / 2$. 

\begin{figure}[htbp]
    \centering
    \includegraphics[width=1.0\linewidth]{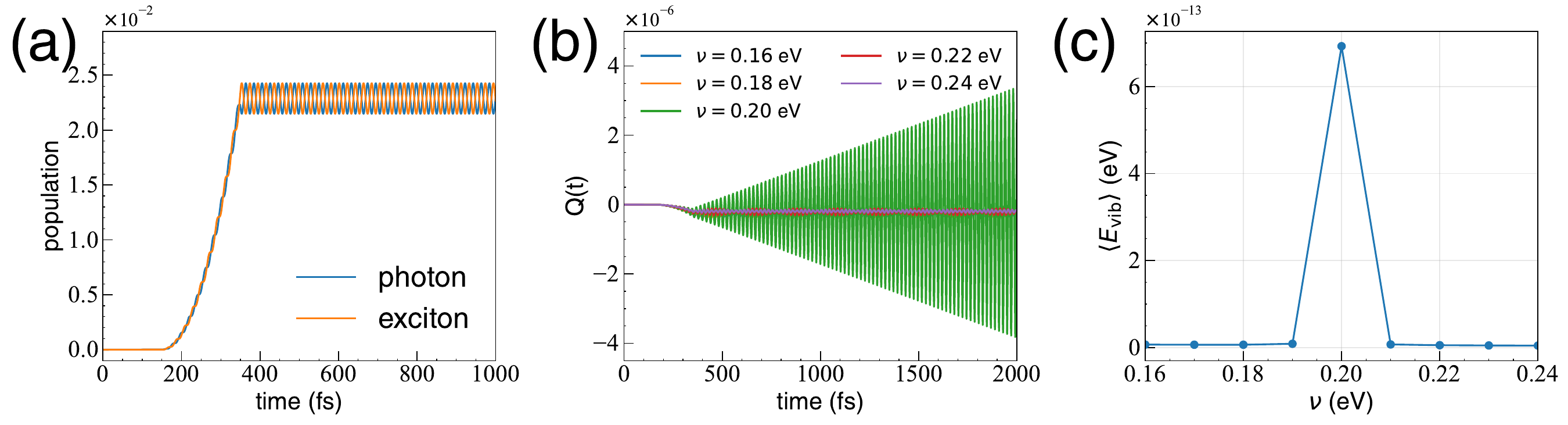}
    \caption{Mean-field resonant vibrational activation with classical nuclei under the same sharp-turn-on pulse as Fig.~\ref{fig:s5}. (a) Mean-field photon and exciton populations. (b) Classical vibrational coordinate $Q(t)$ for different vibrational frequencies $\nu$, showing resonant growth when $\nu=\Omega$. (c) Time-averaged vibrational energy $\langle E_{\mathrm{vib}}\rangle$ as a function of $\nu$, displaying a pronounced resonance at $\nu\sim\Omega$. }
    \label{fig:s6}
\end{figure}

Figure~\ref{fig:s6} presents the corresponding mean-field dynamics when the vibrational mode is treated classically, while keeping the same sharp-turn-on (Tukey-window) pulse protocol as in Fig.~\ref{fig:s5}. 
Fig.~\ref{fig:s6}a shows the mean-field photon and exciton populations (similar as Fig.~\ref{fig:s5}b), which display pronounced oscillations at the MF Rabi frequency. Fig.~\ref{fig:s6}b reports the classical vibrational coordinate $Q(t)$ for several vibrational frequencies $\nu$, revealing a resonant growth only near $\nu\simeq \Omega$. Fig.~\ref{fig:s6}c summarizes this behavior by plotting the time-averaged vibrational energy $\langle E_{\mathrm{vib}}\rangle$ versus $\nu$, which exhibits a sharp peak at $\nu=\Omega$. These results demonstrate that the resonance does not rely on a quantum description of the nuclei, but instead arises from resonant classical driving induced by light--matter beating in the MF dynamics. Moreover, this behavior is consistent with the sharp resonance obtained using Ehrenfest nuclear dynamics in TD-DFTB+Maxwell simulations reported in Ref.~\citenum{bustamante_2026}.

\subsection{Effect of cavity loss} \label{sec:MF-loss}
\begin{figure}[htbp]
    \centering
    \includegraphics[width=0.7\linewidth]{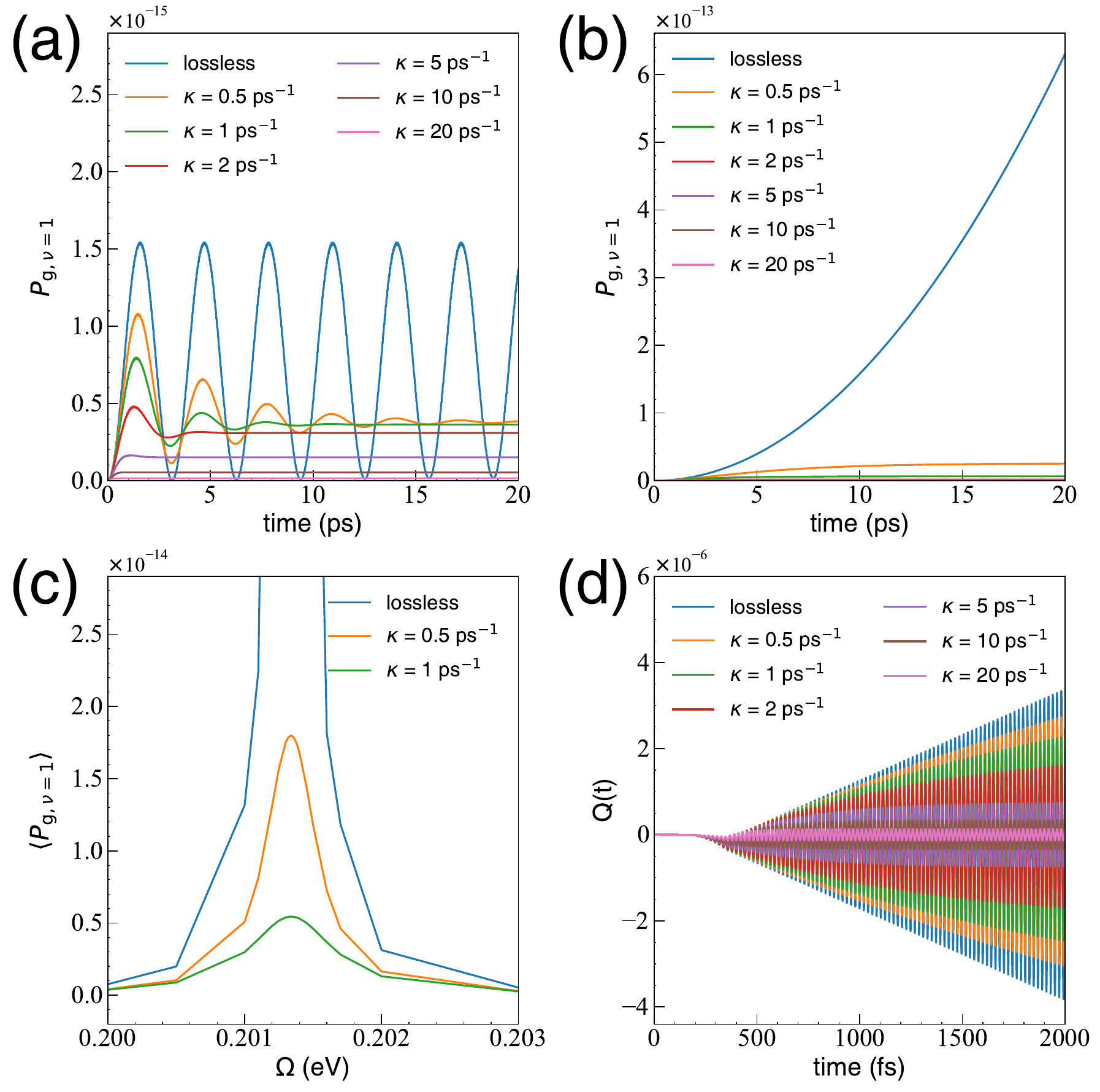}
    \caption{Effect of cavity loss on polariton-driven vibrational excitation under quantum and classical descriptions of nuclear motion. (a) Time-dependent vibration population in the ground electronic state, $P_{\mathrm{g},\nu=1}(t)$, under ultrashort-pulse excitation and $\Omega=\nu = 0.20$ eV, for different cavity loss rates $\kappa$ (the lossless case corresponds to $\kappa=0$). All other parameters are identical to those used in Fig.~\ref{fig:s3}a. (b) The same as (a) but with $\Omega = 0.20134$ eV (accurate resonance condition read from Fig.~\ref{fig:s4}c). (c) Time-averaged vibration population in the ground electronic state $\langle P_{\mathrm{g},\nu=1} \rangle$ as a function of the Rabi splitting $\Omega$ under different cavity loss rates. (d) Corresponding classical nuclear coordinate $Q(t)$ obtained from MF calculations analogous to those in Fig.~\ref{fig:s6}b, showing progressive suppression of the driven nuclear motion and convergence to a steady-state response as $\kappa$ increases. }
    \label{fig:s7}
\end{figure}

Figure~\ref{fig:s7} illustrates the effect of cavity loss on field-driven vibrational activation under both quantum and classical descriptions of nuclear motion. 
Fig.~\ref{fig:s7}a shows the time-dependent ground electronic state vibrational population $P_{g,\nu=1}(t)$ under ultrashort-pulse excitation at the condition $\Omega=\nu = 0.20$ eV, for different cavity loss rates $\kappa$. 
In the absence of cavity loss, the population dynamics exhibits coherent oscillations with a long-time period $\sim 3$ ps (as explained by Eq.~\ref{eq:MF-detuning} and discussions therein). 
Including cavity loss quenches the population dynamics: as $\kappa$ increases, both the amplitude of vibrational excitation and the long-time oscillatory behavior are progressively suppressed, and the $P_{g,\nu=1}(t)$ dynamics eventually approach a stationary value. 

Fig.~\ref{fig:s7}b is the same as Fig.~\ref{fig:s7}a but with $\Omega =$ 0.20134 eV (accurate resonance condition read from Fig.~\ref{fig:s4}c). One sees that the vibration population $P_{g,\nu=1}$ growth monotonically in time rather than off-resonant oscillation shown in Fig.~\ref{fig:s7}a; the inclusion of $\kappa$ further quenches the response, reaching to a steady-state behavior. Increasing $\kappa$ leads to a less significant steady-state population of $P_{g,\nu=1}$. 

Fig.~\ref{fig:s7}c presents the time-averaged vibration population in the ground electronic state, $\langle P_{\mathrm{g},\nu=1} \rangle$, as a function of the Rabi splitting $\Omega$ for different cavity loss rates $\kappa$. As $\kappa$ increases, the resonance peak is progressively suppressed in amplitude and significantly broadened. 
One would expect that when $\kappa$ becomes comparable to $\Omega$--as is typical for plasmonic cavities--the dissipative linewidth exceeds the small vibronic renormalization of the resonance condition. In this regime, the intrinsic shift of the resonance becomes indistinguishable within the broadened spectral response. 
Accordingly, while lossless simulations reveal a subtle vibronic-induced shift of the resonance condition, under experimentally relevant dissipative conditions the peak position becomes largely insensitive to this fine detuning.

Furthermore, Fig.~\ref{fig:s7}d presents the corresponding classical nuclear dynamics $Q(t)$ obtained from mean-field calculations analogous to those in Fig.~\ref{fig:s6}. Consistent with the quantum-nuclear results, increasing cavity loss reduces the amplitude of the driven nuclear motion and leads to a steady-state response at long times.

Together, these results show that photon leakage from the cavity suppresses resonant vibrational activation by damping the underlying light--matter beating and associated coherence.

\subsection{Scaling behaviors with respect to the driving field amplitude}
\begin{figure}[htbp]
    \centering
    \includegraphics[width=1.0\linewidth]{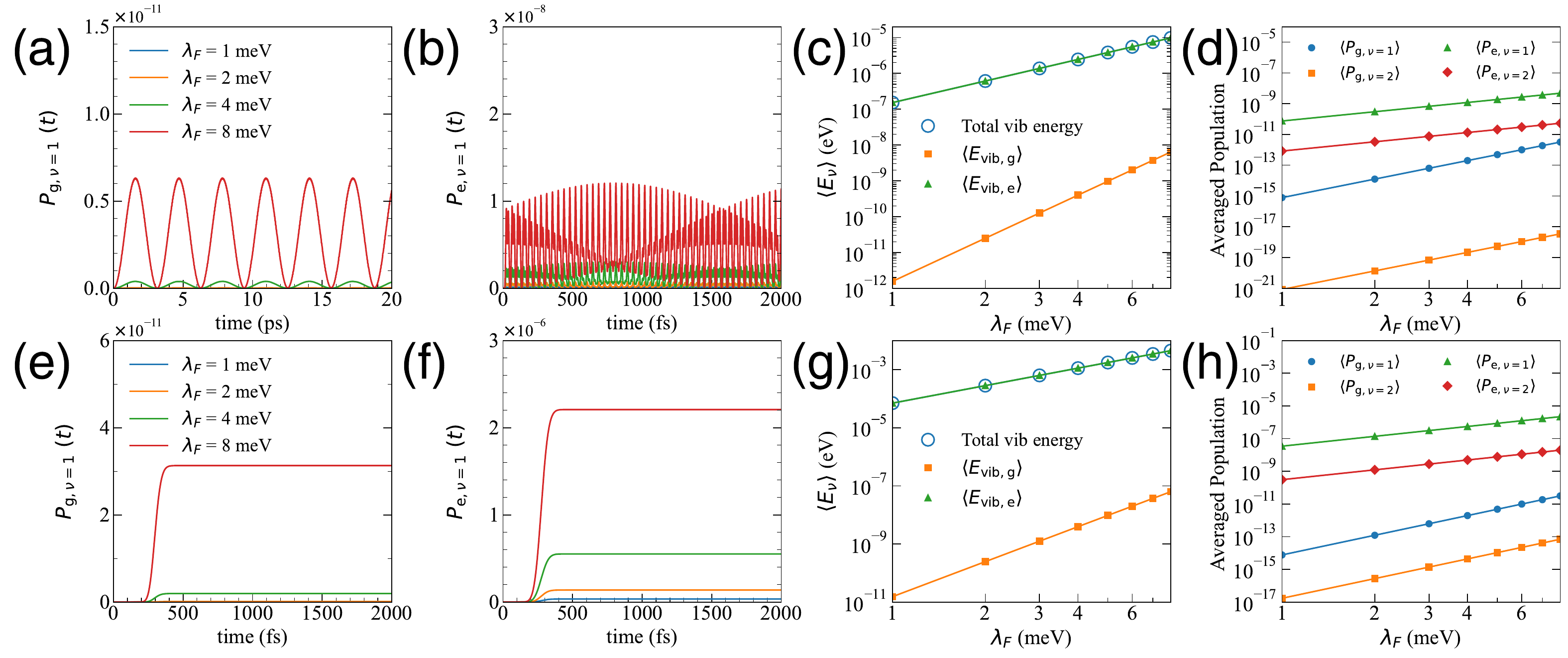}
    \caption{The same as Fig.~4 of the main text but obtained with MF dynamics. }
    \label{fig:s8}
\end{figure}

Figure~\ref{fig:s8} presents the same analysis as Fig.~4 of the main text, but obtained using the MF description. 
Throughout, we fix the condition $\hbar\Omega=\hbar\nu=0.20~\mathrm{eV}$, and change the field amplitude $\lambda_F$. 
One sees that for both ultrashort- (upper panels (a)-(d)) and long-pulse (lower panels (e)-(h)) excitation, the total vibrational energy is dominated by contributions from the excited electronic state, in qualitative agreement with the SE results.

Moreover, the dependence of the time-averaged vibrational energy and populations on $\lambda_F$ in Figs.~\ref{fig:s8}c-d and \ref{fig:s8}g-h follows the same scaling relations as in the SE framework, namely $\langle E_{\mathrm{vib, e}} \rangle$, $\langle P_{\mathrm{e},\nu} \rangle \propto \lambda_F^2$ (or equivalently, $\propto I$ in terms of the field intensity) and $\langle E_{\mathrm{vib, g}} \rangle$, $\langle P_{\mathrm{g},\nu} \rangle \propto \lambda_F^4$ (or equivalently, $\propto I^2$ in terms of the field intensity). 
The linear fitting results for Figs.~\ref{fig:s8}c, \ref{fig:s8}d, \ref{fig:s8}g, and \ref{fig:s8}h on a logarithmic scale are summarized below, with the scaling exponents given by the slopes of the fitted lines.

\begin{table}[!htbp]
    \caption{Linear fitting results for Fig.~\ref{fig:s8}c ($\log\langle E_\nu \rangle$ vs. $\log \lambda_F$).}
    \begin{tabular*}{0.7\columnwidth}{c @{\extracolsep{\fill}} c @{\extracolsep{\fill}} c}
        \hline\hline
        Observable & Fitting Result & Coefficient of Determination \\
        \hline
        Total vib energy & $y=2.000x-6.816$ & $R^2=1.0000$ \\
        Ground vib energy & $y=3.999x-11.807$ & $R^2=1.0000$ \\
        Excited vib energy & $y=2.000x-6.816$ & $R^2=1.0000$ \\ [0.5ex]
        \hline\hline
    \end{tabular*}
    \label{tabs1}
\end{table}

\begin{table}[!htbp]
    \caption{Linear fitting results for Fig.~\ref{fig:s8}d ($\log\mathrm{[Averaged~Population]}$ vs. $\log \lambda_F$).}
    \begin{tabular*}{0.7\columnwidth}{c @{\extracolsep{\fill}} c @{\extracolsep{\fill}} c}
        \hline\hline
        Observable & Fitting Result & Coefficient of Determination \\
        \hline
        $\langle P_{\mathrm{g}, \nu = 1} \rangle$ & $y=3.999x-15.108$ & $R^2=1.0000$ \\
        $\langle P_{\mathrm{g}, \nu = 2} \rangle$ & $y=4.001x-21.078$ & $R^2=1.0000$ \\
        $\langle P_{\mathrm{e}, \nu = 1} \rangle$ & $y=2.000x-10.127$ & $R^2=1.0000$ \\ 
        $\langle P_{\mathrm{e}, \nu = 2} \rangle$ & $y=2.000x-12.080$ & $R^2=1.0000$ \\ [0.5ex]
        \hline\hline
    \end{tabular*}
    \label{tabs2}
\end{table}

\begin{table}[!htbp]
    \caption{Linear fitting results for Fig.~\ref{fig:s8}g ($\log\langle E_\nu \rangle$ vs. $\log \lambda_F$).}
    \begin{tabular*}{0.7\columnwidth}{c @{\extracolsep{\fill}} c @{\extracolsep{\fill}} c}
        \hline\hline
        Observable & Fitting Result & Coefficient of Determination \\
        \hline
        Total vib energy & $y=2.000x-4.153$ & $R^2=1.0000$ \\
        Ground vib energy & $y=4.000x-10.814$ & $R^2=1.0000$ \\
        Excited vib energy & $y=2.000x-4.153$ & $R^2=1.0000$ \\ [0.5ex]
        \hline\hline
    \end{tabular*}
    \label{tabs3}
\end{table}

\begin{table}[!htbp]
    \caption{Linear fitting results for Fig.~\ref{fig:s8}h ($\log\mathrm{[Averaged~Population]}$ vs. $\log \lambda_F$).}
    \begin{tabular*}{0.7\columnwidth}{c @{\extracolsep{\fill}} c @{\extracolsep{\fill}} c}
        \hline\hline
        Observable & Fitting Result & Coefficient of Determination \\
        \hline
        $\langle P_{\mathrm{g}, \nu = 1} \rangle$ & $y=4.000x-14.117$ & $R^2=1.0000$ \\
        $\langle P_{\mathrm{g}, \nu = 2} \rangle$ & $y=4.000x-16.769$ & $R^2=1.0000$ \\
        $\langle P_{\mathrm{e}, \nu = 1} \rangle$ & $y=2.000x-7.462$ & $R^2=1.0000$ \\ 
        $\langle P_{\mathrm{e}, \nu = 2} \rangle$ & $y=2.000x-9.512$ & $R^2=1.0000$ \\ [0.5ex]
        \hline\hline
    \end{tabular*}
    \label{tabs4}
\end{table}

Notably, although $\langle P_{\mathrm{g},\nu=1} \rangle$ exhibits a pronounced enhancement at $\Omega=\nu$ under ultrashort-pulse excitation, as discussed in Fig.~\ref{fig:s3}, this resonance feature is absent for long-pulse excitation. 
Nevertheless, the underlying power-law scaling with $\lambda_F$ remains unchanged, indicating that the observed scaling behavior is a robust consequence that does not rely on resonance conditions or pulse bandwidth.

\subsection{$N$-independence of the vibrational population oscillation period}
Following Sec.~III-D and Fig.~5 of the main text, we further demonstrate that the MF dynamical results are independent of the number of molecules $N$ --- in particular, the oscillation period of the vibrational population --- highlighting a clear discrepancy with the SE description.

\begin{figure}[htbp]
    \centering
    \includegraphics[width=0.5\linewidth]{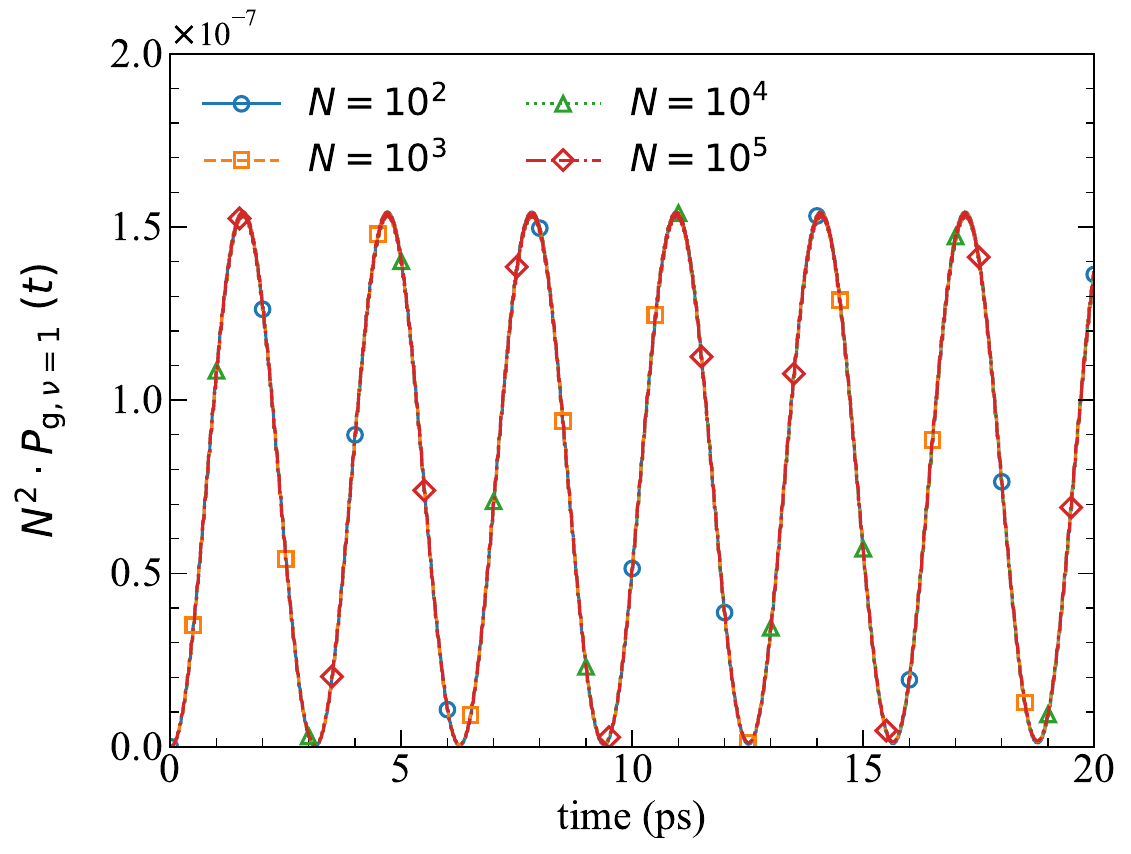}
    \caption{Dynamics of the vibrational population in the ground electronic state multiplied by $N^2$, $N^2 \cdot P_\mathrm{g,\nu=1}(t)$, under a varying number of molecules $N$ ($10^2$, $10^3$, $10^4$, and $10^5$). Here, an ultrashort Gaussian pulse is used (with all other parameters kept the same as Fig.~\ref{fig:s3}), with the field amplitude fixed as $\lambda_F = 1$ meV. All four curves are overlapping with each other. }
    \label{fig:s9}
\end{figure}

Figure~\ref{fig:s9} shows the time-dependent vibration population in the ground electronic state --- multiplied by $N^2$ to balance the amplitude, $N^2 \cdot P_{\mathrm{g},\nu=1}(t)$ --- for different numbers of molecules $N = 10^2, 10^3, 10^4,$ and $10^5$. An ultrashort Gaussian pulse is applied, with the field amplitude held fixed at $\lambda_F = 1$~meV. All four curves are overlapping with each other.  
Apparently, the individual quantity --- oscillation amplitude of $P_\mathrm{g,\nu=1}(t)$ is suppressed as $1/N^2$ as $N$ increases (because $\langle P_{\mathrm{g},\nu} \rangle \propto \lambda_F^4$, and $\lambda_F \propto 1/\sqrt{N}$). 
Within the range of $N$ considered, the oscillation period (as well as the total vibration energy, not shown) remain unchanged as $N$ increases. The observed $N$-independence of the vibrational population oscillation period within the MF framework deviates from the behavior observed in the SE results.
Similarly, one expects $N$-independence of the $N \cdot P_{\mathrm{e},\nu=1}(t)$ dynamics.

\newpage
\section{Additional Information on the Dynamics within the SE Subspace}
In this section, we provide additional information on the dynamics within the SE subspace, including details on the numerical simulations, linear fitting results in Figs.~4 and 5 of the main text, non-resonant behavior of the ground electronic state vibrational population, effect of the number of molecules $N$ under a long-pulse excitation, and resonance peak shift due to vibronic coupling induced frequency renormalization. 

\subsection{Computational details}

\subsubsection{Details on quantum dynamics propagation within the SE subspace}
Within the SE framework, the full system dynamics is described by a time-dependent wavefunction $|\Psi(t)\rangle$ propagated under the field-driven HTC Hamiltonian in Eq.~1 of the main text (within the zero- and one-excitation manifold). 
We perform direct time-dependent Schr\"odinger equation (TDSE) simulations for the total wavefunction $|\Psi(t)\rangle$ to obtain the polariton and vibrational dynamics, {\it i.e.},
\begin{align}
    \hat{H}_\text{HTC}(t) |\Psi(t)\rangle = i\hbar \frac{\partial}{\partial t} |\Psi(t)\rangle. 
\end{align}
Here, $|\Psi(t)\rangle$ is spanned by excitonic, photonic, and vibrational basis. The excitonic and photonic basis is just the TC basis in the single excitation subspace mentioned before, and we use a finite number of vibrational Fock states to represent the $\hat{b}_{n}$ ($\hat{b}^\dagger_{n}$) operators. Furthermore, we initialize the system at the global ground electronic state $|\text{G},0\rangle$, with all vibrations at their ground state, $|\Psi(0)\rangle = |\text{G},0\rangle \bigotimes_{n=1}^N |0_n\rangle$, corresponding to an empty
cavity and molecules prepared in their electronic and vibrational ground states, respectively. 

For the molecular DOF, we take the electronic excitation energy $\hbar\omega_0 = 2.0~\mathrm{eV}$, the vibrational energy $\hbar\nu = 0.20~\mathrm{eV}$, and the electron--vibration coupling strength $c_\nu = 20~\mathrm{meV}$, unless otherwise specified. 
For the cavity mode DOF, we impose the resonance condition $\hbar\omega_\mathrm{c} = \hbar\omega_0 = 2.0~\mathrm{eV}$ and set the collective light--matter coupling strength to $\hbar\Omega = 0.20~\mathrm{eV}$ by default unless specified. 
Dissipative processes, including cavity loss, electronic relaxation, and vibrational damping, are neglected throughout.
Time-dependent driving is introduced via coupling to the cavity field using a Gaussian optical pulse. 
Two pulse protocols are considered: an ultrashort pulse with width $\Delta^{-1} = 2~\mathrm{fs}$ centered at $t_0 = 10~\mathrm{fs}$, and a long pulse with width $\Delta^{-1} = 50~\mathrm{fs}$ centered at $t_0 = 250~\mathrm{fs}$ (as is also stated in the main text). 
Unless otherwise stated, the peak field strength is fixed at $\lambda_F = 1~\mathrm{meV}$.

All TDSE simulations are performed using a Trotter decomposition of the time-evolution operator,
\[
e^{-i\hat{H}t_\text{max}} = \prod_{j} e^{-i\hat{H}\Delta t},
\]
with a time step $\Delta t = 0.025~\mathrm{fs}$. 
The total propagation time is $t_\text{max} = 2~\mathrm{ps}$, which is sufficient to capture both long-time vibrational dynamics and coherence-induced oscillations. 
Unless otherwise specified, the number of molecules is set to $N = 4$. 
The vibrational Hilbert space is truncated at $N_\nu$ vibrational states per molecule, chosen to ensure numerical convergence. 
We find that $N_\nu = 3$ is sufficient near the primary resonance condition $\Omega = \nu$, while $N_\nu = 5$ is required near the secondary resonance $\Omega = 2\nu$.
As a consequence of the direct TDSE formulation, the total Hilbert space dimension grows rapidly with the number of molecules, scaling as $(N+2)\,N_\nu^{\,N}$, which limits simulations to relatively small $N$.

\subsubsection{Observables of interest}
All observables are evaluated as expectation values or overlap probabilities with respect to $|\Psi(t)\rangle$. In particular, we are interested in the following observables,\\

\paragraph{Polaritonic and dark-state populations.}
The two polaritonic eigenstates $|+\rangle$ and $|-\rangle$ are defined in Eq.~7 of the main text.
The corresponding polariton populations are evaluated as reduced overlap probabilities by tracing over the vibrational DOF,
\begin{equation}
P_{\mathrm{UP}}(t)
= \langle\Psi(t)|
\bigl(|+\rangle\langle +|\otimes \hat{\mathcal I}_\nu\bigr)\,
|\Psi(t)\rangle,
\qquad
P_{\mathrm{LP}}(t)
= \langle\Psi(t)|
\bigl(|-\rangle\langle -|\otimes \hat{\mathcal I}_\nu\bigr)\,
|\Psi(t)\rangle.
\end{equation}
Here $\hat{\mathcal I}_\nu$ denotes the identity operator in the full vibrational Hilbert space,
\begin{equation}
\hat{\mathcal I}_\nu
=
\sum_{\nu_1=0}^{N_\nu-1}
\sum_{\nu_2=0}^{N_\nu-1}
\cdots
\sum_{\nu_N=0}^{N_\nu-1}
\bigotimes_{n=1}^N
|\nu_n\rangle\langle \nu_n |.
\end{equation}
Similarly, the populations in the dark-state manifold, ground electronic state are computed as
\begin{equation}
P_{\mathrm{DS}}(t)
=
\sum_k
\mathrm{Tr}\!\left[
\bigl(|\mathrm{D}_k\rangle\langle \mathrm{D}_k|\otimes \hat{\mathcal I}_\nu\bigr)\,
|\Psi(t)\rangle\langle\Psi(t)|
\right],
\end{equation}
\begin{equation}
P_{\mathrm{GS}}(t)
=
\mathrm{Tr}\!\left[
\bigl(|\mathrm{G},0\rangle\langle \mathrm{G},0|\otimes \hat{\mathcal I}_\nu\bigr)\,
|\Psi(t)\rangle\langle\Psi(t)|
\right],
\end{equation}
Representative examples of these population dynamics are shown in Fig.~2 of the main text.

One can alternatively report the excitonic and photonic populations as
\begin{equation}
P_{\mathrm{ex}}(t)
=
\sum_{n=1}^N
\mathrm{Tr}\!\left[
\bigl(|\mathrm{E}_n,0\rangle\langle \mathrm{E}_n,0|\otimes \hat{\mathcal I}_\nu\bigr)\,
|\Psi(t)\rangle\langle\Psi(t)|
\right],
\end{equation}
\begin{equation}
P_{\mathrm{ph}}(t)
=
\mathrm{Tr}\!\left[
\bigl(|\mathrm{G},1\rangle\langle \mathrm{G},1|\otimes \hat{\mathcal I}_\nu\bigr)\,
|\Psi(t)\rangle\langle\Psi(t)|
\right].
\end{equation}
Examples are shown in Fig.~\ref{fig:s1}.

\paragraph{Vibrational energies on the ground and excited electronic and photonic states.}
Let $\hat{n}_j=b_j^\dagger b_j$ denote the vibrational number operator for mode $j$. 
The total vibrational energy is evaluated as
\begin{equation}
E_{\mathrm{vib}}(t)
= \hbar\nu \sum_{j=1}^{N}\langle \Psi(t)|\hat{n}_j|\Psi(t)\rangle .
\end{equation}
To resolve vibrational energy contributions associated with different electronic states, we further define
\begin{align}
E_{\mathrm{vib,G}}(t)
&= \hbar\nu \sum_{j=1}^{N}
\langle \Psi(t)|\hat{P}_{\mathrm{G}} \otimes \hat{n}_j|\Psi(t)\rangle,\\
E_{\mathrm{vib,E}}(t)
&= \hbar\nu \sum_{j=1}^{N}
\langle \Psi(t)|(\hat{P}_{\mathrm{c}}+\hat{P}_{\mathrm{ex}})\otimes \hat{n}_j|\Psi(t)\rangle,
\end{align}
where $\hat{P}_{\mathrm{G}} = |\mathrm{G},0\rangle\langle \mathrm{G},0|$ projects onto the global ground electronic state, and
\(
\hat{P}_{\mathrm{c}}+\hat{P}_{\mathrm{ex}}
= |\mathrm{G},1\rangle\langle \mathrm{G},1|
+ \sum_{n} |\mathrm{E}_n,0\rangle\langle \mathrm{E}_n,0|
\)
spans the excited electronic and photonic states within the single-excitation subspace.
By construction, the total vibrational energy satisfies
\begin{equation}
E_{\mathrm{vib}}(t)=E_{\mathrm{vib,G}}(t)+E_{\mathrm{vib,E}}(t).
\end{equation}

\paragraph{State-resolved vibronic populations.}
The vibrational populations (per molecule) resolved by electronic states and vibrational quantum number are computed as
\begin{equation}
P_{\mathrm{G},\nu}(t)
=\frac{1}{N}\langle \Psi(t)|\hat{\Pi}_{\mathrm{G},\nu}|\Psi(t)\rangle,
\qquad
P_{\mathrm{E},\nu}(t)
=\frac{1}{N}\langle \Psi(t)|\hat{\Pi}_{\mathrm{E},\nu}|\Psi(t)\rangle,
\end{equation}
where the prefactor $1/N$ denotes an average over the $N$ molecular vibrational modes.
Here, the projectors onto vibrational level $\nu$ on the ground, excited electronic and photonic states are defined as
\begin{align}
\hat{\Pi}_{\mathrm{G},\nu}
&= \hat{P}_{\mathrm{G}} \otimes \sum_{j=1}^{N} |\nu_j\rangle\langle \nu_j|,\\
\hat{\Pi}_{\mathrm{E},\nu}
&= (\hat{P}_{\mathrm{c}}+\hat{P}_{\mathrm{ex}}) \otimes \sum_{j=1}^{N} |\nu_j\rangle\langle \nu_j|,
\end{align}
where $|\nu_j\rangle\langle \nu_j|$ projects onto the vibrational Fock state with $\nu$ quanta on mode $j$.
With this definition, $P_{\mathrm{G},\nu}(t)$ and $P_{\mathrm{E},\nu}(t)$ represent vibrational populations per molecule, marginalised over molecular sites, and satisfy $\sum_{\nu} \big(P_{\mathrm{G},\nu}+P_{\mathrm{E},\nu}\big)=\langle \hat{n}\rangle/N$, where $\langle \hat{n}\rangle = \sum_{j=1}^{N}\langle \Psi(t)|\hat{n}_j|\Psi(t)\rangle$.
Representative examples of these quantities are shown in Figs.~3a, 4a of the main text, as well as Figs.~\ref{fig:s10}a, \ref{fig:s10}c. 

\subsubsection{Time-averaging}
Furthermore, time-averaged observables are obtained by averaging the corresponding time-dependent quantities over a post-pulse time window $[t_{\mathrm{p}},t_\text{max}]$, where $t_\text{max}=2~\mathrm{ps}$ is the total propagation time and $t_{\mathrm{p}}$ denotes a plateau time after which the external driving field has sufficiently decayed. 
In practice, we choose $t_{\mathrm{p}}=100~\mathrm{fs}$ for the ultrashort-pulse excitation and $t_{\mathrm{p}}=1000~\mathrm{fs}$ for the long-pulse excitation.
Specifically, the time-averaged total vibrational energy and its decomposition into ground- and excited-state contributions are evaluated as
\begin{align}
\langle E_{\mathrm{vib}} \rangle
&=\frac{1}{t_\text{max}-t_{\mathrm{p}}}\int_{t_{\mathrm{p}}}^{t_\text{max}} dt\,E_{\mathrm{vib}}(t), \\
\langle E_{\mathrm{vib,G/E}} \rangle
&=\frac{1}{t_\text{max}-t_{\mathrm{p}}}\int_{t_{\mathrm{p}}}^{t_\text{max}} dt\,E_{\mathrm{vib,G/E}}(t).
\end{align}
Representative results are shown in Figs.~4b, 4e, and 5c of the main text, as well as Fig.~\ref{fig:s11}c.
Similarly, the time-averaged state-resolved vibrational populations are computed as
\begin{equation}
\langle P_{\mathrm{G/E},\nu} \rangle
=\frac{1}{t_\text{max}-t_{\mathrm{p}}}\int_{t_{\mathrm{p}}}^{t_\text{max}} dt\,P_{\mathrm{G/E},\nu}(t).
\end{equation}
Representative examples are shown in Figs.~4c, 4f of the main text.

\subsection{Linear fitting results in Figs.~4 and 5 of the main text}
We present details of the linear fitting results in Figs.~4 and 5 of the main text. Tables~\ref{tabs5}-\ref{tabs8} present the linear fitting results for Fig.~4, panels (b), (c), (e), (f) of the main text, respectively, with the scaling exponents given by the slopes of the fitted lines. 

\begin{table}[!htbp]
    \caption{Linear fitting results for Fig.~4b of the main text ($\log\langle E_\nu \rangle$ vs. $\log \lambda_F$).}
    \begin{tabular*}{0.7\columnwidth}{c @{\extracolsep{\fill}} c @{\extracolsep{\fill}} c}
        \hline\hline
        Observable & Fitting Result & Coefficient of Determination \\
        \hline
        Total vib energy & $y=2.000x-6.142$ & $R^2=1.0000$ \\
        Ground vib energy & $y=4.000x-16.536$ & $R^2=1.0000$ \\
        Excited vib energy & $2.000x-6.142$ & $R^2=1.0000$ \\ [0.5ex]
        \hline\hline
    \end{tabular*}
    \label{tabs5}
\end{table}

\begin{table}[!htbp]
    \caption{Linear fitting results for Fig.~4c of the main text ($\log\mathrm{[Averaged~Population]}$ vs. $\log \lambda_F$).}
    \begin{tabular*}{0.7\columnwidth}{c @{\extracolsep{\fill}} c @{\extracolsep{\fill}} c}
        \hline\hline
        Observable & Fitting Result & Coefficient of Determination \\
        \hline
        $\langle P_{\mathrm{G}, \nu = 1} \rangle$ & $y=4.000x-16.441$ & $R^2=1.0000$ \\
        $\langle P_{\mathrm{G}, \nu = 2} \rangle$ & $y=4.000x-19.303$ & $R^2=1.0000$ \\
        $\langle P_{\mathrm{E}, \nu = 1} \rangle$ & $y=2.000x-6.046$ & $R^2=1.0000$ \\ 
        $\langle P_{\mathrm{E}, \nu = 2} \rangle$ & $y=2.000x-8.853$ & $R^2=1.0000$ \\ [0.5ex]
        \hline\hline
    \end{tabular*}
    \label{tabs6}
\end{table}

\begin{table}[!htbp]
    \caption{Linear fitting results for Fig.~4e of the main text ($\log\langle E_\nu \rangle$ vs. $\log \lambda_F$).}
    \begin{tabular*}{0.7\columnwidth}{c @{\extracolsep{\fill}} c @{\extracolsep{\fill}} c}
        \hline\hline
        Observable & Fitting Result & Coefficient of Determination \\
        \hline
        Total vib energy & $y=1.962x-3.456$ & $R^2=0.9999$ \\
        Ground vib energy & $y=3.972x-23.477$ & $R^2=1.0000$ \\
        Excited vib energy & $y=1.962x-3.456$ & $R^2=0.9999$ \\ [0.5ex]
        \hline\hline
    \end{tabular*}
    \label{tabs7}
\end{table}

\begin{table}[!htbp]
    \caption{Linear fitting results for Fig.~4f of the main text ($\log\mathrm{[Averaged~Population]}$ vs. $\log \lambda_F$).}
    \begin{tabular*}{0.7\columnwidth}{c @{\extracolsep{\fill}} c @{\extracolsep{\fill}} c}
        \hline\hline
        Observable & Fitting Result & Coefficient of Determination \\
        \hline
        $\langle P_{\mathrm{G}, \nu = 1} \rangle$ & $y=3.972x-23.380$ & $R^2=1.0000$ \\
        $\langle P_{\mathrm{G}, \nu = 2} \rangle$ & $y=3.945x-27.726$ & $R^2=0.9999$ \\
        $\langle P_{\mathrm{E}, \nu = 1} \rangle$ & $y=1.962x-3.361$ & $R^2=0.9999$ \\ 
        $\langle P_{\mathrm{E}, \nu = 2} \rangle$ & $y=1.960x-6.230$ & $R^2=0.9999$ \\ [0.5ex]
        \hline\hline
    \end{tabular*}
    \label{tabs8}
\end{table}

Tables~\ref{tabs9}--\ref{tabs10} present the linear fitting results for Fig.~5, panels (b)--(c) of the main text, respectively, with the scaling exponents given by the slopes of the fitted lines. 
Note that for total vibration energy and $\langle E_{\mathrm{vib, E}} \rangle$ in Table~\ref{tabs10}, the fitted slope is close to zero, consistent with an $N$-independent behavior; in this case the small variance of the data renders the associated $R^2$ value uninformative, being not a meaningful measure of goodness of fit.

\begin{table}[!htbp]
    \caption{Oscillation periods $T$ under different number of molecules $N$ and its linear fitting result for Fig.~5b of the main text ($\log T$ vs. $\log N$).}
    \begin{tabular*}{0.8\columnwidth}{c @{\extracolsep{\fill}} c | @{\extracolsep{\fill}} c @{\extracolsep{\fill}} c}
        \hline\hline
        $N$ & $T$ (fs) \qquad\qquad & Fitting Result & Coefficient of determination \\
        \hline
        1 & 208.48 \qquad\qquad & & \\
        2 & 293.05 \qquad\qquad & & \\
        3 & 358.05 \qquad\qquad & & \\
        4 & 413.84 \qquad\qquad & & \\
        5 & 462.55 \qquad\qquad & $y = 0.495 x + 2.319$ & $R^2 = 1.0000$ \\
        6 & 505.88 \qquad\qquad & & \\
        7 & 546.60 \qquad\qquad & & \\
        8 & 580.46 \qquad\qquad & & \\[0.5ex]
        \hline\hline
    \end{tabular*}
    \label{tabs9}
\end{table}

\begin{table}[!htbp]
    \caption{Linear fitting results for Fig.~5c of the main text ($\log\langle E_\nu \rangle$ vs. $\log N$).}
    \begin{tabular*}{0.7\columnwidth}{c @{\extracolsep{\fill}} c @{\extracolsep{\fill}} c}
        \hline\hline
        Observable & Fitting Result & Coefficient of Determination \\
        \hline
        Total vib energy & $y=-0.001x-6.153$ & $R^2=0.0007$ \\
        Ground vib energy & $y=-1.000x-15.934$ & $R^2=1.0000$ \\
        Excited vib energy & $y=-0.001x-6.153$ & $R^2=0.0007$ \\ [0.5ex]
        \hline\hline
    \end{tabular*}
    \label{tabs10}
\end{table}

\subsection{Non-resonant behavior of the vibration population in the ground electronic state}
\begin{figure}[htbp]
    \centering
    \includegraphics[width=0.7\linewidth]{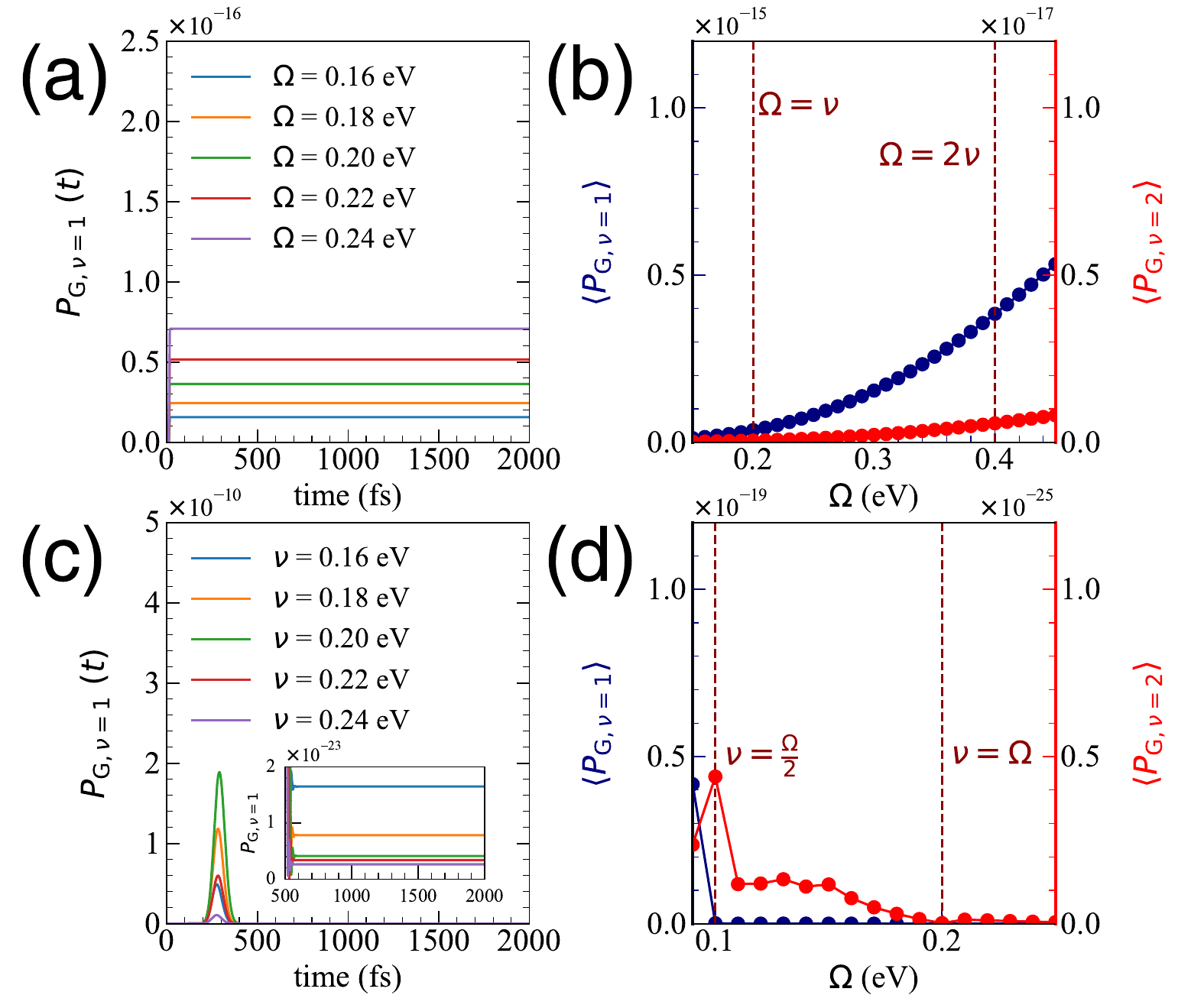}
    \caption{The same as Fig.~3 of the main text, but showing the vibrational population in the ground electronic state. }
    \label{fig:s10}
\end{figure}

Figure~\ref{fig:s10} presents the same type of plot as Fig.~3 of the main text, but focusing on the time-averaged vibrational population in the ground electronic state. 
Specifically, Figs.~\ref{fig:s10}a and \ref{fig:s10}c presents the population dynamics of $P_\mathrm{G, \nu = 1}(t)$, while Figs.~\ref{fig:s10}b and \ref{fig:s10}d display the corresponding averaged vibration population in the ground electronic state, $\langle P_{\mathrm{G},\nu} \rangle$, for ultrashort- (upper panels (a)--(b)) and long-pulse (lower panels (c)--(d)) excitation, respectively.

In contrast to the resonance behaviors of the excited-state vibrational population, $\langle P_{\mathrm{G},\nu} \rangle$ does not exhibit pronounced resonance features at either $\Omega=\nu$ or $\Omega=2\nu$ under either excitation protocol. 
This behavior indicates that vibrational excitation in the ground electronic state remains largely insensitive to polaritonic resonance conditions or Rabi-driving.

We further note that, under long-pulse excitation, the magnitude of $\langle P_{\mathrm{G},\nu} \rangle$ is several orders of magnitude smaller than in the ultrashort-pulse case (as weak as $10^{-23}$ in terms of numerical value). 
As a result, the data in Fig.~\ref{fig:s10}d are more susceptible to numerical noise, leading to a less smooth dependence on $\Omega$ compared to Fig.~\ref{fig:s10}b. 
Importantly, this numerical sensitivity does not affect the qualitative conclusion regarding the absence of resonance enhancement in the vibration population in the ground electronic state.

\subsection{Effect of the number of molecules $N$ under a long-pulse excitation}
\begin{figure}[htbp]
    \centering
    \includegraphics[width=1.0\linewidth]{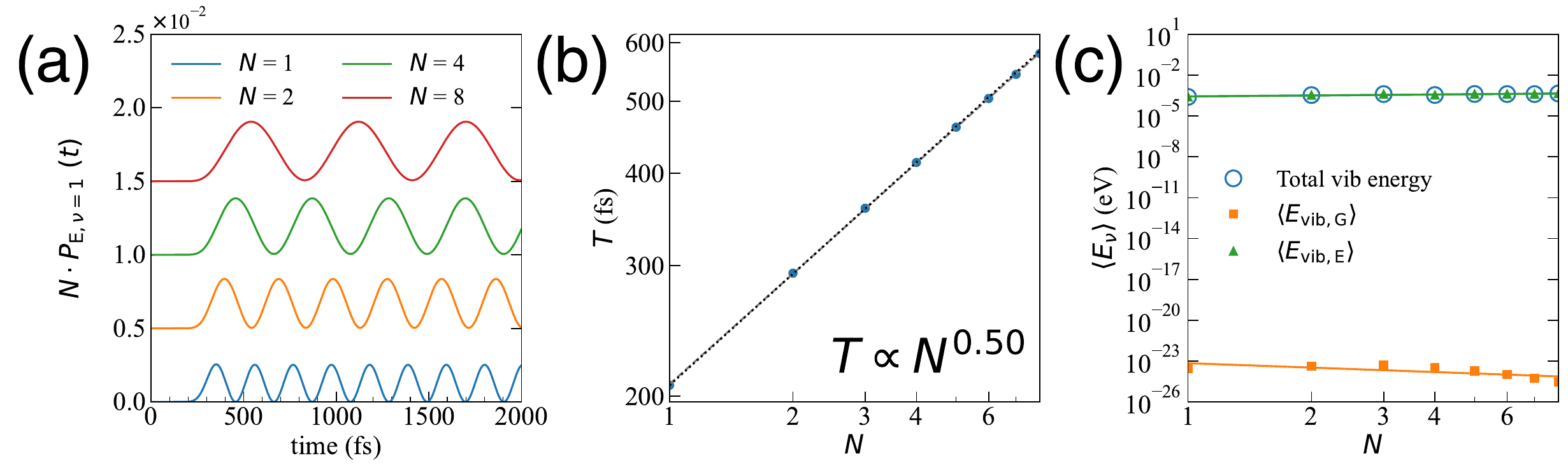}
    \caption{The same as Fig.~5 of the main text, but with a long pulse excitation. }
    \label{fig:s11}
\end{figure}

Figure~\ref{fig:s11} presents the same analysis as Fig.~5 of the main text, but performed under long-pulse excitation. 
As shown in Fig.~\ref{fig:s11}b, the same scaling relation for the oscillation period $T \propto \sqrt{N}$ is observed.
Furthermore, Fig.~\ref{fig:s11}c also exhibit the same scaling relations with respect to the number of molecules $N$. 
The corresponding linear fitting results are summarized in Tables~\ref{tabs11}--\ref{tabs12} as follows, with the scaling exponents given by the slopes of the fitted lines.

\begin{table}[!htbp]
    \caption{Oscillation periods $T$ under different number of molecules $N$ and its linear fitting result for Fig.~\ref{fig:s11}b ($\log T$ vs. $\log N$).}
    \begin{tabular*}{0.8\columnwidth}{c @{\extracolsep{\fill}} c | @{\extracolsep{\fill}} c @{\extracolsep{\fill}} c}
        \hline\hline
        $N$ & $T$ (fs) \qquad\qquad & Fitting Result & Coefficient of determination \\
        \hline
        1 & 206.9 \qquad\qquad & & \\
        2 & 293.1 \qquad\qquad & & \\
        3 & 358.7 \qquad\qquad & & \\
        4 & 413.5 \qquad\qquad & & \\
        5 & 461.5 \qquad\qquad & $y = 0.496 x + 2.317$ & $R^2 = 1.0000$ \\
        6 & 504.6 \qquad\qquad & & \\
        7 & 544.0 \qquad\qquad & & \\
        8 & 580.3 \qquad\qquad & & \\[0.5ex]
        \hline\hline
    \end{tabular*}
    \label{tabs11}
\end{table}

\begin{table}[!htbp]
    \caption{Linear fitting results for Fig.~\ref{fig:s11}c ($\log\langle E_\nu \rangle$ vs. $\log N$).}
    \begin{tabular*}{0.7\columnwidth}{c @{\extracolsep{\fill}} c @{\extracolsep{\fill}} c}
        \hline\hline
        Observable & Fitting Result & Coefficient of Determination \\
        \hline
        Total vib energy & $y=0.236x-3.569$ & $R^2=0.8197$ \\
        Ground vib energy & $y=-1.080x-23.152$ & $R^2=0.5455$ \\
        Excited vib energy & $y=0.236x-3.569$ & $R^2=0.8197$ \\ [0.5ex]
        \hline\hline
    \end{tabular*}
    \label{tabs12}
\end{table}

We note that, under long-pulse excitation, the vibrational populations in the ground electronic state are extremely small in magnitude. 
As a consequence, the corresponding time-averaged ground-electronic state vibrational energy and population are more susceptible to numerical noise, which manifests as a weaker apparent linear dependence on $1/N$ in Fig.~\ref{fig:s11}c (orange symbols). Importantly, this numerical sensitivity does not affect the validity of the scaling relations or the conclusions drawn in Sec.~III-D of the main text.
In contrast, the corresponding quantities in the excited electronic and photonic state exhibit a more robust scaling -- roughly being independent with $N$ (green symbols in Fig.~\ref{fig:s11}c).

\subsection{Resonance peak shift due to vibronic coupling induced frequency renormalization}
Finally, we demonstrate that the tiny resonance peak shift observed in the MF simulations (see Sec.~\ref{sec:MF-resonance} and Fig.~\ref{fig:s4}) also appears in the SE simulations, due to the same reason of vibronic coupling induced frequency renormalization. 

\begin{figure}[htbp]
    \centering
    \includegraphics[width=0.4\linewidth]{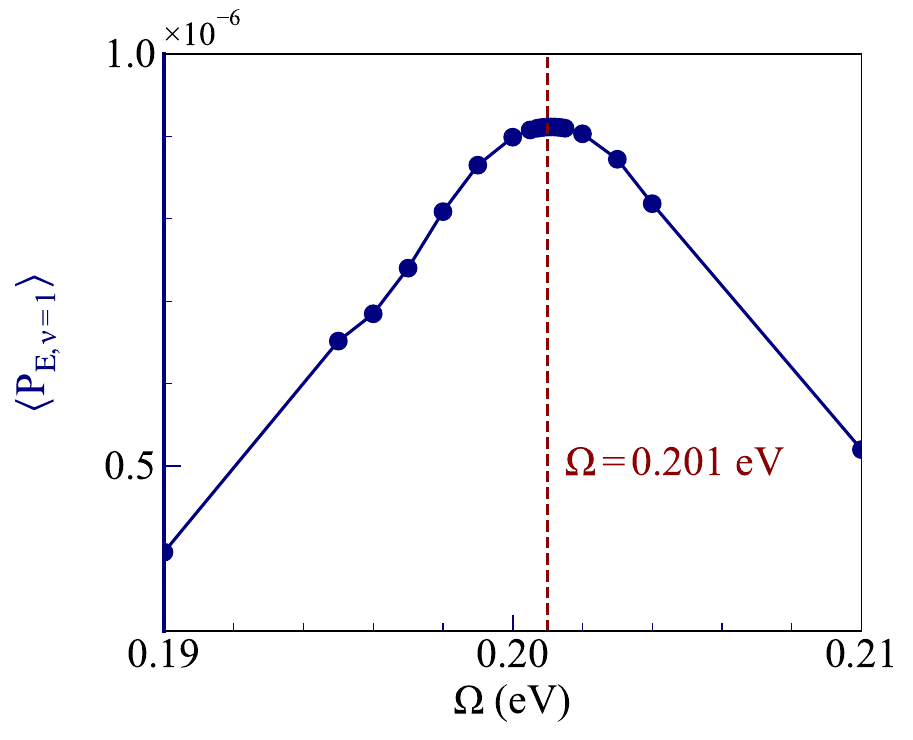}
    \caption{The resonance peak of time-averaged vibration population in the excited electronic and photonic state $\langle P_\mathrm{E,\nu=1} \rangle$, under a fine tuning of the Rabi splitting $\Omega$, in which an ultrashort pulse is applied (the same as Figs.~3a-b of the main text). }
    \label{fig:s12}
\end{figure}

Fig.~\ref{fig:s12} shows the resonance peak of time-averaged vibration population in the excited electronic and photonic state $\langle P_\mathrm{E,\nu=1} \rangle$ under a fine tuning of the Rabi splitting near $\Omega = \nu$, all parameters are kept the same as Figs.~3a-b of the main text. One sees that resonance peak is accurately positioned at $\Omega = 0.201$ eV, slightly blue-shifted from $\nu = 0.20$ eV and is consistent with the amount of shift observed in MF simulations (Fig.~\ref{fig:s4}c).  
Nevertheless, the peak intensity is less sensitive to fine tuning around $\Omega = \nu$ compared to the MF results in Fig.~\ref{fig:s4}c. 

\subsection{Effect of cavity loss on polariton and vibrational population dynamics} \label{sec:SE-loss}

To complement the MF analysis of cavity loss in Sec.~\ref{sec:MF-loss}, we investigate the effect of cavity photon loss directly at the SE level.
We solve the time-dependent Schr\"{o}dinger equation with a non-Hermitian effective Hamiltonian,
\begin{equation}
    \hat{H}_{\mathrm{eff}} = \hat{H}_{\mathrm{HTC}} - \frac{i\kappa}{2}\hat{a}^\dagger\hat{a},
\end{equation}
where $\hat{H}_{\mathrm{HTC}}$ is the Holstein--Tavis--Cummings Hamiltonian defined in the main text and $\kappa$ is the cavity photon loss rate.
The imaginary term causes the norm of the state vector to decay, reflecting photon leakage from the cavity.
All other parameters are kept identical to those in Figs.~3a--b of the main text: $N = 4$ molecules, vibrational truncation $N_v = 4$, cavity and molecular transition frequencies $\omega_\mathrm{c} = \omega_0 = 2.0~\mathrm{eV}$, vibrational frequency $\nu = 0.2~\mathrm{eV}$, bare Rabi splitting $\Omega = 2g_c\sqrt{N} = 0.20~\mathrm{eV}$ (at the Rabi resonance condition $\Omega = \nu$), Holstein coupling $c_{\nu} = 0.02~\mathrm{eV}$, and an ultrashort Gaussian pulse centered at $t_0 = 10~\mathrm{fs}$ with standard deviation $\sigma = 2~\mathrm{fs}$, carrier frequency $\omega_\mathrm{p} = 2.0~\mathrm{eV}$, and field amplitude $\lambda_F = 1~\mathrm{meV}$.
Seven values of the cavity loss rate are considered: $\kappa = 0$ (lossless), $0.5$, $1$, $2$, $5$, $10$, and $20~\mathrm{ps}^{-1}$.

\begin{figure}[htbp]
    \centering
    \includegraphics[width=0.9\linewidth]{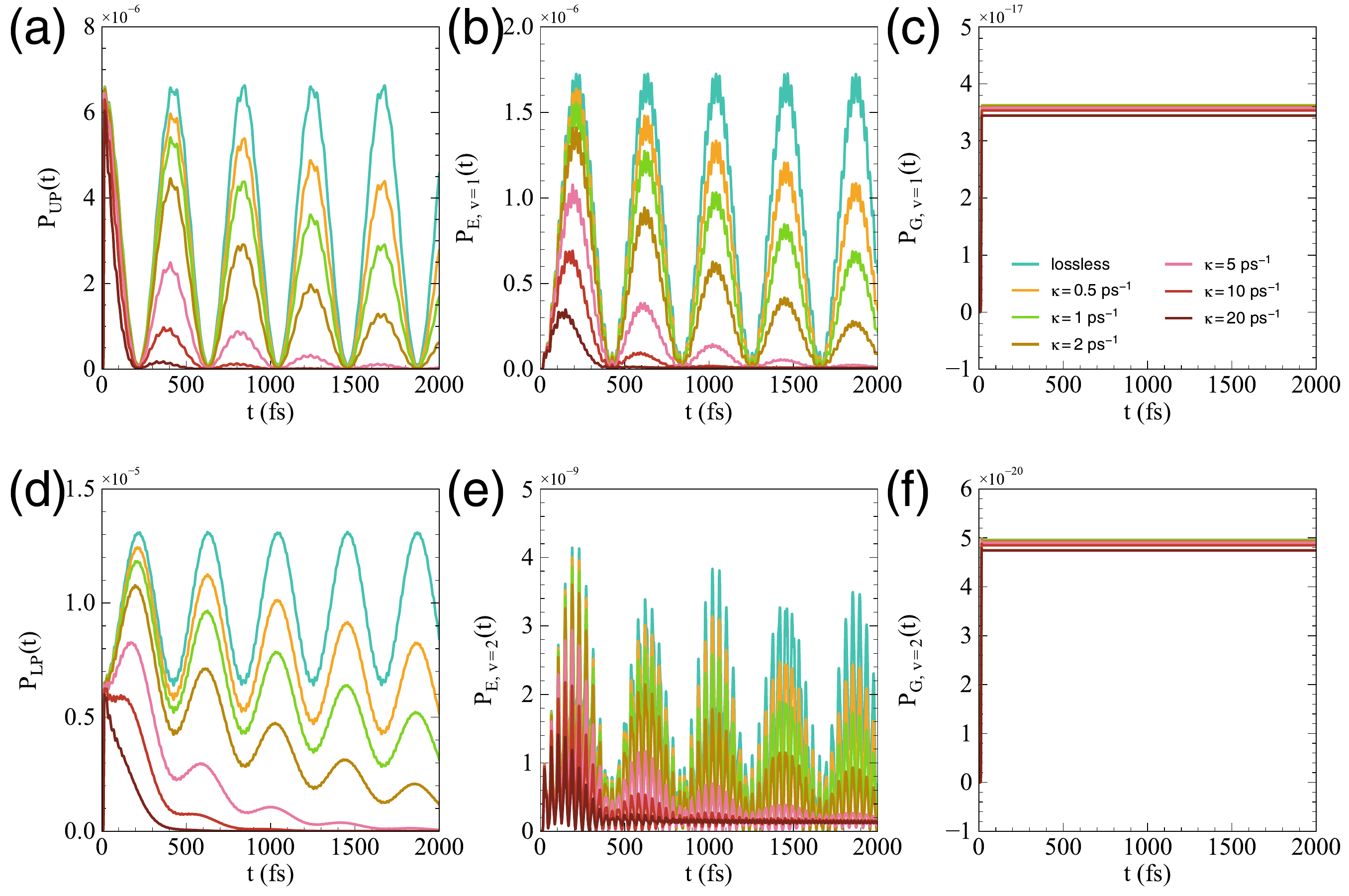}
    \caption{
    Effect of cavity loss on polariton and vibrational population dynamics at the SE level, under ultrashort-pulse excitation with parameters identical to Figs.~3a--b of the main text.
    Seven cavity loss rates $\kappa = 0$ (lossless), $0.5$, $1$, $2$, $5$, $10$, and $20~\mathrm{ps}^{-1}$ are compared.
    (a) Upper-polariton population $P_{\mathrm{UP}}(t)$.
    (b) Excited-state vibrational population at level $\nu=1$, $P_\mathrm{E,\nu=1}(t)$.
    (c) Ground-state vibrational population at level $\nu=1$, $P_\mathrm{G,\nu=1}(t)$.
    (d) Lower-polariton population $P_{\mathrm{LP}}(t)$.
    (e) Excited-state vibrational population at level $\nu=2$, $P_\mathrm{E,\nu=2}(t)$.
    (f) Ground-state vibrational population at level $\nu=2$, $P_\mathrm{G,\nu=2}(t)$.
    }
    \label{fig:s13}
\end{figure}

Figure~\ref{fig:s13} reveals a clear asymmetry in the sensitivity of different observables to cavity loss.
The excited-state and photonic populations --- $P_{\mathrm{UP}}$, $P_{\mathrm{LP}}$, $P_\mathrm{E,\nu=1}(t)$, and $P_\mathrm{E,\nu=2}(t)$ --- are all systematically suppressed as $\kappa$ increases.
Among the polariton populations, $P_{\mathrm{LP}}$ is more strongly quenched: in the lossless case $P_{\mathrm{LP}}^{\max} \approx 1.3\times10^{-5}$ is about twice $P_{\mathrm{UP}}^{\max} \approx 6.6\times10^{-6}$, whereas at $\kappa = 20~\mathrm{ps}^{-1}$ both converge to a comparable value of $\sim\!6\times10^{-6}$ (Figs.~\ref{fig:s13}a,d).
The vibrational populations in the electronically excited manifold, $P_\mathrm{E,\nu=1}(t)$ and $P_\mathrm{E,\nu=2}(t)$, also decrease monotonically with $\kappa$ (Figs.~\ref{fig:s13}b,e).
In sharp contrast, the ground-state vibrational populations $P_\mathrm{G,\nu=1}(t)$ and $P_\mathrm{G,\nu=2}(t)$ remain negligible throughout the entire dynamics, irrespective of $\kappa$ (Figs.~\ref{fig:s13}c,f).
This result is physically transparent: within the HTC model augmented with cavity photon loss only, the Holstein coupling $c_\nu$ drives nuclear motion exclusively in the electronically excited manifold, and there is no pathway to populate ground-state vibrational levels without electronic relaxation.
Consequently, cavity loss alone --- even at large rates $\kappa \gg \Omega$ --- is insufficient to induce ground-state vibrational excitation; molecular dephasing or non-radiative decay channels would be required to achieve this.

Taken together, these results establish that cavity photon loss significantly affects polariton-state and excited-state vibrational population dynamics, while leaving the electronically ground-state nuclear populations essentially unaffected.

\newpage
\bibliography{Ref}